\documentclass[preprint,amssymb,amsmath,aip,cha]{revtex4-1}

\usepackage{graphicx}
\usepackage{amsfonts}
\usepackage{appendix}
\usepackage{bbm}
\usepackage{amsmath}
\usepackage{gensymb}
\usepackage{amssymb}
\usepackage{upgreek}
\usepackage{array}

\begin{document}

\title{The velocity gradient tensor for homogeneous, isotropic turbulence ({HIT}), with explicit consideration of local and non-local effects using a {S}chur decomposition}

\author{Christopher J. Keylock}

\affiliation{$^1$Sheffield Fluid Mechanics Group and Department of Civil and Structural Engineering, University of Sheffield, Mappin Street, Sheffield, S1 3JD, UK}

%
%

\begin{abstract}
A {S}chur decomposition of the velocity gradient tensor (VGT) for homogeneous, isotropic turbulence (HIT) is undertaken and its physical consequences examined. This decomposition permits the normal parts of the tensor (represented by the eigenvalues) to be separated explicitly from the non-normal effects. Given the restricted {E}uler approximation to the VGT dynamics is written in terms of the isotropic part of the pressure Hessian and the invariants of the characteristic equation of the VGT (in turn expressed in terms of the eigenvalues), the non-normal terms are related to the non-local aspects of the dynamics and the anisotropic part of the pressure Hessian. Using a direct numerical simulation of HIT, we show that the norm of the non-normal part of the tensor is of a similar order to the normal part, highlighting the importance of non-local effects. In fact, beneath the discriminant function in a $\mbox{Q}-\mbox{R}$ plot, all enstrophy arises from the non-normal term, meaning that vorticity and intermediate strain eigenvector alignment in this region is an immediate consequence of non-normality. A non-normal term appears in the expressions for both enstrophy and total strain and cancels when calculating the second invariant of the VGT, while the self-amplification of non-normality and the normal straining of non-normality appear in the strain production and enstrophy production equations and cancel when calculating the third invariant. However, these terms are significant for understanding the full VGT dynamics, explaining how flow structures evolve to a disc-like state despite the strain eigenvalues sometimes indicating opposite (rod-like) behaviour, as well as explaining vorticity and strain alignments in HIT.
\end{abstract}

\maketitle

\section{Introduction}
\label{intro}
\subsection{Brief overview}
This paper is concerned with properties of the velocity gradient tensor (VGT) for incompressible, homogeneous, isotropic turbulence (HIT). This is a classical topic in turbulence fluid mechanics as HIT is the testing ground for a great deal of turbulence theory. A recent review paper by \citet{meneveau11} provides a great deal of information on the properties of the VGT and serves as both a basis and a point of departure for this study. The key difference between this study and previous work concerns the form of decomposition of the tensor that underpins the analysis. Conventional studies employ the clearly physically interpretable Hermitian/skew-Hermitian decomposition into strain, $\mathsf{S}$, and rotation, $\mathsf{\Omega}$, components, with the latter often then modified to a vorticity vector. However, the eigenvalues are also important for delimiting different topological states of the tensor, and therefore there is some history to studying the  invariants of the characteristic equation for the VGT. Our approach is to prioritize the eigenvalue-based approach and to then deploy a subsequent Hermitian/skew-Hermitian decomposition. However, the first thing we note is that, just examining the eigenvalues is insufficient. While the Hermitian/skew-Hermitian decomposition gives an additive decomposition of the velocity gradient tensor, $\mathsf{A}$: 
\begin{equation}
\mathsf{A} = \mathsf{S} + \mathsf{\Omega},
\label{eq.additive1}
\end{equation}
to form the equivalent with an eigenvalue-based approach, we need a decomposition of the tensor into a normal tensor, $\mathsf{B}$, (characterized by the eigenvalues) and a non-normal part, $\mathsf{C}$, (characterizing the local torques acting on the tensor resulting from tensor asymmetries). Hence, we may write an alternative to (\ref{eq.additive1}) as
\begin{equation}
\mathsf{A} = \mathsf{B} + \mathsf{C}.
\label{eq.additive2}
\end{equation}
The tool from matrix algebra we use to accomplish this is the Schur transform \citep{schur1909}.

Having commenced our analysis of $\mathsf{A}$ from this starting point, we can then use the Hermitian/skew-Hermitian decompositions of $\mathsf{B}$ and $\mathsf{C}$ to elucidate the relative importance of strain and rotation for both the normal and non-normal parts of the tensor. We use this framework to re-interrogate and shed further light on a number of properties of the VGT. 

\subsection{The velocity gradient tensor and the invariants of its characteristic equation} 
\label{sect.VGTinv}
The velocity gradient tensor, $\mathsf{A} \in \Re^{3 \times 3}$, is given by $A_{ij} = \partial u_{i}/\partial x_{j}$, where $u$ is a velocity component and $x$ is a spatial direction, and is directly related to the Navier-Stokes equations, 
\begin{equation}
\frac{\partial}{\partial t}\mathbf{u} + \mathbf{u} \cdot \nabla \mathbf{u} = -\frac{1}{\rho}\nabla p + \nu \Delta \mathbf{u},
\label{eq.NS}
\end{equation}
where $t$ is time, $p$ is the pressure, $\rho$ is the density and $\nu$ is the kinematic viscosity. This may be made explicit by taking the spatial gradient of the Navier-Stokes equations:
\begin{equation}
\frac{\partial}{\partial t}\mathsf{A} + \mathbf{u} \cdot \nabla \mathsf{A} = -\mathsf{A}^{2} - \mathsf{H} + \nu \Delta \mathsf{A},
\label{eq.VGT}
\end{equation}
where $\mathsf{H}$ is the Hessian of the kinematic pressure field, i.e. $H_{ij} = \frac{\partial^{2} p}{\partial x_{i} \partial x_{j}}$. The characteristic equation for $\mathsf{A}$ is
\begin{equation}
\lambda_{i} ^{3} + \mbox{P}_{A} \lambda_{i}^{2} + \mbox{Q}_{A} \lambda_{i} + \mbox{R}_{A} = 0,
\label{eq.lambdaPQR}
\end{equation}
where the $\lambda_{i}$ are the eigenvalues of the tensor. Clearly, the three invariants of this equation may be expressed in terms of the eigenvalues and for an incompressible flow, because there is zero trace, $\text{tr}(\mathsf{A})=0$, and $\mbox{P}_{A} = \sum \lambda_{i}$, it follows that $\mbox{P}_{A} = 0$. The second and third invariants, $\mbox{Q}_{A}$ and $\mbox{R}_{A}$ are given by:
\begin{eqnarray}
\label{eq.Q}
\mbox{Q}_{A} &=& -\frac{1}{2}\text{tr}(\mathsf{A}^{2}) \equiv (1 - \delta_{ij})\sum \lambda_{i}\lambda_{j}\\
\mbox{R}_{A} &=&  -\mbox{det}(\mathsf{A}) \equiv \prod \lambda_{i},
\label{eq.R}
\end{eqnarray} 
where $\delta_{ij}$ is the Kronecker delta. These expressions for $\mbox{Q}_{A}$ and $\mbox{R}_{A}$ are of significance topologically, because the sign of the discriminant function for incompressible flow
\begin{equation}
\Delta_{L} = \mbox{Q}_{A}^{3} + \frac{27}{4}\mbox{R}_{A}^{2},
\label{eq.discrim}
\end{equation}
separates regions where the eigenvalues form a conjugate pair ($\Delta_{L} > 0$) and where they are all real ($\Delta_{L} < 0$). In the Lagrangian frame of a moving fluid element, the former results in closed streamlines, explaining the use of $\Delta_{L} > 0$ as a local, practical tool for coherent structure identification \citep{chong90}, although because $\mbox{Q}$ is raised to an odd power in (\ref{eq.discrim}), it follows that $\mbox{Q}_{A} > 0$ is a more restrictive definition \citep{hunt88,dubief00}. Because of this physical meaning to the change in eigenvalue behaviour, it follows that particular regions of joint $\mbox{Q}_{A}-\mbox{R}_{A}$ space have topological interpretation \citep{perry87}:
\begin{itemize}
\item $\Delta_{L} > 0, \mbox{R}_{A} > 0$ - compressing of the flow towards an unstable focus region;
\item $\Delta_{L} > 0, \mbox{R}_{A} < 0$ - stretching of the flow away from a stable focus region;
\item $\Delta_{L} < 0, \mbox{R}_{A} > 0$ - two saddles with an unstable node; 
\item $\Delta_{L} < 0, \mbox{R}_{A} < 0$ - two saddles with a stable node.
\end{itemize}

\begin{figure*}
  \includegraphics[width=0.85\textwidth]{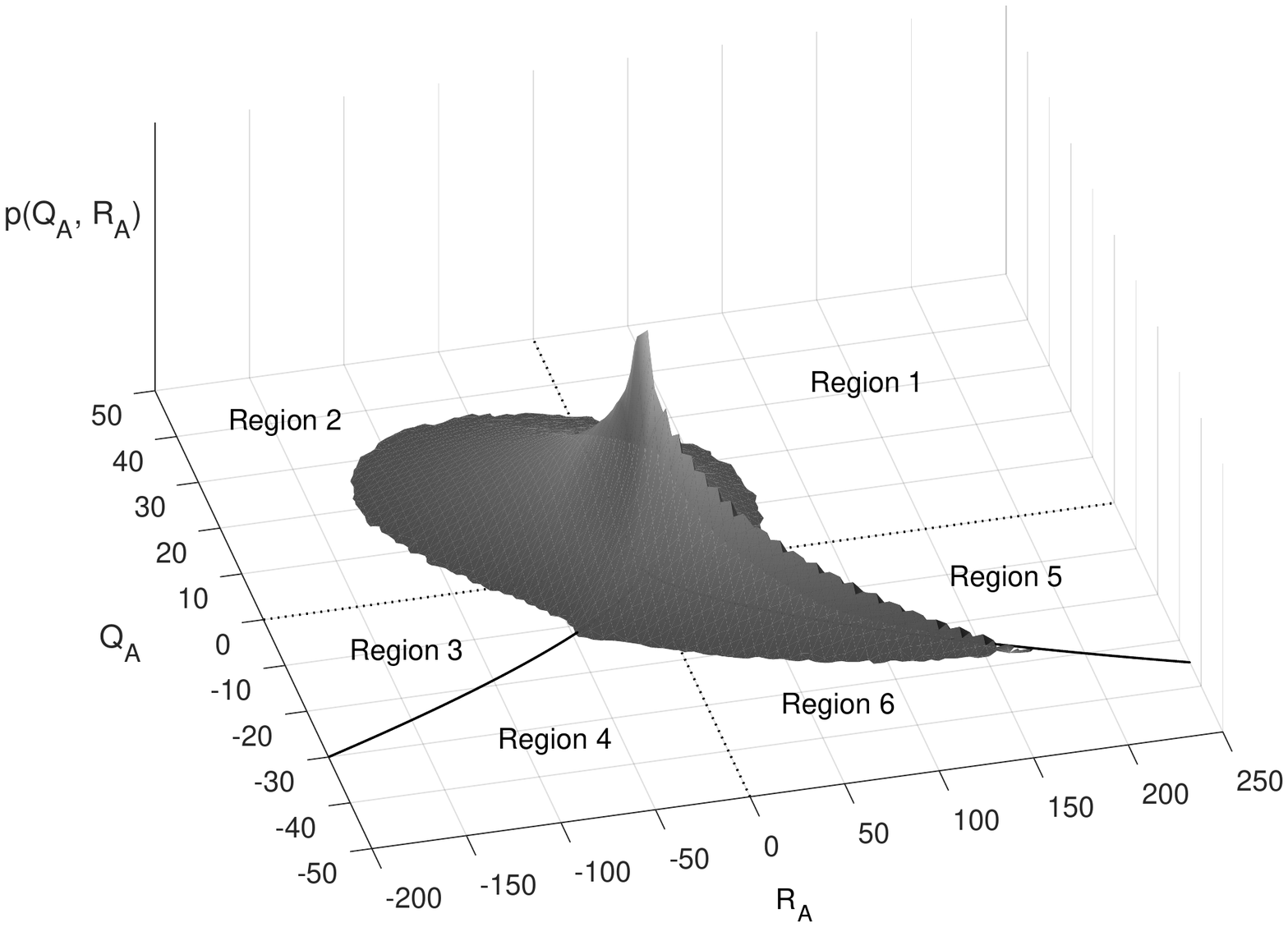}
\caption{A schematic of the joint distribution function that forms the $\mbox{Q}_{A}-\mbox{R}_{A}$ diagram, plotted as a surface, with the discriminant function, $\Delta_{L} = 0$ plotted as a solid line and the six regions used throughout this study highlighted. The Vieillefosse tail is the elongated feature in the bottom-right region where $\Delta_{L} = 0$ \citep{vieillefosse84}. This figure was constructed using the results stored in the Johns Hopkins database \citep{yili}.}
\label{fig.QR}       
\end{figure*}

Alternatively, $\mbox{Q}_{A}$ and $\mbox{R}_{A}$ can be defined in terms of strain and rotation tensors. A Hermitian-skew Hermitian decomposition into strain and rotation is given by 
\begin{eqnarray}
\mathsf{S}_{A} &=& \frac{1}{2}(\mathsf{A} + \mathsf{A}^{*}) \\
\mathsf{\Omega}_{A} &=& \frac{1}{2}(\mathsf{A} - \mathsf{A}^{*})
\label{eq.SOmegavort}
\end{eqnarray}
where the $*$ superscript is the conjugate transpose. This leads to 
\begin{align}
\label{eq.Q2}
\notag
\mbox{Q}_{A} &= \frac{1}{2} (||\mathsf{\Omega}_{A}||^{2} - ||\mathsf{S}_{A}||^{2}) \\
&\equiv \mbox{Q}^{(\Omega)}_{A} - \mbox{Q}^{(S)}_{A}\\
\notag
\mbox{R}_{A} &= -\mbox{det}(\mathsf{S}_{A}) - \mbox{tr}(\mathsf{\Omega}_{A}^{2}\mathsf{S}_{A}) \\
&\equiv \mbox{R}^{(S)}_{A} - \mbox{tr}(\mathsf{\Omega}_{A}^{2}\mathsf{S}_{A}),
\label{eq.R2}
\end{align} 
where, e.g. $||\mathsf{S}_{A}|| = \sqrt{\mbox{tr}(\mathsf{S}_{A}\mathsf{S}_{A}^{*})}$ is the Frobenius norm and one may also choose to work with the vorticity, $\omega_{i} = -\epsilon_{ijk}\Omega_{jk}$ where $\epsilon_{ijk}$ is the Levi-Cevita symbol. These latter expressions provide physical interpretations of the invariants as the excess of enstrophy with respect to total strain (\ref{eq.Q2}) and the excess of strain production with respect to enstrophy production (\ref{eq.R2}). It should be noted that while $\mbox{Q}_{A}$ is therefore the difference between two positive quantities, $\mbox{R}_{A}$ reflects a balance between enstrophy production and strain production, for which both mean values are positive \citep{taylor38,betchov56}, but the instantaneous values may take either sign. This provided the motivation for L\"{u}thi and co-workers to examine the velocity gradient tensor from the perspective of a $\mbox{Q}_{A}; -\mbox{det}(\mathsf{S}_{A}); \mbox{tr}(\mathsf{\Omega}_{A}^{2}\mathsf{S}_{A})$ decomposition \citep{luthi09}. 

The well-known $\mbox{Q}-\mbox{R}$ diagram ($\mbox{Q}_{A}-\mbox{R}_{A}$ in our notation) is shown for homogeneous, isotropic turbulence (HIT) at a Taylor Reynolds number of 433 in Fig. \ref{fig.QR}, with the discriminant function, $\Delta_{L}$ as a solid line and the six regions that are delimited throughout this study also labelled. The features of this diagram are well-known and of particular prominence is the Vieillefosse tail \citep{vieillefosse84} that forms a `ridge' to the joint distribution function on the positive $R_{A}$ side. The degree of mass along the tail, close to the origin and in the opposite, ($\mbox{Q} > 0, \mbox{R} > 0$), quadrant can be shown to be statistically significant features of HIT both with respect to Gaussian, random tensors \citep{tsinober01} and random tensors constrained to local properties \citep{k17}. It was shown by Cantwell that the restricted Euler equations that consider the dynamics of the velocity gradient tensor only in terms of $\mbox{Q}_{A}$ and $\mbox{R}_{A}$ (see below) have two possible solutions \citep{cantwell92}:
\begin{itemize}
\item Given the timescale, $t_{0} = 1 / \sqrt{|\mbox{Q}_{A}^{0}|}$, for $\mbox{Q}_{A}^{0} = 0, \mbox{R}_{A} < 0$ the solutions evolve to the fixed point, $\mbox{Q}_{A} = \mbox{R}_{A} = 0$;
\item For any other initial conditions the solution evolves towards the Vieillefosse tail, defined as the bottom right quadrant of Fig. \ref{fig.QR} along the line $\Delta_{L} = 0$. 
\end{itemize}
That the majority of points lie close to the origin, or along the Vieillefosse tail, highlights the importance of the restricted Euler approximation. However, it is clearly necessary to go beyond this approach to derive a workable model for the velocity gradient tensor dynamics that includes terms that prevent the mass of the distribution function over-accumulating at the extreme of the Vieillefosse tail. There have been a number of papers that have proposed models for this behavior and these are discussed briefly below. However, a contribution to this area is not the intent of this study, which focuses instead on refining understanding of terms associated with the evolution of the VGT.

\begin{table}
\caption{Occupany (expressed as a \% of realizations) of different states for HIT for the velocity gradient tensor, $\mathsf{A}$. For compactness, the $A$ subscript is not used in the table headings, and $\mbox{R}^{(S)} \equiv -\mbox{det}(\mathsf{S}_{A})$ and $\Omega^{2}S \equiv \mbox{tr}(\mathsf{\Omega}_{A}^{2}\mathsf{S}_{A})$.}
\centering
\begin{tabular}{llccccc}
\noalign{\smallskip}
Region & & $\mbox{R}^{(S)} > 0$, & $\mbox{R}^{(S)} > 0$, & $\mbox{R}^{(S)} < 0$, & $\mbox{R}^{(S)} < 0$, & Total \\
 & & $\Omega^{2} S > 0$ & $\Omega^{2}S < 0$ & $\Omega^{2}S < 0$ & $\Omega^{2}S > 0$ &\\
\noalign{\smallskip}
1 & $Q_{A} > 0, R_{A} > 0$ & 1.8 & 4.0 & 5.3 & 0 & 11.1\\
2 & $Q_{A} > 0, R_{A} < 0$ & 20.0 & 0 & 0.8 & 5.7 & 26.5\\
3 & $Q_{A} < 0, \Delta_{L} > 0, R_{A} < 0$ & 5.3 & 0 & 1.0 & 3.6 & 9.9\\
4 & $Q_{A} < 0, \Delta_{L} < 0, R_{A} < 0$ & 4.0 & 0 & 3.0 & 2.0 & 9.0\\
5 & $Q_{A} < 0, \Delta_{L} > 0, R_{A} > 0$ & 7.0 & 5.0 & 1.0 & 0 & 13.0\\
6 & $Q_{A} < 0, \Delta_{L} < 0, R_{A} > 0$ & 26.2 & 3.3 & 1.0 & 0 & 30.5\\
\noalign{\smallskip}
\label{table.Q_wws_sss}
\end{tabular}
\end{table}

Table \ref{table.Q_wws_sss} summarizes how the velocity gradient tensors for HIT extracted from the Johns Hopkins database used to populate Fig. \ref{fig.QR} are distributed over the six regions of the $\mbox{Q}_{A}-\mbox{R}_{A}$ diagram and the four feasible states for enstrophy production and strain production based on their signs. What is very clear is that the two most frequently occupied regions in terms of their relative occupany, regions 2 and 6, with opposite signs for $Q_{A}$ and $R_{A}$, are both dominated by positive strain production and enstrophy production. Clearly the relative magnitudes of these terms differ significantly in these two cases to establish the change in sign for $R_{A}$. The only region where tensors with negative values for both enstophy production and strain production are more likely than tensors with both terms positive is region 1. The opposite region to this (region 4) has a nearly equal probability of the two strain/enstrophy production states, while in the regions adjacent to, but above, the discriminant function (regions 3 and 5) it is seen that the physically feasible combination with different signs for the two production terms is nearly as likely as the case where both are positive. Hence, these latter two regions exhibit intermediate characteristics compared to regions 2 and 4, and 1 and 6, respectively.

\subsection{Models for the velocity gradient tensor}
\label{sect.VGTmodels}
The restricted Euler model for the velocity gradient tensor already discussed \citep{cantwell92} has been followed by a number of studies that have sought to improve the physical representation of the Lagrangian dynamics of the velocity gradient tensor \citep{pope90, martin98, chevillard08, wilczek14, johnson16}. These studies have followed a stochastic approach where the unclosed terms in the Lagrangian evolution equation (the deviatoric part of the pressure Hessian and the viscous term) are split into fluctuations that are modelled as Gaussian white noise, and mean quantities conditioned on the velocity gradient tensor itself. Such models result in behaviour in $\mbox{Q}_{A} - \mbox{R}_{A}$ space that is a much better representation of the true PDF than seen in the restricted Euler formalism (see Fig. 6 of \citet{johnson16} by way of example).

More related to the theme of this paper is the formulation of the full set of coupled ODEs that describe the Lagrangian evolution of $\mathsf{A}$ \citep{martin98,nomurapost98}. Such an approach also benefits from a decomposition of the presure Hessian, $\mathsf{H}$, in (\ref{eq.VGT}) into isotropic and deviatoric parts \citep{ohkitani95}:
\begin{align}
\mathsf{H} &= \mathsf{H}^{iso} + \mathsf{H}^{dev}\nonumber \\ 
\mathsf{H}^{iso} &= -\frac{2}{3}\mbox{Q}_{A}\mathsf{I} \equiv  \frac{1}{3}\mbox{tr}(\mathsf{H})\mathsf{I},
\label{eq.Hessian}
\end{align}
where $\mathsf{I}$ is the $3 \times 3$ identity matrix and the substitution of $\mathsf{Q}_{A}$ for the trace of the Hessian comes from the Poisson equation for the pressure. 

The isotropic term acts to preserve the volume of the fluid element without directional preference, and the deviatoric term includes all the non-local effects from the surrounding flow (hence, requiring an integration over the whole flow). Thus, where $\Delta_{L} > 0$ and enstrophy dominates total strain, the isotropic part acts to reduce the enstrophy growth rate, so that amplification is either due to stretching, i.e. the enstrophy production $\mbox{tr}(\mathsf{\Omega}_{A}^{2}\mathsf{S}_{A})$, or the action of $\mathsf{H}^{dev}$.

In the restricted Euler formulation, working in the frame of reference of the moving fluid  and with viscous effects and $\mathsf{H}^{dev}$ set to zero, we have just two coupled ODEs that give the evolution of the scalars $\mbox{Q}_{A}$ and $\mbox{R}_{A}$ \citep{cantwell92}:
\begin{eqnarray}
\label{eq.QEuler}
\frac{d \mbox{Q}_{A}}{d t} &=& -3 \mbox{R}_{A} \\
\frac{d \mbox{R}_{A}}{d t} &=& \frac{2}{3} \mbox{Q}_{A}^{2}.
\label{eq.REuler}
\end{eqnarray}
Because there are 8 independent terms in $\mathsf{A}$ and three scalars are needed to specify orientation in three dimensions \citep{meneveau11}, five orientation-free scalars can be used to determine the system. If we consider the invariants for the strain and rotation, which we indicate with a bracketed superscript, then:
\begin{itemize}
\item $\mbox{P}_{A}^{(\mathsf{S})} = 0$ because of incompressibility;
\item The first and third invariants for $\mathsf{\Omega}_{A}$ are zero because of the skew-symmetric nature of $\mathsf{\Omega}_{A}$;
\item This leaves three terms, $\mbox{Q}_{A}^{({S})} = \frac{1}{2} ||\mathsf{S}_{A}||^{2}$, $\mbox{R}_{A}^{({S})} = - \mbox{det}(\mathsf{S}_{A})$, and $\mbox{Q}_{A}^{({\Omega})} = \frac{1}{2} ||\mathsf{\Omega}_{A}||^{2}$. 
\end{itemize}
From, (\ref{eq.Q2}) and (\ref{eq.R2}), these are three of the four constitutive terms for $\mbox{Q}_{A}$ and $\mathsf{R}_{A}$, and we can eliminate one of these using $\mbox{Q}_{A} = \mbox{Q}_{A}^{({S})} + \mbox{Q}_{A}^{({\Omega})}$. The remaining term needed to close the system is the square of the stretching vector which in its rotation (rather than vorticity form) is $V^{2} = -\frac{1}{2}\mbox{tr}[(\mathsf{\Omega}_{A}\mathsf{S}_{A}+\mathsf{S}_{A}\mathsf{\Omega}_{A})^{2}]$. The following evolution equations may then be derived \citep{martin98}:
\begin{eqnarray}
\label{eq.ExtendedEuler1}
\frac{d \mbox{Q}_{A}^{(S)}}{d t} &=& -2 \mbox{R}_{A}^{(S)} - \mbox{R}_{A} \\
\label{eq.ExtendedEuler2}
\frac{d \mbox{R}_{A}^{(S)}}{d t} &=& \frac{2}{3}  \mbox{Q}_{A}\mbox{Q}_{A}^{(S)} + \frac{1}{4}V^{2} \\
\frac{d V^{2}}{d t} &=& -\frac{16}{3} ( \mbox{R}_{A}^{(S)} - \mbox{R}_{A} ) \mbox{Q}_{A}.
\label{eq.ExtendedEuler3}
\end{eqnarray}
We revisit the component terms in these equations in the next sections, where they are re-cast and studied from the perspective of the formulation developed in this paper. First, however, we develop our alternate approach to tensorial decomposition and examine some of its implications and consequences.
 
\section{A formulation of velocity gradient tensor analysis resolving normal and non-normal effects explicitly}
\subsection{Tensor non-normality and the {S}chur transform}
The primary innovation in this study is to undertake an additive decomposition of the velocity gradient tensor, $\mathsf{A}$, into normal, $\mathsf{B}$, and non-normal, $\mathsf{C}$, components before any subsequent decomposition into rotation or straining aspects. This permits us to unpack a number of phenomena commonly lumped together and thereby clarify the behaviour of the tensor. If $\mathsf{A}$ is normal, then 
\begin{equation}
\mathsf{A}\mathsf{A}^{*} = \mathsf{A}^{*}\mathsf{A}.
\label{eq.norm}
\end{equation} 
The eigenvalue decomposition is given by 
\begin{equation}
\mathsf{L}\mathsf{\Lambda}\mathsf{L}^{-1} = \mathsf{A},
\label{eq.eig} 
\end{equation}
where $\mathsf{L}$ contains the eigenvectors and $\mathsf{\Lambda}$ is a diagonal matrix of eigenvalues ($\Lambda_{1,1} = \lambda_{1}$). We now state the Schur transform \citep{schur1909}:
\begin{equation}
\mathsf{U}\mathsf{T}\mathsf{U}^{*} = \mathsf{A},
\label{eq.schur}
\end{equation}
where $\mathsf{U}$ is unitary and $\mathsf{T} = \mathsf{\Lambda} + \mathsf{N}$, with $\mathsf{N}$ an upper triangular tensor. We note that because the $\lambda_{i}$ may contain a conjugate pair (where $\Delta_{L} > 0$), to ensure $\mathsf{N}$ is upper triangular rather than quasi-upper triangular, a complex Schur decomposition is used throughout this study \citep{GolubVanLoan13}. The appendix reviews the distinction between the complex and real forms for the decomposition. 

Note that the Schur transform contains a stronger constraint than the eigen decomposition on the form of the rotation matrix. Because $\mathsf{U}$ is unitary, $\mathsf{U}\mathsf{U}^{*} = \mathsf{I}$, where $\mathsf{I}$ is the identity matrix, and this means that $\mathsf{U}\mathsf{U}^{-1} = \mathsf{I}$ as well. For the eigen decomposition, while $\mathsf{L}\mathsf{L}^{-1} = \mathsf{I}$, $\mathsf{L}\mathsf{L}^{*} = \mathsf{I}$ is only true when $\mathsf{T} = \mathsf{\Lambda}$, i.e. $\mathsf{N} = \mathsf{O}$ and, therefore, $\mathsf{L} = \mathsf{U}$. From (\ref{eq.eig}) and (\ref{eq.schur}), when $\mathsf{N}$ is a zero tensor, $\mathsf{A}$ is normal and described thoroughly by its eigenvalues. It therefore follows that $\mathsf{A}\mathsf{A}^{*} \ne \mathsf{A}^{*}\mathsf{A}$ and $||\mathsf{N}|| > 0$ are both measures of non-normality and there are a number of papers that give bounds for $||\mathsf{N}||$ given $||\mathsf{A}\mathsf{A}^{*} - \mathsf{A}^{*}\mathsf{A}||$ \citep{henrici62, eberlein65, lee95}. Thus, some `residual' dynamics exist independent of the eigenvalues of $\mathsf{A}$ and the advantage of the Schur decomposition is that it moves these effects out of the eigenvectors (orientations) and into a tensor $\mathsf{N}$ that, with the eigenvalues, contributes to $\mathsf{T}$.
 
\subsection{Normal and non-normal velocity gradient tensors and properties of the second invariant of the velocity gradient tensor}
Returning to our alternative additive decomposition postulated in (\ref{eq.additive2}), we may now write that:
\begin{eqnarray}
\label{eq.B}
\mathsf{B} &=& \mathsf{U}\mathsf{\Lambda}\mathsf{U}^{*} \\
\mathsf{C} &=& \mathsf{U}\mathsf{N}\mathsf{U}^{*},
\label{eq.C}
\end{eqnarray}
and this is the key conceptual innovation in this paper. Thus, $\mathsf{A} = \mathsf{B} + \mathsf{C}$ provides an explicit means to separate the normal and non-normal contributions to the dynamics. That is, we have an additive decomposition into a tensor, $\mathsf{B}$ containing the dynamics driven by the eigenvalues, which from the restricted Euler formulation are preferentially associated with the local dynamics, and a tensor, $\mathsf{C}$ that contains dynamics that are a result of asymmetric structure in the VGT induced by non-local effects.

Taking the rotation and strain tensors for $\mathsf{B}$ and $\mathsf{C}$, we have
\begin{align}
\mathsf{B} &= \mathsf{S}_{B} + \mathsf{\Omega}_{B} \\
\mathsf{C} &= \mathsf{S}_{C} + \mathsf{\Omega}_{C}.
\end{align}
In terms of the Frobenius norms, non-normality is partitioned equally across $\mathsf{S}_{C}$ and $\mathsf{\Omega}_{C}$, which means that the constitutive terms for $\mbox{Q}_{A}$ (\ref{eq.Q2}) may be written as
\begin{eqnarray}
||\mathsf{S}_{A}||^{2} &=& ||\mathsf{S}_{B}||^{2} + ||\mathsf{S}_{C}||^{2} \equiv ||\mathsf{S}_{B}||^{2} + ||\mathsf{\Omega}_{C}||^{2} \\
||\mathsf{\Omega}_{A}||^{2} &=& ||\mathsf{\Omega}_{B}||^{2} + ||\mathsf{\Omega}_{C}||^{2} \equiv ||\mathsf{\Omega}_{B}||^{2} + ||\mathsf{S}_{C}||^{2}
\label{eq.newOmSA}
\end{eqnarray}
The importance of this is seen with respect to eq. (\ref{eq.Q2}): the second invariant may now be written as
\begin{equation}
\mbox{Q}_{A} \equiv \mbox{Q}_{B} = \frac{1}{2}\left(||\mathsf{\Omega}_{B}||^{2} - ||\mathsf{S}_{B}||^{2}\right),
\label{eq.newQ}
\end{equation}
where the component terms are smaller by a factor of $||\mathsf{S}_{C}||^{2}=||\mathsf{\Omega}_{C}||^{2}$.

Recalling that $\mbox{R}_{A} = -\mbox{det}(\mathsf{A}) \equiv \prod \lambda_{i}$ and the eigenvalues of $\mathsf{A}$ and $\mathsf{B}$ are identical, then it must also follow that $\mbox{R}_{A} = \mbox{R}_{B}$. Thus, the third invariant becomes
\begin{align}
\label{eq.newR}
\mbox{R}_{B} &= -\mbox{det}(\mathsf{S}_{B}) - \mbox{tr}(\mathsf{\Omega}_{B}^{2}\mathsf{S}_{B})\\
\notag{}
&= \mbox{R}^{(S)}_{B} - \mbox{tr}(\mathsf{\Omega}_{B}^{2}\mathsf{S}_{B}),
\end{align}
where we still have to establish the additional terms involving $\mathsf{C}$ that appear on both sides of the difference in (\ref{eq.newR}) so that $\mbox{R}_{A} = \mbox{R}_{B}$ but $\mbox{R}^{(S)}_{A} \ne \mbox{R}^{(S)}_{B}$ except for where $\mathsf{A} = \mathsf{B}$.

\subsection{Some physical aspects of this decomposition}
\label{sect.physical}
From the above section we see that, in terms of the first two, autonomous ODEs for the VGT dynamics (\ref{eq.ExtendedEuler1},\ref{eq.ExtendedEuler2}), we may directly substitute expressions written for $\mathsf{A}$ with those for $\mathsf{B}$. This is physically intuitive because the restricted Euler formulation is a local and inviscid model involving the isotropic part of the pressure Hessian only. Studying the component terms of the second and third invariants when $||\mathsf{C}|| \ne 0$ introduces non-local effects into consideration. 

A further important aspect of our approach is that the discriminant, $\Delta_{L}$ has explicit physical consequences in the analysis of rotation, which is not the case when one studies $||\mathsf{S}_{A}||$ and $||\mathsf{\Omega}_{A}||$. Thus, while the discriminant function partitioning real eigenvalue regions from a conjugate pair and  closed streamlines is super-imposed on Fig. \ref{fig.QR}, the total strain and enstrophy for $\mathsf{A}$ are defined continuously over the $\mathsf{Q}_{A}$ axis (although, of course, their relative magnitudes change). The advantage of our approach is that because $\mathsf{B}$ is an eigenvalue-based tensor, $\Delta_{L}=0$ demarcates a change in physical behaviour. Mathematically, this arises because the eigenvalues for $\mathsf{S}_{B}$ and $\mathsf{\Omega}_{B}$ are the real and imaginary parts of the eigenvalues for $\mathsf{A}$. Hence, where $\Delta_{L} < 0$ there is no imaginary part and we have $||\mathsf{\Omega}_{B}|| = 0$, i.e. there is no rotation in the normal part of the tensor. Therefore, $\mbox{Q}_{B} = -\frac{1}{2}||\mathsf{S}_{B}||^{2}$ and all enstrophy comes from $\mathsf{C}$, a contribution to $\mbox{Q}_{B}$ that is equal to the strain contribution from $\mathsf{C}$. As a consequence of $||\mathsf{\Omega}_{B}|| = 0$, $\mbox{tr}(\mathsf{\Omega}_{B}^{2}\mathsf{S}_{B}) = 0$ and the eigenvalues. for $\mbox{R}_{B}$ and $\mbox{R}^{(S)}_{B}$ are identical. Hence, the third invariant also has a simple expression beneath the discriminant function in this case: $\mbox{R}_{A} \equiv \mbox{R}_{B} = \mbox{R}^{(S)}_{B}$. 

Above the discriminant function, while we can no longer equate the values for the third invariant of $\mathsf{A}$ with those for $\mbox{R}^{(S)}_{B}$, we can state that $\mbox{sgn}(\mbox{R}^{(S)}_{B}) = \mbox{sgn}(\mbox{R}_{B})$. From the eigenvalue structure, it follows that
\begin{equation}
\mbox{tr}(\mathsf{\Omega}_{B}^{2}\mathsf{S}_{B}) = \mbox{Im}(\lambda_{c})^2 \lambda_{r},
\label{eq.enstroprodB}
\end{equation}
where the $r$ and $c$ subscripts indicate the real and conjugate pair eigenvalues for $\mathsf{A}$ (and $\mathsf{B}$). Thus, $\mbox{sgn}(\mbox{tr}(\mathsf{\Omega}_{B}^{2}\mathsf{S}_{B})) = -\mbox{sgn}(\mbox{R}^{(S)}_{B})$. Therefore, the basic nature of the normal contributions to strain production and enstrophy production are known from inspection of the $\mbox{Q}_{A}-\mbox{R}_{A}$ diagram and observed departures from such relations for $\mbox{tr}(\mathsf{\Omega}_{A}^{2}\mathsf{S}_{A})$ and $-\mbox{det}(\mathsf{S}_{A})$, as identified in Table \ref{table.Q_wws_sss}, are a consequence of the over-riding influence of the non-normal terms or those representing the interaction between normal and non-normal effects.

\subsection{Evolution equations for the strain and rotation of $\mathsf{B}$ and $\mathsf{C}$}
In order to determine the terms that yield the difference between $-\mbox{det}(\mathsf{S}_{A})$ and $-\mbox{det}(\mathsf{S}_{B})$ in (\ref{eq.R2}) and (\ref{eq.newR}) it is helpful to write down the equations for the Lagrangian evolution of strain and rotation for $\mathsf{B}$ and $\mathsf{C}$:
\begin{align}
\notag
\frac{\partial \mathsf{S}_{B}}{\partial t} &+ \mathsf{S}_{B}^{2} + \mathsf{\Omega}_{B}^{2} = -\frac{1}{\rho} \mathsf{H}^{iso} + \nu \nabla^{2} \mathsf{S}_{B}  \\ 
\notag
\frac{\partial \mathsf{S}_{C}}{\partial t} &+ \mathsf{S}_{C}^{2} + \mathsf{\Omega}_{C}^{2}+\mathsf{S}_{B}\mathsf{S}_{C} + \mathsf{S}_{C}\mathsf{S}_{B} + \mathsf{\Omega}_{B}\mathsf{\Omega}_{C} + \mathsf{\Omega}_{C}\mathsf{\Omega}_{B} = -\frac{1}{\rho} \mathsf{H}^{dev} + \nu \nabla^{2} \mathsf{S}_{C} \\
\notag
\frac{\partial \mathsf{\Omega}_{B}}{\partial t} &+ \mathsf{\Omega}_{B}\mathsf{S}_{B} + \mathsf{S}_{B}\mathsf{\Omega}_{B} = \nu \nabla^{2} \mathsf{\Omega}_{B}\\ 
\frac{\partial \mathsf{\Omega}_{C}}{\partial t} &+ \mathsf{\Omega}_{C}\mathsf{S}_{C} + \mathsf{S}_{C}\mathsf{\Omega}_{C} + \mathsf{\Omega}_{B}\mathsf{S}_{C} + \mathsf{S}_{C}\mathsf{\Omega}_{B} + \mathsf{\Omega}_{C}\mathsf{S}_{B} + \mathsf{S}_{B}\mathsf{\Omega}_{C} = \nu \nabla^{2} \mathsf{\Omega}_{C}
\label{eq.my}
\end{align}

If we multiply the first equation in (\ref{eq.my}) by $\mathsf{S}_{B}$, take the trace and divide by -2, then using the Cayley-Hamilton theorem, $\mbox{det}(\mathsf{A}) = \frac{1}{3}\mbox{tr}(\mathsf{A}^{3})$, and (\ref{eq.R2}), and adopting the superscript notation introduced in section \ref{sect.VGTmodels}, we obtain 
\begin{equation}
\frac{\partial \mbox{Q}^{(S)}_{B}}{\partial t} +\frac{1}{2}\mbox{R}_{B} + \mbox{R}^{(S)}_{B} =  -\frac{1}{2}\nu\mbox{tr}(\mathsf{S}_{B} \nabla^{2} \mathsf{S}_{B}),
\label{eq.mySB}
\end{equation} 
where the zero trace for $\mathsf{S}_{B}$ sets $-\frac{1}{\rho} \mbox{tr}(\mathsf{H}^{iso}\mathsf{S}_{B}) = 0$. 

Multiplying the evolution equation for $\mathsf{S}_{C}$ in (\ref{eq.my}) by $\mathsf{S}_{C}$ and undertaking similar operations gives
\begin{equation}
\frac{\partial \mbox{Q}^{(S)}_{C}}{\partial t} + \mbox{R}^{(S)}_{C} -\mbox{tr}(\mathsf{S}_{C}^{2}\mathsf{S}_{B}) = -\frac{1}{2}\left[\frac{1}{\rho} \mbox{tr}(\mathsf{H}^{dev}\mathsf{S}_{C}) + \nu \mbox{tr}(\mathsf{S}_{C} \nabla^{2} \mathsf{S}_{C})\right],
\label{eq.mySC}
\end{equation}
where we have used $\mbox{R}^{(S)}_{C} \equiv -\mbox{det}(\mathsf{S}_{C}) = \mbox{tr}(\mathsf{\Omega}_{C}^{2}\mathsf{S}_{C})$ to eliminate the latter term. Hence, we highlight that the evolution of non-normal total strain is a consequence of stretching by both $\mathsf{S}_{C}$ and $\mathsf{S}_{B}$. Note also that our decomposition highlights the importance of the deviatoric part of the pressure Hessian in the evolution equations for non-normal total strain.

Taking the evolution equation for $\mathsf{\Omega}_{B}$  in (\ref{eq.my}), multiplying by $\mathsf{\Omega}_{B}$, taking the trace and dividing by -2 gives
\begin{equation}
\label{eq.OmegaB}
\frac{\partial \mbox{Q}^{(\Omega)}_{B}}{\partial t} + \mbox{R}_{B} + \mbox{R}^{(S)}_{B} = -\frac{1}{2} \nu \mbox{tr}(\mathsf{\Omega}_{B} \nabla^{2} \mathsf{\Omega}_{B}).
\end{equation}
Undertaking a similar set of operations for $\mathsf{\Omega}_{C}$ 
gives
\begin{equation}
\frac{\partial \mbox{Q}^{(\Omega)}_{C}}{\partial t} - \mbox{R}^{(S)}_{C} - \mbox{tr}(\mathsf{\Omega}_{C}^{2}\mathsf{S}_{B}) = -\frac{1}{2}\nu \mbox{tr}(\mathsf{\Omega}_{C} \nabla^{2} \mathsf{\Omega}_{C}).
\label{eq.OmegaC}
\end{equation}

Given that $\mbox{Q}^{(S)}_{C} = -\mbox{Q}^{(\Omega)}_{C}$ as can be shown using (\ref{eq.Q}), (\ref{eq.Q2}) and  (\ref{eq.newOmSA}), it follows that we can remove either (\ref{eq.mySC}) or (\ref{eq.OmegaC}). Adding these two equations together gives 
\begin{equation}
\frac{1}{\rho} \mbox{tr}(\mathsf{H}^{dev}\mathsf{S}_{C}) = \nu [\mbox{tr}(\mathsf{\Omega}_{C} \nabla^{2} \mathsf{\Omega}_{C} - \mbox{tr}(\mathsf{S}_{C} \nabla^{2} \mathsf{S}_{C})].
\end{equation}
In other words, non-normal straining of the non-local effects in the pressure Hessian equates to the dissipation due to the action of $\mathsf{C}$.

\begin{table}
\caption{The strain  production and enstrophy production terms in our framework.}
\centering
\begin{tabular}{lll}
\noalign{\smallskip}
Term & Equivalent & Interpretation\\
\noalign{\smallskip}
$\mbox{R}^{(S)}_{B}$ & - & self-amplification of normal strain\\
$\mbox{tr}(\mathsf{\Omega}_{B}^{2}\mathsf{S}_{B})$ & - & normal enstrophy production by normal straining\\
$\mbox{R}^{(S)}_{C}$ & $\mbox{tr}(\mathsf{\Omega}_{C}^{2}\mathsf{S}_{C})$ & self-amplification of non-normality \\
$\mbox{tr}(\mathsf{\Omega}_{C}^{2}\mathsf{S}_{B})$ & $-\mbox{tr}(\mathsf{S}_{C}^{2}\mathsf{S}_{B})$ & normal straining of non-normality \\
\noalign{\smallskip}
\label{table.terms}
\end{tabular}
\end{table}

\subsection{The third invariant of the velocity gradient tensor}
Table \ref{table.terms} provides a conceptual summary of the production terms that arise in our equations. Based on what we have now established, it is straightforward to return to the equations for the third invariant and show that the component terms in the strain production - enstrophy production balance for $\mbox{R}_{B}$ in (\ref{eq.newR}) differ from those for $\mbox{R}_{A}$ in (\ref{eq.R}) by $\mbox{R}^{(S)}_{C} +\mbox{tr}(\mathsf{\Omega}_{C}^{2}\mathsf{S}_{B})$. Thus, the strain production and enstrophy production terms in (\ref{eq.R}) are
\begin{align}
\label{eq.RSA}
\mbox{R}_{A}^{(S)} &\equiv -\mbox{det}(\mathsf{S}_{A}) = \mbox{R}^{(S)}_{B} + \mbox{R}^{(S)}_{C} +\mbox{tr}(\mathsf{\Omega}_{C}^{2}\mathsf{S}_{B}) \\
\mbox{tr}(\mathsf{\Omega}_{A}^{2}\mathsf{S}_{A}) &= \mbox{tr}(\mathsf{\Omega}_{B}^{2}\mathsf{S}_{B}) + \mbox{R}^{(S)}_{C} +\mbox{tr}(\mathsf{\Omega}_{C}^{2}\mathsf{S}_{B}). 
\label{eq.enstroprodA}
\end{align}

\subsection{The square of the stretching vector}
The square of the stretching vector is used to close the ODE system for the velocity gradient tensor (\ref{eq.ExtendedEuler3}), where 
\begin{equation}
V^{2} = -\frac{1}{2}\mbox{tr}[(\mathsf{\Omega}_{A}\mathsf{S}_{A}+\mathsf{S}_{A}\mathsf{\Omega}_{A})^{2}].
\end{equation}
We define 
\begin{align}
\notag
\mathsf{V}_{BB} &= \mathsf{\Omega}_{B}\mathsf{S}_{B} + \mathsf{S}_{B}\mathsf{\Omega}_{B}\\
\notag
\mathsf{V}_{BC} &= \mathsf{\Omega}_{C}\mathsf{S}_{B} + \mathsf{S}_{B}\mathsf{\Omega}_{C} \\
\notag
\mathsf{V}_{CB} &= \mathsf{\Omega}_{B}\mathsf{S}_{C} + \mathsf{S}_{C}\mathsf{\Omega}_{B}\\
\mathsf{V}_{CC} &= \mathsf{\Omega}_{C}\mathsf{S}_{C} + \mathsf{S}_{C}\mathsf{\Omega}_{C}.
\label{eq.Vterms}
\end{align}
It then follows that
\begin{equation}
V^{2} = -\frac{1}{2}\mbox{tr}(\mathsf{V}_{BB}^{2} + [\mathsf{V}_{BC}^{2} + \mathsf{V}_{CB}^{2} + \mathsf{V}_{CC}^{2} + 2\mathsf{V}_{BC}\mathsf{V}_{CC} + 2\mathsf{V}_{CB}\mathsf{V}_{CC}]),
\label{eq.newV}
\end{equation}
where the square brackets partition the terms involving $\mathsf{C}$ from that solely in terms of $\mathsf{B}$. Thus, the first term of the right-hand side appears in the evolution equation for $\mathsf{\Omega}_{B}$ in (\ref{eq.my}), and those in square brackets appear in the evolution equation for $\mathsf{\Omega}_{C}$. Note also, that when $\Delta_{L} < 0$, the above expression simplifies to
\begin{equation}
V^{2} = -\frac{1}{2}\mbox{tr}( [\mathsf{V}_{BC}^{2} + \mathsf{V}_{CC}^{2} + 2\mathsf{V}_{BC}\mathsf{V}_{CC}]),
\label{eq.newV2}
\end{equation}
and that the only terms that can contribute in a negative fashion to $V^{2}$ are the two interaction terms on the far right-hand end of (\ref{eq.newV}). Furthermore, because both of these terms are part of the evolution equation for $\mathsf{\Omega}_{C}$, it follows that $0 \le \mbox{tr}(\mathsf{V}_{BB}^{2})/V^{2} \le 1$ with values identically 0 when $\Delta_{L} < 0$. Thus, in order to evaluate the typical size of negative interaction terms, it is logical to study $V^{2} / V_{abs}$ where 
\begin{equation}
V_{abs} = -\frac{1}{2}[\mbox{tr}(\mathsf{V}_{BB}^{2} + \mathsf{V}_{BC}^{2} + \mathsf{V}_{CB}^{2} + \mathsf{V}_{CC}^{2}) + |\mbox{tr}(2\mathsf{V}_{BC}\mathsf{V}_{CC})| +  |\mbox{tr}(2\mathsf{V}_{CB}\mathsf{V}_{CC})|].
\label{eq.Vabs}
\end{equation}

\subsection{The second strain eigenvalue and its {L}und and {R}ogers normalization}
\label{sect.eLRABC}
One property of HIT that was observed in the early simulations was a strong preference for a positive second eigenvalue of the strain rate tensor \citep{kerr85,ashurst87}. This may be inferred from the shape of the $\mbox{Q}_{A}-\mbox{R}_{A}$ diagram in Fig. \ref{fig.QR} and, in particular, the values in Table \ref{table.Q_wws_sss} that indicate a preference for positive strain production and, thus, two positive eigenvalues. Topologically, this means that flow packets are more prone to evolve to disc-like features than rod-like features. 

The Lund and Rogers normalization of the second eigenvalue of the strain rate tensor is given by \citep{LR94}:
\begin{equation}
e^{(LR)}_{A} = \frac{3\sqrt{6}\mbox{R}^{(S)}_{A}}{\left(-2\mbox{Q}^{(S)}_{A} \right)^{\frac{3}{2}}},
\end{equation}
which is bounded to $-1 \le e^{(LR)}_{A} \le 1$. Similar relative measures for the second strain eigenvalue for $\mathsf{B}$ and $\mathsf{C}$ then follow. However, we noted in (\ref{eq.RSA}) that our formulation for $\mbox{R}^{(S)}_{A}$ also contains an interaction term: $\mbox{R}_{A}^{(S)} = \mbox{R}^{(S)}_{B} + \mbox{R}^{(S)}_{C} +\mbox{tr}(\mathsf{\Omega}_{C}^{2}\mathsf{S}_{B})$. Hence, we can also study
\begin{align}
e^{(LR)}_{B|A} &= \frac{3\sqrt{6}\mbox{R}^{(S)}_{B}}{\left(-2\mbox{Q}^{(S)}_{A} \right)^{\frac{3}{2}}}\\
e^{(LR)}_{C|A} &= \frac{3\sqrt{6}\mbox{R}^{(S)}_{C}}{\left(-2\mbox{Q}^{(S)}_{A} \right)^{\frac{3}{2}}}\\
e^{(LR)}_{C,B|A} &= \frac{3\sqrt{6} \mbox{tr}(\mathsf{\Omega}_{C}^{2}\mathsf{S}_{B})}{\left(-2\mbox{Q}^{(S)}_{A} \right)^{\frac{3}{2}}},
\end{align}
where $e^{(LR)}_{A} = e^{(LR)}_{B|A} + e^{(LR)}_{C|A} + e^{(LR)}_{C,B|A}$. In section \ref{sect.physical} we have described how the strain and enstrophy for $\mathsf{B}$ are constrained by the eigenvalues and, therefore, the regions of the $\mbox{Q}_{A}-\mbox{R}_{A}$ diagram. Thus, we know \emph{a priori} that $0 < e^{(LR)}_{B} \le 1$ on the $\mbox{R}_{A} > 0$ side of the diagram and $-1 \le e^{(LR)}_{B} < 0$ on the negative side.    

\subsection{Alignment properties of the vorticity vector and the strain eigenvectors}
\label{sect.vortstrain}
An important and surprising early result in the study of the velocity gradient tensor was the preferred alignment between the vorticity vector, $\boldsymbol{\omega}_{A}$ and the eigenvector for the intermediate eigenvalue of the strain rate tensor \citep{kerr85,ashurst87,jimenez92}. Our decomposition gives vorticity vectors for $\mathsf{A}$, $\mathsf{B}$, and $\mathsf{C}$, respectively, as well as nine possible strain eigenvectors. Thus we introduce the notation that with $e_{i}^{A}$ indicating an eigenvalue of the strain rate tensor for $\mathsf{A}$, ordered from most positive to most negative, and $\mathbf{e}^{A}_{i}$ its corresponding eigenvector, we can define, as an example:
\begin{equation}
\theta^{A,C}_{i} = \cos(\boldsymbol{\omega}_{A}, \mathbf{e}^{C}_{i}),
\end{equation}
as the angle between the vorticity vector for $\mathsf{A}$ and the \emph{i}th eigenvector for $\mathsf{C}$. 

One aspect of our decomposition is that because $\mathsf{S}_{A} = \mathsf{S}_{B} + \mathsf{S}_{C}$, we can examine the alignment between $\mathbf{e}_{i}^{A}$, and  $\mathbf{e}_{j}^{B}$ or $\mathbf{e}_{k}^{C}$. For example,
\begin{equation}
\phi_{i,j}^{A,B} = \cos(\mathbf{e}_{i}^{A},\mathbf{e}_{j}^{B}).
\end{equation}
Note that the eigenvalues for $\mathsf{S}_{B}$ are always equal to the real part of the eigenvalues for $\mathsf{A}$. Hence, as described in section \ref{sect.physical}, when above the discriminant function, we will always have a pair of equal eigenvalues, meaning that the second eigenvalue for $\mathsf{S}_{B}$ is not properly defined in regions 1, 2, 3, and 5.

Finally, we may also consider vorticity vector alignments, such as
\begin{equation}
\xi^{A,B} = \cos(\boldsymbol{\omega}^{A},\boldsymbol{\omega}^{B}).
\end{equation}
Because the eigenvalues for $\boldsymbol{\omega}_{B}$ are the imaginary part of the eigenvalues for $\mathsf{A}$, beneath the discriminant function, in regions 4 and 6, $\boldsymbol{\omega}_{A} = \boldsymbol{\omega}_{C}$ and $\xi^{A,C} = 1$ with $\xi^{A,B}$ and $\xi^{B,C}$ undefined.

\section{The Numerical Simulation}
This study makes use of velocity gradient tensors extracted from the Johns Hopkins Turbulence Database numerical simulation of forced isotropic turbulence at a Taylor Reynolds number of 433 \citep{yili} as described by \citet{jhu17}. The direct numerical simulation is undertaken on a $1024^{3}$ grid using a pseudo-spectral method. The energy is injected to maintain the total energy in the Fourier modes, and also retaining a wave number magnitude less than or equal to 2 in each mode. The basic properties of the simulation are summarized in Table \ref{tab.HIT} and a number of other studies have made use of this resource for studying turbulence physics \citep{wan2010,lawsondawson15} or for the testing of data post-processing algorithms \citep{higham16}. 
\begin{table}
\caption{Properties of the HIT simulation in the Johns Hopkins database \citep{yili}.}
\label{tab.HIT}
\begin{center}
\begin{tabular}{lc}
 Property & Value\\
Grid & $1024^{3}$ periodic box\\
Domain & $[0, 2\pi]^{3}$\\
Viscosity, $\nu$ & $1.85 \times 10^{-4}$\\
Mean dissipation rate, $\epsilon$ & 0.0928\\
Taylor micro-scale, $\lambda$ & 0.118\\
Taylor Reynolds number, $Re_{\lambda}$ & 433\\
Kolmogorov length, $\eta$ & $2.87 \times 10^{-3}$\\
\end{tabular}
\end{center} 
\end{table}

\section{Results: The role of non-normality}

\subsection{The importance of non-normal effects}
\label{sect.ResKappa}

\begin{figure}
\vspace*{2mm}
\begin{center}
\includegraphics[width=8.8cm]{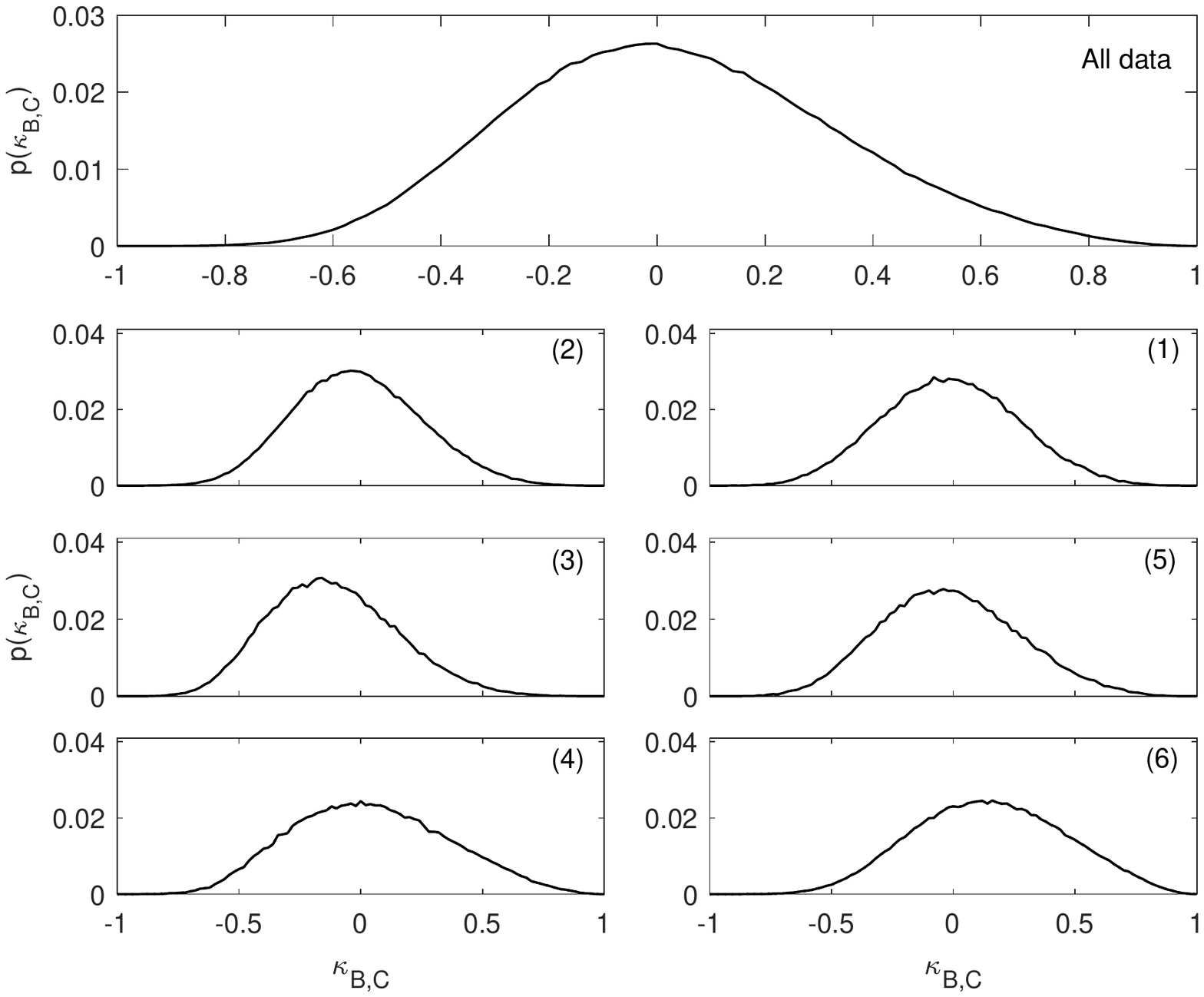}
\end{center}
\caption{The probability curves for $\kappa_{B,C}$, the normalized difference in the Frobenius norms for $\mathsf{B}$ and $\mathsf{C}$, shown for all the data and as a function of the six regions of the $\mbox{Q}_{A}-\mbox{R}_{A}$ diagram  in the various panels.} 
\label{fig.kappaBC}
\end{figure}

Given our decomposition, $\mathsf{A} = \mathsf{B} + \mathsf{C}$, the first thing to establish is the relative importance of the normal and non-normal tensors; clearly if $\mathsf{C}$ is small then it is legitimate to approximate the behaviour of $\mathsf{A}$ with eigenvalue-based formulations. To evaluate this aspect of the behaviour of our decomposition we can define
\begin{align}
\label{eq.kappaBC}
\kappa_{B,C} &= \frac{||\mathsf{B}|| - ||\mathsf{C}||}{||\mathsf{B}|| + ||\mathsf{C}||},
\end{align} 
as a normalized measure of the magnitude of the two tensors. This is shown in Fig. \ref{fig.kappaBC} and it is clear from the upper panel that the overall mode for the distribution is slightly negative, with a median close to $\kappa_{B,C} = 0$. Hence, the  non-normality is as important to the tensor as the part explained by the eigenvalues. Thus, in HIT, asymmetrical forcings on the tensor as a consequence of non-local effects are an important part of the flow dynamics.  When the results in Fig. \ref{fig.kappaBC} are partitioned by the six regions of the $\mbox{Q}_{A}-\mbox{R}_{A}$ diagram, we see that for each pair of diagrams for a given $\mbox{Q}_{A}$ state, it is the left-hand variant, with $\mbox{R}_{A} < 0$, where there is a greater contribution from $||\mathsf{C}||$. 

Region 3 is where $\kappa_{B,C}$ is most strongly negative and region 6 is where $\kappa_{B,C}$ is most positive. This latter result is all the more impressive because for $\Delta_{L} < 0$ (regions 4 and 6), $||\mathsf{\Omega}_{B}|| = 0$, meaning that in the majority of instances, $||\mathsf{S}_{B}|| >||\mathsf{S}_{C}|| + ||\mathsf{\Omega}_{C}||$, or $||\mathsf{S}_{B}|| >2||\mathsf{S}_{C}||$. Thus, an eigenvalue-based description of the flow is particularly effective near the Vieillefosse tail, which is consistent with this being an attractor for the dynamics of the restricted Euler (eigenvalue-based) set of equations for the dynamics of the VGT \citep{cantwell92}.

\subsection{The second invariant and non-normality}
\label{sect.inv2}
\begin{figure}
\vspace*{2mm}
\begin{center}
\includegraphics[width=9.8cm]{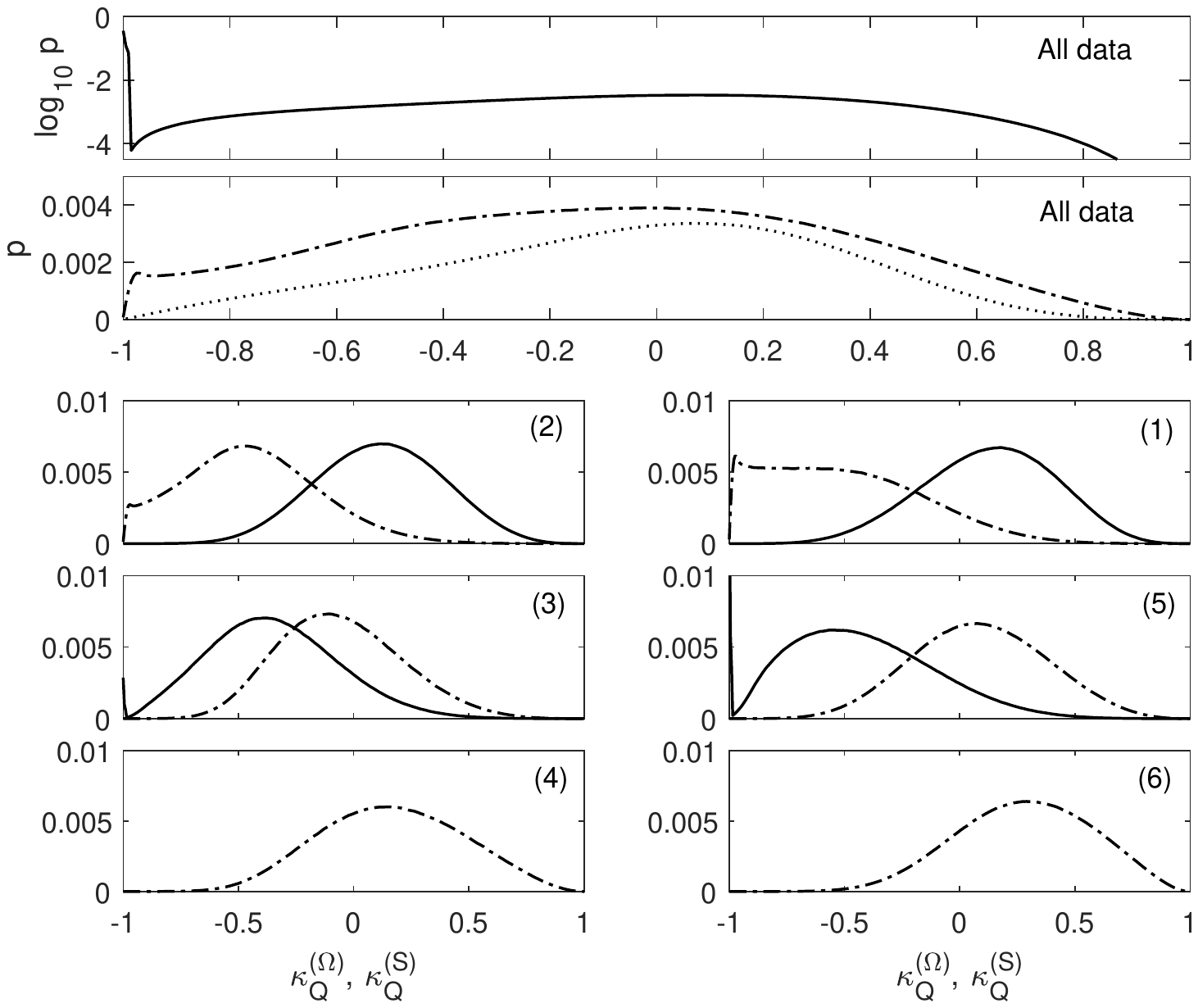}
\end{center}
\caption{The probability curves for $\kappa_{Q}^{(\Omega)}$ (solid lines) and $\kappa_{Q}^{(S)}$ (dot-dashed lines), the normalized differences in the Frobenius norms for $\mathsf{\Omega}_{B}$ and $\mathsf{S}_{B}$, respectively, compared to the non-normal term, $\mathsf{S}_{C}$. Results are shown for all the data and as a function of the six regions of the $\mbox{Q}_{A}-\mbox{R}_{A}$ diagram. The dotted lines are the results for $\kappa_{Q}^{(\Omega)}$ with the data at $\kappa_{Q}^{(\Omega)} = -1$ excluded. The full distribution including such values is shown on a log-scale in the top panel.} 
\label{fig.kappaQ}
\end{figure}

From (\ref{eq.newQ}), we may examine the components of the second invariant with respect to non-normality
\begin{align}
\kappa_{Q}^{(\Omega)} &= \frac{||\mathsf{\Omega}_{B}|| - ||\mathsf{S}_{C}||}{||\mathsf{\Omega}_{B}|| + ||\mathsf{S}_{C}||} \\
\kappa_{Q}^{(S)} &= \frac{||\mathsf{S}_{B}|| - ||\mathsf{S}_{C}||}{||\mathsf{S}_{B}|| + ||\mathsf{S}_{C}||}.
\label{eq.kappaQ}
\end{align} 
Beneath the discriminant function, $\kappa_{Q}^{(\Omega)} = -1$ by definition and such results are not shown in the panels for regions (4) and (6) in Fig. \ref{fig.kappaQ}. Their contribution to the overall distribution function for $\kappa_{Q}^{(\Omega)}$ is shown in the top panel, which has a log-scale because of the dominance of the -1 limit. The second panel is on a linear scale and the results for $\kappa_{Q}^{(\Omega)}$ are shown as a dotted line and exclude the values at -1, providing a better means to evaluate the overall shape of the distribution.

While an increase in the mean values for $\kappa_{Q}^{(\Omega)}$ and decrease for $\kappa_{Q}^{(S)}$ with $\mbox{Q}_{A}$ is very distinct, the differences as a function of the sign of $\mbox{R}_{A}$ are more obvious than in Fig. \ref{fig.kappaBC}. For example, while we anticipate a difference between regions 4 and 6 in their values for $\kappa_{Q}^{(S)}$ from the discussion of Fig. \ref{fig.kappaBC}, their respective modes of 0.16 and 0.28 indicate this very clearly. In regions 3 and 5, where $\Delta_{L} > 0$ and $\mbox{Q}_{A} < 0$, we have a positive mode for region 5 at $\kappa_{Q}^{(S)} = 0.07$ and a mode of $\kappa_{Q}^{(S)} = -0.12$ for region 3. Furthermore, the probability of $\kappa_{Q}^{(\Omega)} < -0.995$ is much higher in region 5 at $p = 0.018$ (it is the distribution's mode) than in region 3 ($p = 0.003$). This highlights the extent to which the Vieillefosse tail is an attractor for the dynamics as there is a concentration of values lying very close to $\Delta_{L} = 0$ in region 5, but not region 3.

Where $\mbox{Q}_{A} > 0$, we see that the mode in region 1 for $\kappa_{Q}^{(\Omega)} = 0.18$ is more positive than for region 2 ($\kappa_{Q}^{(\Omega)} = 0.12$), and there is also a stronger tendency for $\kappa_{Q}^{(S)} \to -1$. Hence, a simplified and approximate view of the right-hand side of the $\mbox{Q}_{A}-\mbox{R}_{A}$ diagram is that $\kappa_{Q}^{(S)} = -1$ for $\mbox{Q}_{A} > 0$ and $\kappa_{Q}^{(\Omega)} = -1$ for $\mbox{Q}_{A} < 0$. In both regions 1 and 2 there are a similar proportion of positive $\kappa_{Q}^{(S)}$ occurrences (8.5\% and 7.1\%, respectively). In such cases the degree of compression or extension due to $\mathsf{S}_{B}$ is greater than from the non-normal part, on average, implying there is a strong and coherent motion orthogonal to the plane of rotation driven by normal straining. This particular hypothesis is considered in Section \ref{sect.alignsumm} where we summarize the various alignment properties of the tensor. 

\subsection{The third invariant and non-normality}
\label{sect.inv3}
The components of the third invariant are listed and interpreted in Table \ref{table.terms}. In a similar fashion to the non-normal contribution cancelling when evaluating the second invariant because $\mbox{R}_{A} = \mbox{R}_{B}$, the non-normal and interaction terms appear in a similar fashion in both the strain production (\ref{eq.RSA}) and enstrophy production (\ref{eq.enstroprodA}) equations. Thus, in this section, we determine the relative significance of non-normal production, $\mbox{R}_{C}^{(S)}$ and interaction production, $\mbox{tr}(\mathsf{\Omega}_{C}^{2}\mathsf{S}_{B})$ relative to the two normal terms: normal strain production, $\mbox{R}_{B}^{(S)}$, and normal enstrophy production, $\mbox{tr}(\mathsf{\Omega}_{B}^{2}\mathsf{S}_{B})$. 

First, we examine the extent to which the sum of the component terms, i.e. $\mbox{R}^{S}_{A}$ and $\mbox{tr}(\mathsf{\Omega}_{A}^{2}\mathsf{S}_{A})$ equates to the sum of the absolute values for each term as a means to assess the extent to which component terms are opposite in sign to $\mbox{R}^{(S)}_{A}$ and $\mbox{tr}(\mathsf{\Omega}_{A}^{2}\mathsf{S}_{A})$:
\begin{align}
[\mbox{R}^{(S)}_{A}]^{*} &= \frac{\mbox{R}^{(S)}_{A}}{|\mbox{R}_{B}^{(S)}| + |\mbox{R}_{C}^{(S)}| + |\mbox{tr}(\mathsf{S}_{C}^{2}\mathsf{S}_{B})|} \\
[\mbox{tr}(\Omega^{2}_{A}{S}_{A})]^{*} &= \frac{\mbox{tr}(\Omega^{2}_{A}{S}_{A})}{|\mbox{tr}(\Omega^{2}_{B}{S}_{B})| + |\mbox{R}_{C}^{(S)}| + |\mbox{tr}(\mathsf{S}_{C}^{2}\mathsf{S}_{B})|}
\end{align}

\begin{figure*}
  \includegraphics[width=0.85\textwidth]{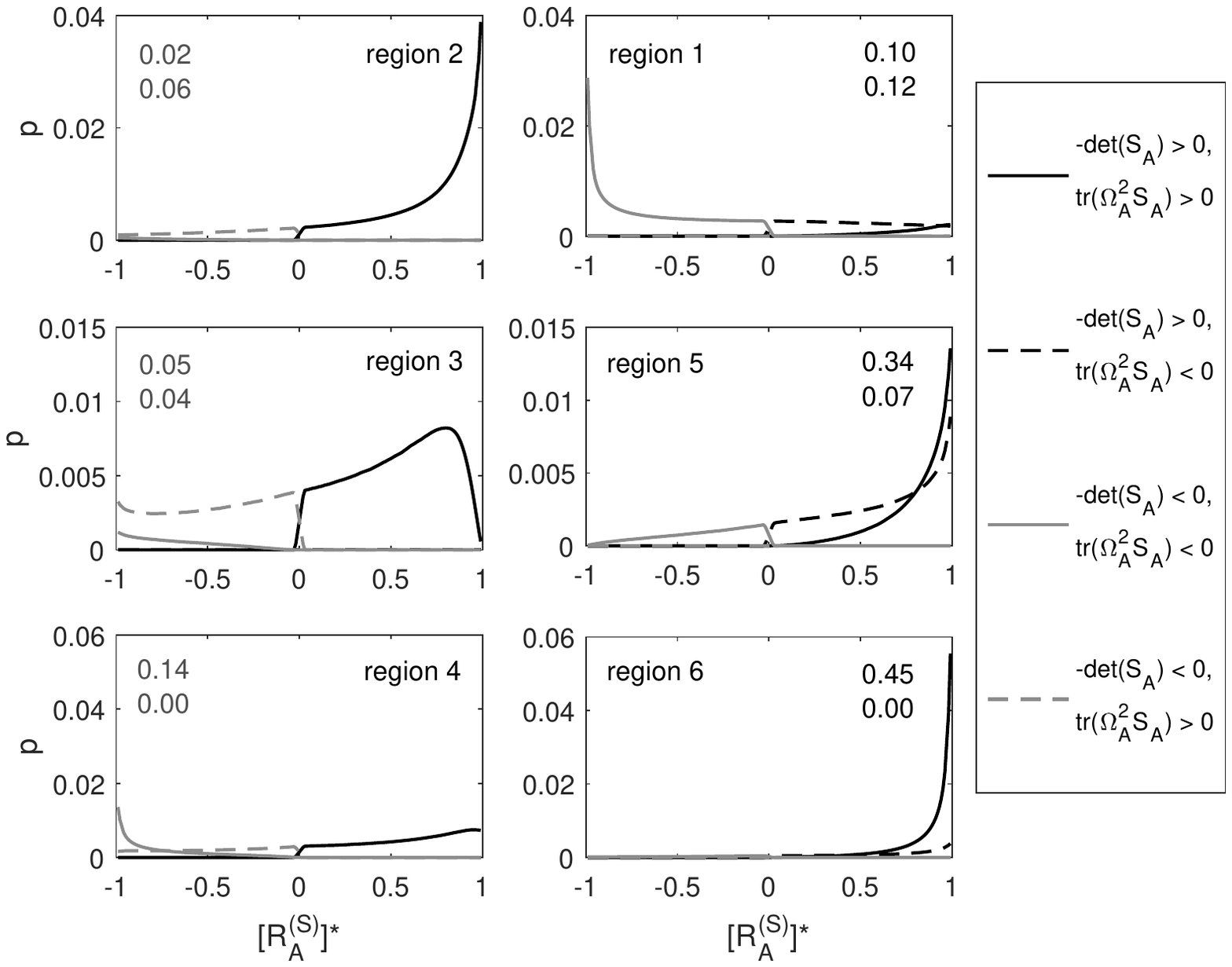}
\caption{Probability curves for $[\mbox{R}^{(S)}_{A}]^{*}$ as a function of the six regions of the $Q_{A}-R_{A}$ diagram and the four possible states of the strain production and enstrophy production for $\mathsf{A}$ and truncated just inside the limits of $[\mbox{R}^{(S)}_{A}]^{*} = \pm 1$. Results are normalized such that the integrated probability in each of the panels is unity. The numbers in each panel are the values for $ [\mbox{R}^{(S)}_{A}]^{*} = -1$ (left-hand panels) and $ [\mbox{R}^{(S)}_{A}]^{*} = 1$ (right-hand panels), where the grey numbers correspond to $\mbox{R}^{(S)}_{A} < 0$ and the black $\mbox{R}^{(S)}_{A} > 0$, where the top numbers in each panel are where the sign for $\mbox{R}^{(S)}_{A}$ and for $\mbox{tr}(\Omega^{2}_{A}{S}_{A})$ are the same.}
\label{fig.sssAbs}       
\end{figure*}

Figure \ref{fig.sssAbs} shows the values for $[\mbox{R}^{(S)}_{A}]^{*}$ as a function of the region of the $Q_{A}-R_{A}$ diagram and the signs of $\mbox{R}^{(S)}_{A}$ and for $\mbox{tr}(\Omega^{2}_{A}{S}_{A})$. The results reflect those seen in Fig. \ref{fig.kappaBC} in that for each pair of panels, departures from $[\mbox{R}^{(S)}_{A}]^{*} = \pm 1$ are always more probable on the $\mbox{R}_{A} < 0$ side.

\begin{figure*}
  \includegraphics[width=0.85\textwidth]{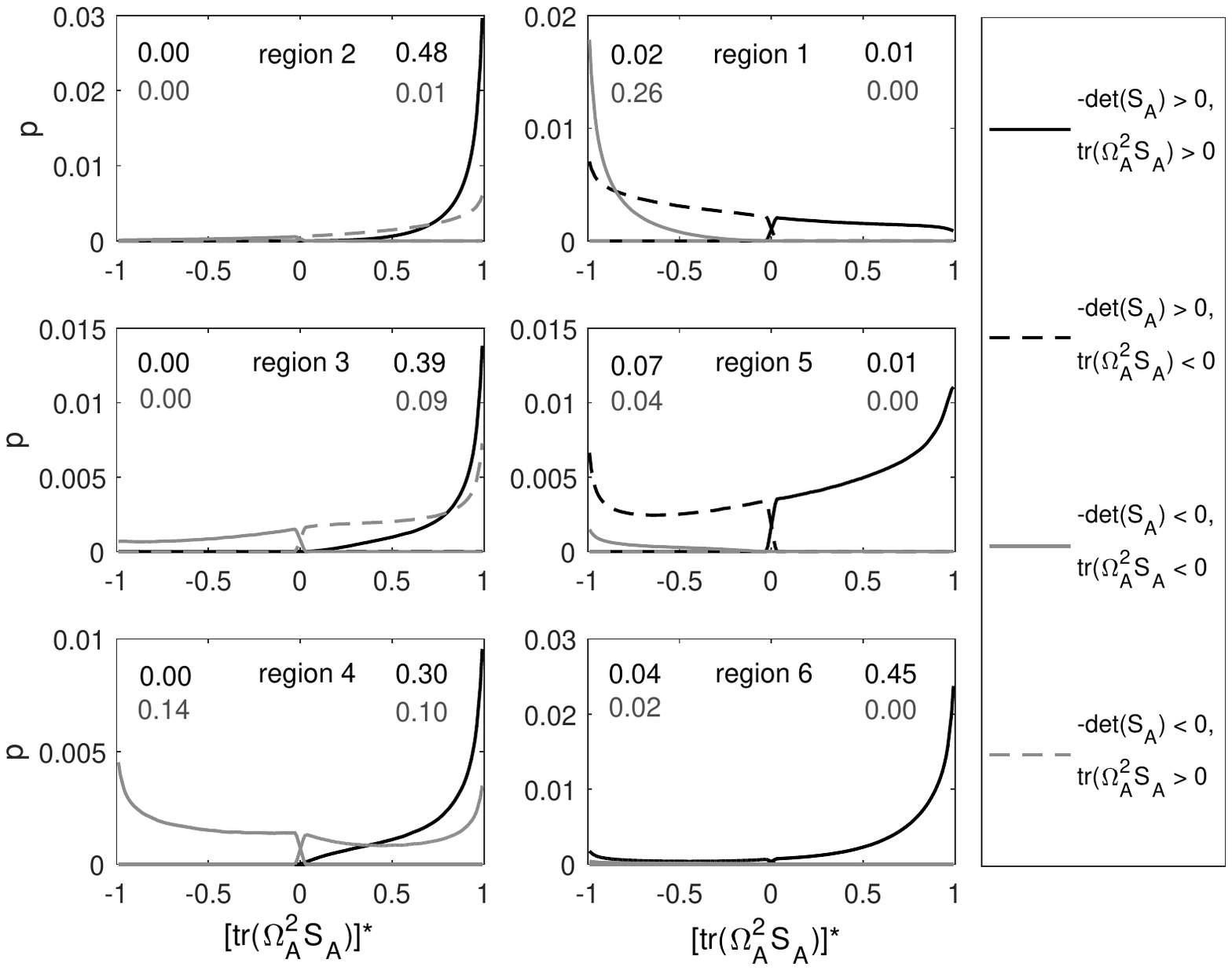}
\caption{Probability curves for $[\mbox{tr}(\mathsf{\Omega}^{2}_{A}\mathsf{S}_{A}]^{*}$ as a function of the six regions of the $Q_{A}-R_{A}$ diagram and the four possible states of the strain production and enstrophy production for $\mathsf{A}$ and truncated just inside the limits of $[\mbox{tr}(\mathsf{\Omega}^{2}\mathsf{S}_{A}]^{*} = \pm 1$. Results are normalized such that the integrated probability in each of the panels is unity. The left hand values in each panel are the probabilities for $[\mbox{tr}(\mathsf{\Omega}^{2}_{A}\mathsf{S}_{A}]^{*} = -1$ with grey for the $R^{(S)}_{A} < 0$, $\mbox{tr}(\mathsf{\Omega}^{2}_{A}\mathsf{S}_{A}) < 0$ curve and black for the $R^{(S)}_{A} > 0$, $\mbox{tr}(\mathsf{\Omega}^{2}_{A}\mathsf{S}_{A}) < 0$ curve. The right hand values in each panel are the probabilities for $[\mbox{tr}(\mathsf{\Omega}^{2}_{A}\mathsf{S}_{A}]^{*} = +1$ with grey for the $R^{(S)}_{A} < 0$, $\mbox{tr}(\mathsf{\Omega}^{2}_{A}\mathsf{S}_{A}) > 0$ curve and black for the $R^{(S)}_{A} > 0$, $\mbox{tr}(\mathsf{\Omega}^{2}_{A}\mathsf{S}_{A}) > 0$ curve.}
\label{fig.wwsAbs}       
\end{figure*}

We might have inferred from Table \ref{table.Q_wws_sss} that the cause of the different degree of importance on the $\mbox{R}_{A} < 0$ side of the $\mbox{Q}_{A} - \mbox{R}_{A}$ diagram was related to the negative strain production - positive enstrophy production state, i.e. the extension of rod-like structures. However, we see in Fig. \ref{fig.sssAbs} that the state where both terms are positive is also more likely to have a significant non-normal contribution to $[\mbox{R}^{(S)}_{A}]^{*}$ than is the case on the $\mbox{R}_{A} > 0$ side. That it is the strain production that is preferentially influenced by non-local effects is shown by comparing the results for  $[\mbox{R}^{(S)}_{A}]^{*}$ with those for  $[\mbox{tr}(\mathsf{\Omega}^{2}_{A}\mathsf{S}_{A}]^{*}$ in Fig. \ref{fig.wwsAbs}, where a large proportion of the values are at $[\mbox{tr}(\mathsf{\Omega}^{2}_{A}\mathsf{S}_{A}]^{*} = 1$. The exceptions to this are regions 1 and 5, where the two states with positive strain production are particularly prone to non-normal interactions. This bias in favour of non-normal interactions for strain production is expected based on the pressure Hessian appearing in the dynamic equation for total strain, but not enstrophy. Hence, it is an indirect mechanism that leads to an impact of non-normality on enstrophy production. For example, we see from Table \ref{table.Q_wws_sss} that negative strain production and enstrophy production is particularly prevalent in region 1. Hence, dissipation by the compression of rod-like turbulent structures \citep{tsinober01} in this region, itself affected significantly by non-normality for strain production (Fig. \ref{fig.sssAbs}) provides a means to induce non-normal effects on the enstrophy production.

\subsection{The four production terms}
\begin{figure*}
  \includegraphics[width=0.85\textwidth]{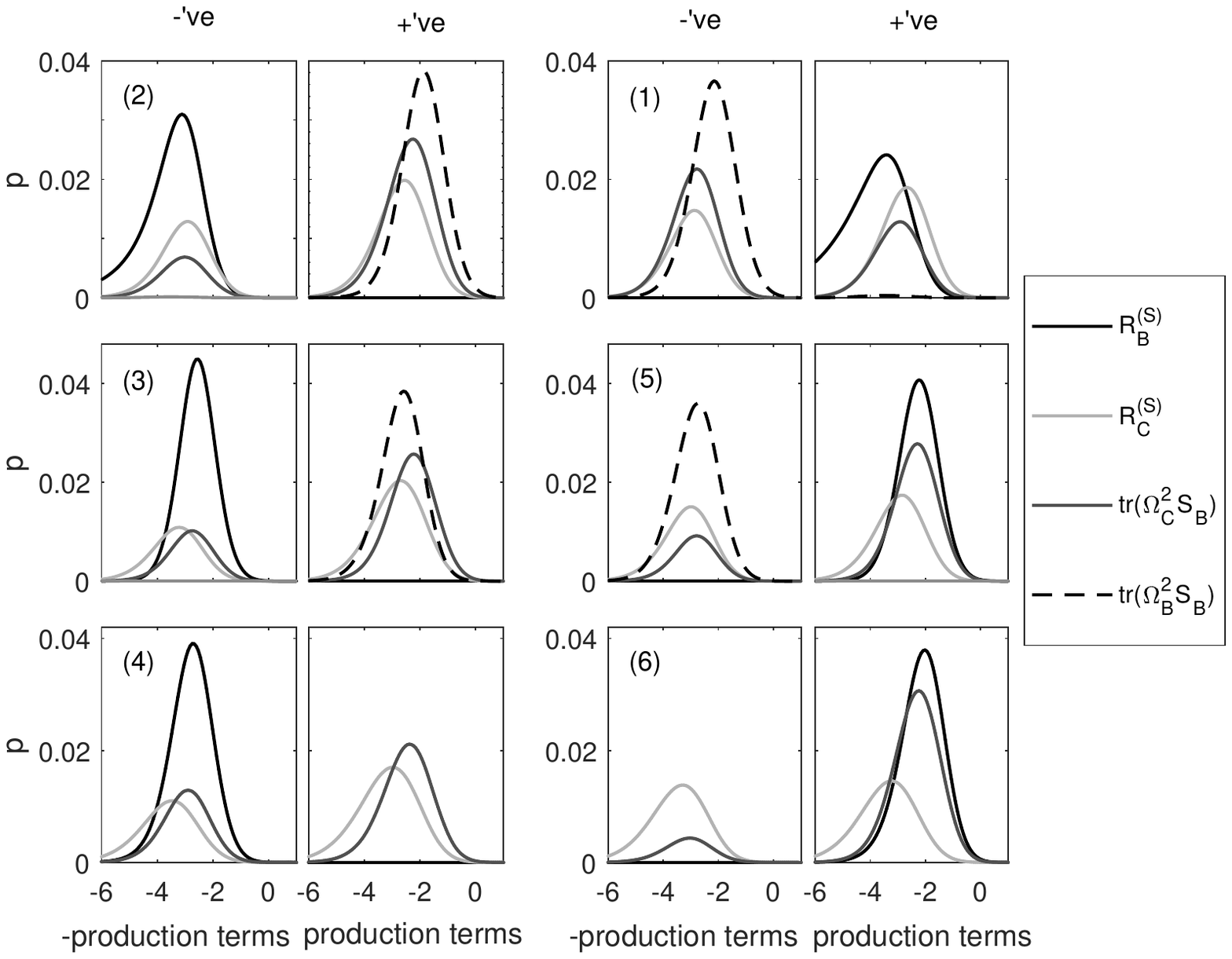}
\caption{Probability curves (on $\mbox{log}_{10}$ axes) for all the production terms conditioned on the region of the $Q_{A}-R_{A}$ diagram and the signs of the terms. Results are non-dimensionalized by the Kolmogorov time cubed and for each pair of panels, the left-hand panels are the distribution functions for the negative values (results multiplied by -1 before taking the logarithm) and the right-hand panels are the positive values.}
\label{fig.prodterms}       
\end{figure*}

The probability curves for our four production terms are given in Fig. \ref{fig.prodterms} conditioned on both the sign of the term and the regions of the $\mbox{Q}_{A}-\mbox{R}_{A}$ diagram. As explained in section \ref{sect.physical}, $\mbox{sgn}(\mbox{R}_{B}^{(S)}) =  \mbox{sgn}(\mbox{R}_{A})$ and $\mbox{sgn}(\mbox{tr}(\mathsf{\Omega}^{2}_{B}\mathsf{S}_{B})) =  -\mbox{sgn}(\mbox{R}_{A})$, while $\mbox{tr}(\mathsf{\Omega}^{2}_{B}\mathsf{S}_{B}) = 0$ beneath the discriminant function (regions 4 and 6). Thus, the basic properties of these terms is known from our formulation.

While $\mbox{Q}_{A} > 0$ is a useful visualization tool, it is of less obvious physical significance than $\Delta_{L} > 0$. However, it is fairly clear from Fig. \ref{fig.prodterms} that apart from the obvious difference that there is no curve for $\mbox{tr}(\mathsf{\Omega}^{2}_{B}\mathsf{S}_{B})$ where $\Delta_{L} < 0$, the other curves exhibit a clearer difference between positive and negative $\mbox{Q}_{A}$ states than positive and negative $\Delta_{L}$ states. For example, in regions 5 and 6 we see there is very little difference in the distribution functions for $\mbox{R}_{B}^{(S)}$ and $\mbox{R}_{C}^{(S)}$, while there is a small decrease in the probability of negative values for $\mbox{tr}(\mathsf{\Omega}^{2}_{C}\mathsf{S}_{B})$ as we move from region 5 to region 6. Hence, the change in the nature of the balance between strain production and enstrophy production and, thus, $\mbox{R}_{A}$ as a flow parcel moves from region 5 to region 6 is driven almost entirely by the existence of a negative contribution from $\mbox{tr}(\mathsf{\Omega}^{2}_{B}\mathsf{S}_{B})$ in region 5 and its absence in region 6. Otherwise, we see that the positive and negative values for $\mbox{R}_{C}^{(S)}$ approximately cancel while the interaction term is biased towards positive values. Thus, we may simplify the strain production and enstrophy production balance in region 6 to being about positive strain production and positive stretching of non-normality by the normal strain tensor. Region 5 has the same terms acting with similar strength, with an additional negative contribution by the enstrophy production term. In regions 3 and 4 we see something close to the mirror image of the behaviour in regions 5 and 6 for the two normal terms, but with the distributions for the non-normal and interaction terms similar to those in regions 5 and 6. However, $\mbox{R}_{C}^{(S)}$ favours positive values more strongly in regions 3 and 4 than in regions 5 and 6, while $\mbox{tr}(\mathsf{\Omega}^{2}_{C}\mathsf{S}_{B})$ is less biased towards positive values, particularly in region 4. 

Regions 1 and 2 have qualitatively different properties to the other cases:
\begin{itemize}
\item The magnitude of the values for $\mbox{R}_{B}^{(S)}$ is drastically reduced on average in both regions 1 and 2;
\item The magnitude of the values for $\mbox{tr}(\mathsf{\Omega}^{2}_{B}\mathsf{S}_{B})$ increases greatly on average in both regions 1 and 2;
\item In region 1, it is negative values for the interaction term, $\mbox{tr}(\mathsf{\Omega}^{2}_{C}\mathsf{S}_{B})$, that are more probable, the only region in the $\mbox{Q}_{A}-\mbox{R}_{A}$ diagram where this is the case. The positive values are also reduced in magnitude compared to other regions;
\item In region 1 we see positive contributions from $\mbox{R}_{C}^{(S)}$ with magnitudes exceeding all other terms. This is the only region where this is the case and means that where $\mbox{Q}_{A} > 0$, $\mbox{R}_{A} > 0$ states are driven by both positive non-normality and negative normal enstrophy production;
\item In region 2, there is an increase in the bias towards positive values for both $\mbox{R}_{C}^{(S)}$ and $\mbox{tr}(\mathsf{\Omega}^{2}_{C}\mathsf{S}_{B})$, although the magnitude of these positive contributions is still dominated by normal enstrophy production, meaning that where $\mbox{Q}_{A} > 0$, $\mbox{R}_{A} < 0$ states are driven by this positive enstrophy production.
\end{itemize}
It was suggested in the previous subsection that in region 1, compression of vortex tubes would be a means by which non-normal strain production and enstrophy production is significant when both strain production and enstrophy production are negative. How this arises is explained by these features of region 1. Because normal strain production is positive, its eigenvalues can only lead to disc-like structures in this region. Hence, for rod-like structures to develop, the combined negative effect of the non-normal and interaction term must be sufficient to result in $\mbox{R}_{A}^{(S)} < 0$, and the frequency of negative values for the interaction term has already been commented on, and appears to be driving such occurrences. This is confirmed below when we look at the joint properties of these terms, although the extent of this effect is mediated by the positive values for the non-normal term. Given that $\mbox{tr}(\mathsf{\Omega}^{2}_{B}\mathsf{S}_{B})$ is negative and large in magnitude, compression of these tubes that result from non-normal effects is readily facilitated.

\subsection{Joint behaviour of the production terms}

\begin{figure*}
  \includegraphics[width=0.85\textwidth]{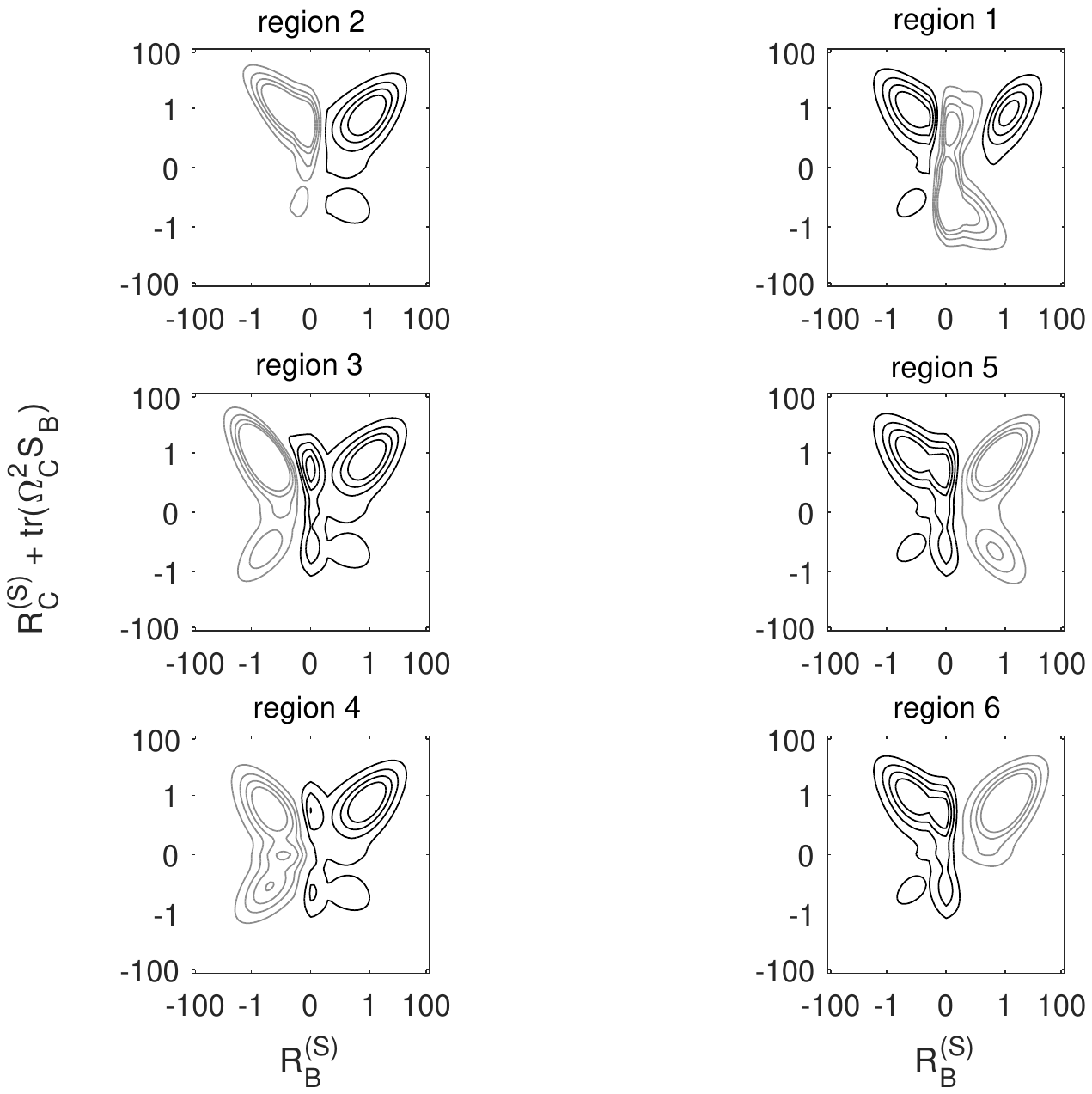}
\caption{Joint probability distributions (on $\mbox{log}_{10}$ axes) for $\mbox{R}_{B}^{(S)}$ and the combined non-normal and interaction terms conditioned on the region of the $Q_{A}-R_{A}$ diagram, and written as the difference between the distribution in this region and the overall distributions function. Positive values indicate an excess for this region and have the lighter probability contours. Contours are in intervals of $2.5 \times 10^{-4}$ ranging from $2.5 \times 10^{-4} \le |p| \le 1.0 \times 10^{-3}$.}
\label{fig.prod_S_combined}       
\end{figure*}

\begin{figure*}
  \includegraphics[width=0.85\textwidth]{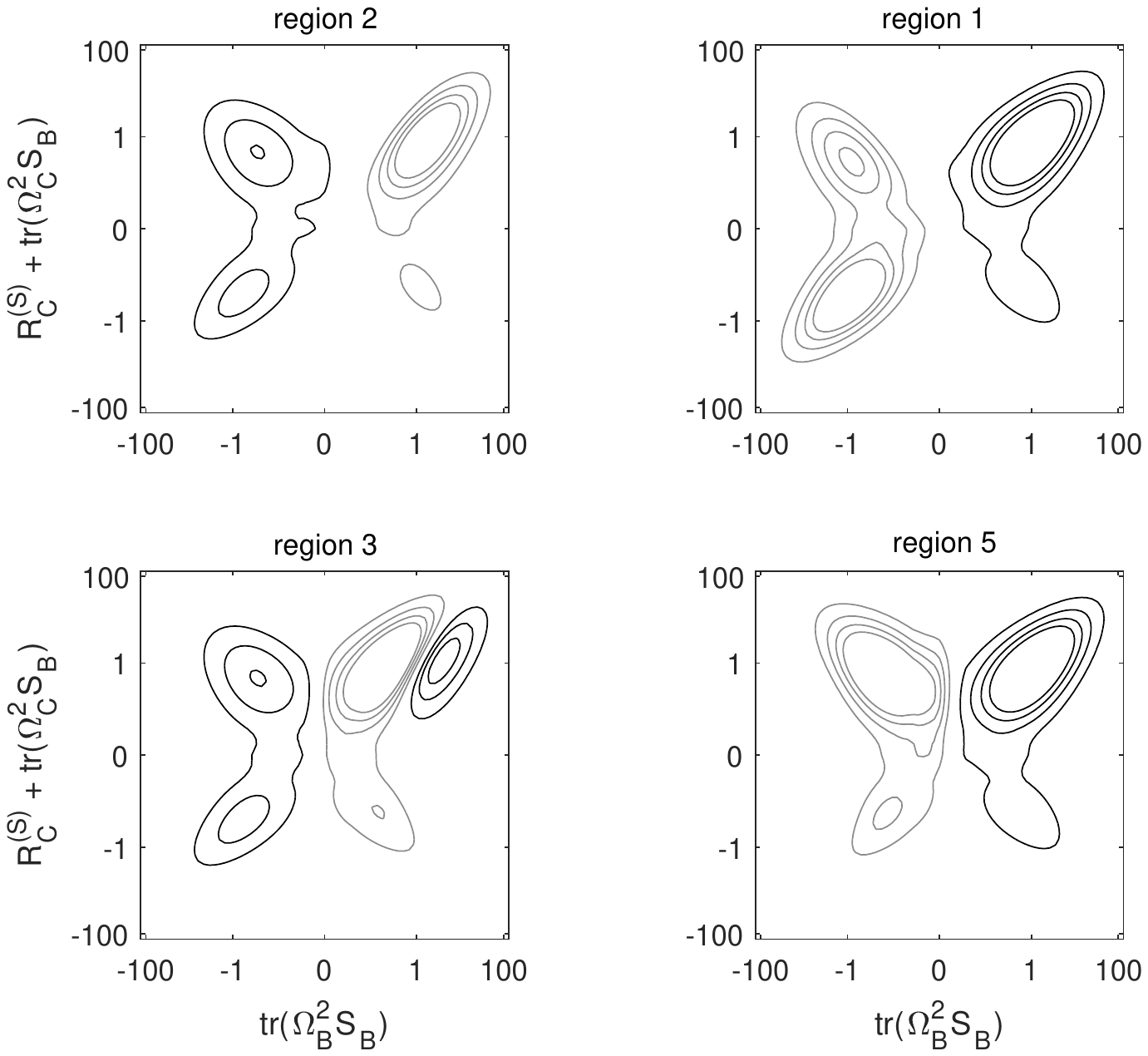}
\caption{Joint probability distributions (on $\mbox{log}_{10}$ axes) for $\mbox{tr}(\mathsf{\Omega}^{2}_{B}\mathsf{S}_{B})$ and the combined non-normal and interaction terms conditioned on the region of the $Q_{A}-R_{A}$ diagram, and written as the difference between the distribution in this region and the overall distributions function. Positive values indicate an excess for this region and have the lighter probability contours. Contours are in intervals of $2.5 \times 10^{-4}$ ranging from $2.5 \times 10^{-4} \le |p| \le 1.0 \times 10^{-3}$.}
\label{fig.prod_Om_combined}       
\end{figure*}

\begin{figure*}
  \includegraphics[width=0.85\textwidth]{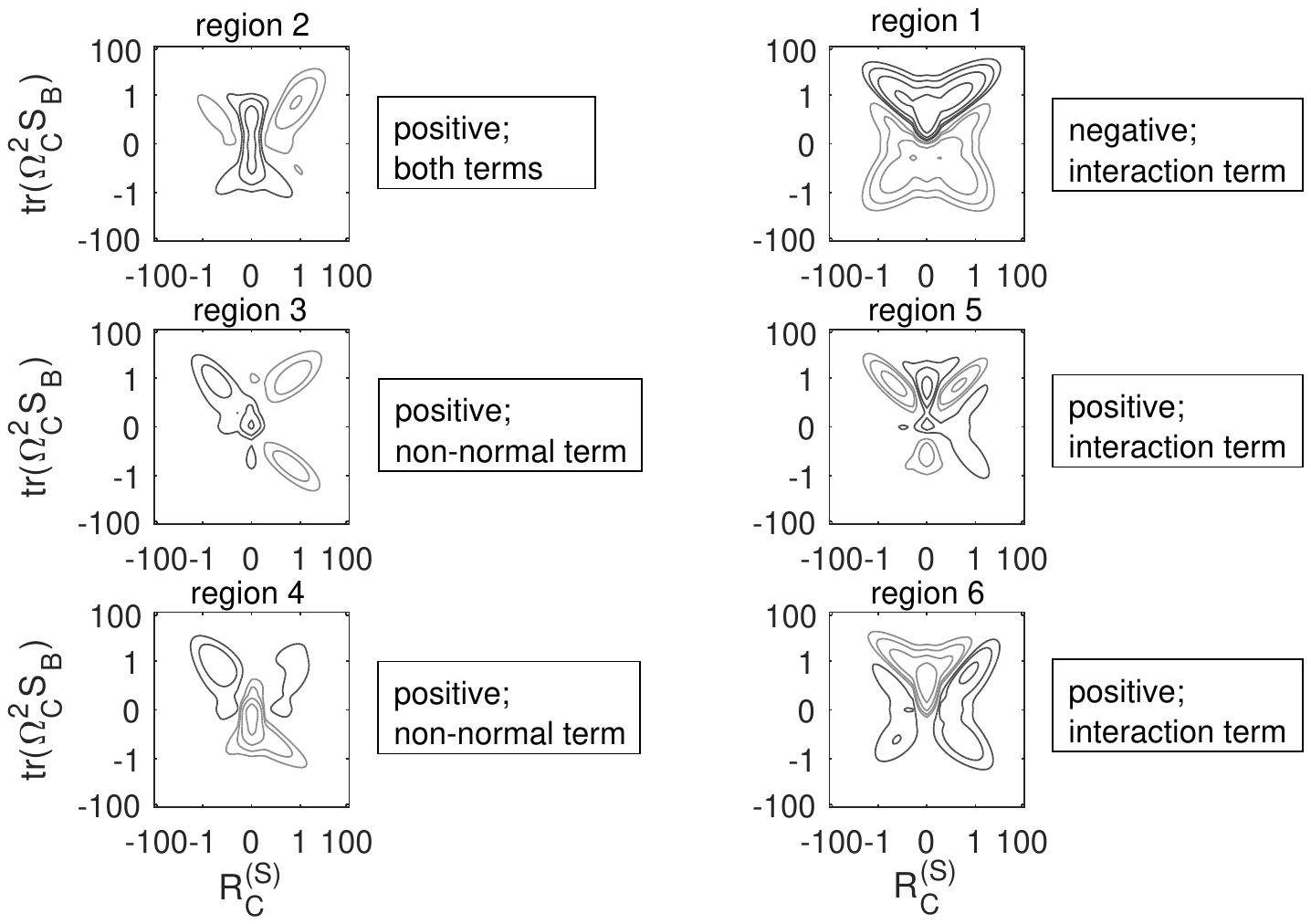}
\caption{Joint probability distributions (on $\mbox{log}_{10}$ axes) for $\mbox{R}_{C}^{(S)}$ and $\mbox{tr}(\mathsf{\Omega}^{2}_{C}\mathsf{S}_{B})$ conditioned on the region of the $Q_{A}-R_{A}$ diagram, and written as the difference between the distribution in this region and the overall distributions function. Positive values indicate an excess for this region and have the lighter probability contours. Contours are at: $\{\pm 1.0 \times 10^{-4}, \pm 2.0 \times 10^{-4}, \pm 4.0 \times 10^{-4}, \pm 8.0 \times 10^{-4}\}$. The text boxes summarize the net effect of the two terms (seen in Fig. \ref{fig.prod_S_combined} and \ref{fig.prod_Om_combined}) and the term(s) that drive this overall pattern.}
\label{fig.prod_C_interact}       
\end{figure*}

Further information on the structure of the production terms can be gleaned from their joint behaviours and we show $\mbox{R}_{B}^{(S)}$ against the combined behaviour of the non-normal and interaction terms (Fig. \ref{fig.prod_S_combined}), $\mbox{tr}(\mathsf{\Omega}^{2}_{B}\mathsf{S}_{B})$ against the combined behaviour of the non-normal and interaction terms (Fig. \ref{fig.prod_Om_combined}), and disaggregate the latter terms into $\mbox{R}_{C}^{(S)}$ and $\mbox{tr}(\mathsf{\Omega}^{2}_{C}\mathsf{S}_{B})$, respectively (Fig. \ref{fig.prod_C_interact}). Each panel in these three figures displays the results for each region of the $\mbox{Q}_{A}-\mbox{R}_{A}$ diagram as a difference from the overall joint PDF for all regions combined, with darker contours a sink for that region of the $\mbox{Q}_{A}-\mbox{R}_{A}$ diagram, and lighter contours an excess. The logarithmic nature of the bins used to generate the PDFs results in a series of ``butterfly'' plots of varying nature. 

In Fig. \ref{fig.prod_S_combined}, the region that most clearly violates the anticipated symmetry is region 1, where the excess is essentially at $\mbox{R}_{B}^{(S)} = 0$, with the bias in the non-normal and interaction terms towards negative values, as anticipated in the previous section. Where $\mbox{R}_{B}^{(S)}$ has larger positive values, in the bottom-right region of this panel, $\mbox{R}_{C}^{(S)} + \mbox{tr}(\mathsf{\Omega}^{2}_{C}\mathsf{S}_{B})$ is also strongly negative, counteracting any tendency from the eigenvalues to drive positive strain production. In Fig. \ref{fig.prod_Om_combined} we see that this negative bias in 
$\mbox{R}_{C}^{(S)} + \mbox{tr}(\mathsf{\Omega}^{2}_{C}\mathsf{S}_{B})$ is strongly correlated with the negative values for normal enstrophy production. We can determine which of the non-normal and interation terms are driving these negative values from Fig. \ref{fig.prod_C_interact}, where we note it is the interaction term that, as already anticipated in the previous section, has a very strong bias to negative values. The non-normal term is very important to the production in region 1, but exhibits both strongly negative and positive contributions with a very weak correlation to the interaction term. To quantify these statements, we determine the cases that lie in region 1 and where $\mbox{tr}(\mathsf{\Omega}^{2}_{C}\mathsf{S}_{B}) < 0$ and then we calculate the mean difference in the absolute production quantities. We find that 
\begin{itemize}
\item $\overline{|\mbox{tr}(\mathsf{\Omega}^{2}_{B}\mathsf{S}_{B})| - |\mbox{R}_{C}^{(S)}|} = 0.027$;
\item $\overline{|\mbox{tr}(\mathsf{\Omega}^{2}_{B}\mathsf{S}_{B})| - |\mbox{tr}(\mathsf{\Omega}^{2}_{C}\mathsf{S}_{B})|} = 0.027$; 
\item $\overline{|\mbox{R}_{B}^{(S)}| - |\mbox{R}_{C}^{(S)}|} = -0.006$; and, 
\item $\overline{|\mbox{R}_{B}^{(S)}| - |\mbox{tr}(\mathsf{\Omega}^{2}_{C}\mathsf{S}_{B})|} = -0.005$.
\end{itemize}
That is, the normal enstrophy production is greater in magnitude on average than either the non-normal or interaction terms, both of which are greater in magnitude than the normal strain production when the interaction term is negative, as may be inferred from Fig. \ref{fig.prodterms}.

With the exception of region 1, the butterfly plots in Fig. \ref{fig.prod_S_combined} exhibit a general symmetry with the combined term strongly biased towards positive values and the sign of $\mbox{R}_{B}^{(S)}$ given by physical constraints, resulting in a positive correlation between the terms on the right-hand side and a negative correlation on the left-hand side. This explains why strain production is highest around the Vieillefosse tail (regions 5 and 6). The results in Fig. \ref{fig.prod_Om_combined} exhibit strong correlations in all four legitimate regions with the sign of this correlation, opposite in sense to that in Fig. \ref{fig.prod_S_combined} because of the opposite sign for $\mbox{tr}(\mathsf{\Omega}^{2}_{B}\mathsf{S}_{B})$. Clearly, region 1 is the unusual case again, with a strong positive correlation in the negative-negative part of this panel. It is also the case that although region 3 exhibits a strong positive correlation in the positive-positive part of the panel, there is a deficit for large values of normal enstrophy production, as can be inferred from the weak marginal distribution for this term in Fig. \ref{fig.prodterms}. The consequence of this is that for positive values of normal enstrophy production, the combined effect of the non-normal and interaction terms outweighs that of the normal term in region 3. An examination of the relevant panel in Fig. \ref{fig.prod_C_interact} shows that, in contrast to region 1, it is $\mbox{R}_{C}^{(S)}$ that drives this behaviour, which is not obvious from the marginal distributions in Fig. \ref{fig.prodterms}.

Given the highly structured nature of the normal terms, the joint distribution function for  the non-normal and interaction terms produces a rather more unique signature in each region than is seen in Figs. \ref{fig.prod_S_combined} and \ref{fig.prod_Om_combined}. This is summarized in the textual descriptions in Fig. \ref{fig.prod_C_interact}, where region 1 is the only region where the net average effect of these two terms is negative. We also see that the interaction term tends to dominate the non-normal contribution on the positive $\mbox{R}_{A}$ side of the $\mbox{Q}_{A}-\mbox{R}_{A}$ diagram, with the non-normal term dominant on the negative side. 

Rather than merely looking at the location of the areas of excess as is summarized in the text boxes in Fig. \ref{fig.prod_C_interact}, we also looked at which term tends to dominate in each of these areas. We extracted the major excess areas for each region in Fig. \ref{fig.prod_C_interact} and determined the mean of the difference of the absolute values, $\overline{|\mbox{tr}(\mathsf{\Omega}^{2}_{C}\mathsf{S}_{B})| - |\mbox{R}_{C}^{(S)}|}$, in each case. These values are given in Table \ref{table.prod_C_interact}. Thus, we can see that the non-normal term is dynamically important in region 1, as has already been reported and is seen in Fig. \ref{fig.prodterms}, but the interaction term is more significant both because the sign of the excess areas is where $\mbox{tr}(\mathsf{\Omega}^{2}_{C}\mathsf{S}_{B} < 0$ and because there is a bias towards the case where both terms are negative (55:45). In contrast, in region 2, we see that the dominant area of excess is where both terms are positive, but that in this region, the interaction term is clearly dominant. In region 3 we see that the strength of the positive net effect comes from a dominant interaction term where both terms are positive, and the magnitude of the  non-normal term exceeding that for the interaction term on average where the former is positive and the latter is negative. This latter situation is what also drives the weak positive net behaviour in region 4, while in regions 5 and 6 it is that the interaction term is positive and much greater in magnitude on average than the non-normal term irrespective of the sign for the latter.

\begin{table}
\caption{Values for $\overline{|\mbox{tr}(\mathsf{\Omega}^{2}_{C}\mathsf{S}_{B})| - |\mbox{R}_{C}^{(S)}|}$ for parts of each region of $\mbox{Q}_{A}-\mbox{R}_{A}$ space where there is a relative excess in Fig. \ref{fig.prod_C_interact}. The identified parts of the distribution function are identified by the signs of the interaction and non-normal terms. Where two states are considered important in a given region, their \% relative frequency is also quoted.}
\centering
\begin{tabular}{ccccl}
\noalign{\smallskip}
Region & $\mbox{sgn}(\mbox{tr}(\mathsf{\Omega}^{2}_{C}\mathsf{S}_{B}))$ & $\mbox{sgn}(\mbox{R}_{C}^{(S)})$ & $\overline{|\mbox{tr}(\mathsf{\Omega}^{2}_{C}\mathsf{S}_{B})| - |\mbox{R}_{C}^{(S)}|}$ & \\
\noalign{\smallskip}
1 & - & - & 0.0015 & 55\%\\
 & - & + & -0.0016 & 45\%\\
2 & + & + & 0.0136 & \\
3 & - & + & -0.0048 & 27\% \\
 & + & + & 0.0160 & 73\%\\
4 & - & + & -0.0009 & \\
5 & + & - & 0.0137 & 44\%\\
 & + & + & 0.0120 & 56\%\\
6 & + & - & 0.0197 & 48\%\\
 & + & + & 0.0188 & 52\%\\
\noalign{\smallskip}
\label{table.prod_C_interact}
\end{tabular}
\end{table}

\subsection{The square of the vortex stretching term}

\begin{figure*}
  \includegraphics[width=0.85\textwidth]{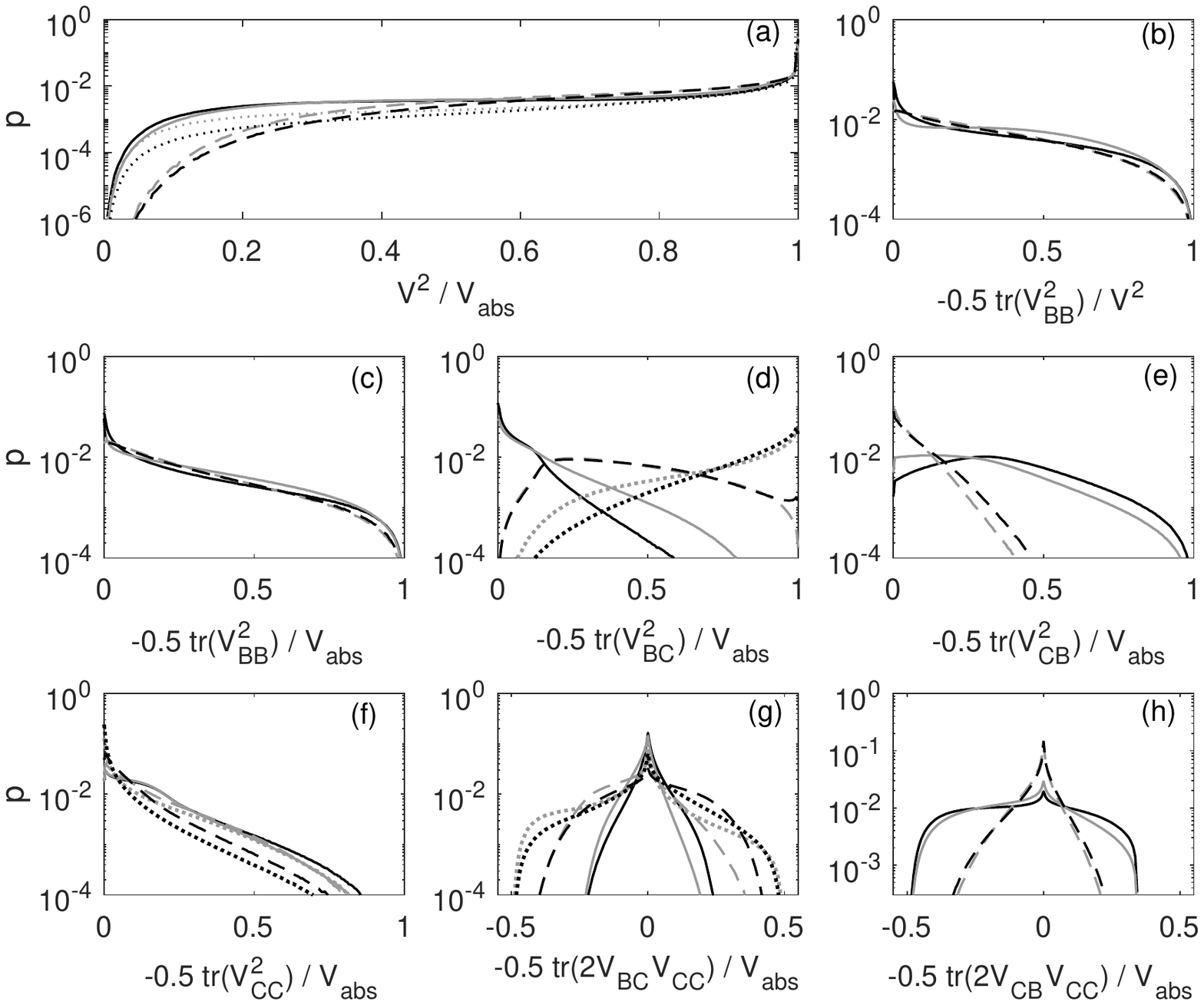}
\caption{Probability curves for the terms contributing to $V^{2}$ normalized by the value for $V_{abs}$. The grey lines indicate regions of $\mbox{Q}_{A} - \mbox{R}_{A}$ space where $\mbox{R}_{A} < 0$ and the black where $\mbox{R}_{A} > 0$. The dotted lines show $\Delta_{L} < 0$, the dashed lines are for $\mbox{Q}_{A} < 0, \Delta_{L} > 0$ and the solid lines are for $\mbox{Q}_{A} > 0$.}
\label{fig:V2hists}       
\end{figure*}

\begin{table}
\caption{Values of the cumulative distribution for $V^{2} / V_{abs}$ for each region of the $\mbox{Q}_{A}-\mbox{R}_{A}$ diagram. For example, $[V^{2} / V_{abs}]_{50}$ is the 50th percentile (median) of the distribution function.}
\centering
\begin{tabular}{lcccc}
\noalign{\smallskip}
Region & $[V^{2} / V_{abs}]_{5}$ & $[V^{2} / V_{abs}]_{25}$ & $[V^{2} / V_{abs}]_{50}$ & $[V^{2} / V_{abs}]_{75}$\\
\noalign{\smallskip}
1 & 0.205 & 0.510 & 0.830 & 0.990\\
2 & 0.225 & 0.515 & 0.810 & 0.970 \\
3 & 0.375 &  0.650 & 0.850 & 0.975\\
4 & 0.300 & 0.770 & 0.975 & > 0.995\\
5 & 0.435 & 0.723 &  0.905 & 0.995\\
6 & 0.425 & 0.855 & 0.995 & >0.999\\
\noalign{\smallskip}
\label{table.V2}
\end{tabular}
\end{table}

As described in (\ref{eq.Vterms}), we decomposed the nonlinear vortex stretching term, $V^{2}$, into four component terms (one purely normal term, $\mathsf{V}_{BB} = \mathsf{\Omega}_{B}\mathsf{S}_{B} + \mathsf{S}_{B}\mathsf{\Omega}_{B}$, a purely non-normal term, $\mathsf{V}_{CC}$, and two interaction terms, $\mathsf{V}_{BC}$ and $\mathsf{V}_{CB}$). In addition, we defined $V_{abs}$ to highlight that the products of the interaction terms with the non-normal term can be negative. In Fig. \ref{fig:V2hists}a we show the full distribution for $V^{2}/V_{abs}$ on a $\mbox{log}_{10}$ scale and some properties of the cumulative distribution function for this term are listed in Table \ref{table.V2}. The two product terms are shown in Fig. \ref{fig:V2hists}g and Fig. \ref{fig:V2hists}h and the values for the fifth percentile of the distribution (in Table \ref{table.V2}) highlight that these product terms have a greater effect on the vortex stretching where $\mbox{Q}_{A} > 0$ and is least important near the Vieillefosse tail, consistent with earlier results. Given that $\mathsf{\Omega}_B$ does not exist beneath the discriminant function, $\mbox{tr}(\mathsf{V}_{CB}\mathsf{V}_{CC})$ will increase in magnitude with $\mbox{Q}_{A}$ as is shown in Fig. \ref{fig:V2hists}h (with very little difference as a function of $\mbox{R}_{A}$). While the magnitude of $\mbox{tr}(\mathsf{V}_{BC}\mathsf{V}_{CC})$ exhibits the expected opposite behaviour in Fig. \ref{fig:V2hists}g, the key difference is the symmetric nature of these distributions compared to Fig. \ref{fig:V2hists}h. This means that it is the high $\mbox{Q}_{A}$ regions that show the greater effect of the two product terms in Fig. \ref{fig:V2hists}a and Table \ref{table.V2}. 

Investigating the other panels in Fig. \ref{fig:V2hists}, we see that the distribution function for $\mbox{tr}(V_{BB}^{2})$ in Fig. \ref{fig:V2hists}b and Fig. \ref{fig:V2hists}c is approximately invariant for regions 1, 3 and 5 with a median of $[\mbox{tr}(V_{BB}^{2})/V^{2}]_{50} \sim 0.20$ in the former case, and $[\mbox{tr}(V_{BB}^{2})/V_{abs}]_{50} \sim 0.15$ in the latter. Normal stretching is of greater importance to the dynamics in region 2 (solid grey line), where the corresponding median values are $[\mbox{tr}(V_{BB}^{2})/V^{2}]_{50} = 0.32$ or $[\mbox{tr}(V_{BB}^{2})/V_{abs}]_{50} = 0.20$. Contributions from the non-normal stretching are tiny in region 6 ($[\mbox{tr}(V_{CC}^{2})/V_{abs}]_{50} = 0.015$) and very small in regions 4 and 5 ($[\mbox{tr}(V_{CC}^{2})/V_{abs}]_{50} \sim 0.055$), while the medians for region 1 to 3 range between 9\% and 12\%. 

Given the relatively small contribution from these terms and the symmetric nature of the $\mbox{tr}(\mathsf{V}_{BC}\mathsf{V}_{CC})$ product term, in particular, it is the two squared interaction terms that dominate vortex stretching. With $\mbox{tr}(\mathsf{V}_{CB}^{2})/V_{abs}$ undefined beneath the discriminant function, it is $\mbox{tr}(\mathsf{V}_{BC}^{2})/V_{abs}$ that drives the stretching dynamics, with $[\mbox{tr}(\mathsf{V}_{CB}^{2})/V_{abs}]_{50} = 0.84$ in region 4, and 0.88 in region 6. For regions 3 and 5, while the values are smaller, this is still the most important single term on average, with $[\mbox{tr}(\mathsf{V}_{CB}^{2})/V_{abs}]_{50} \sim 0.4$. The probability curves for $\mbox{tr}(\mathsf{V}_{CB}^{2})/V_{abs}$ in regions 1 and 2 have medians of $[\mbox{tr}(\mathsf{V}_{CB}^{2})/V_{abs}]_{50} = 0.33$ in region 1 and $[\mbox{tr}(\mathsf{V}_{CB}^{2})/V_{abs}]_{50} = 0.235$ in region 2, this difference being the primary way in which the greater emphasis on the normal stretching vector in region 2 is accommodated for when considering the other terms.

In summary, and in common with the analysis of the production terms, analysis of vortex stretching has shown the dynamical importance of the interaction between normal and non-normal terms, particularly below the discriminant function where the non-normal term remains small and the normal term is undefined as a consequence of the absence of a normal rotation tensor in this region. The negative skewness to the distribution for $\mbox{tr}(\mathsf{V}_{CB}\mathsf{V}_{CC})$ in regions 1 and 2, in particular, means that for $\mbox{Q}_{A} > 0$, the value for $V^{2}$ is less effectively determined by the sum of the squared terms, i.e. $\mbox{tr}(\mathsf{V}_{BB}^{2} + \mathsf{V}_{BC}^{2} + \mathsf{V}_{CB}^{2} + \mathsf{V}_{CC}^{2})$, than is the case below the discriminant function as can be seen in Table \ref{table.V2}.

\section{Results: The second eigenvalue of the strain rate tensor for $\mathsf{A}$}
\label{sect.res_eLRA}
\begin{figure*}
  \includegraphics[width=0.85\textwidth]{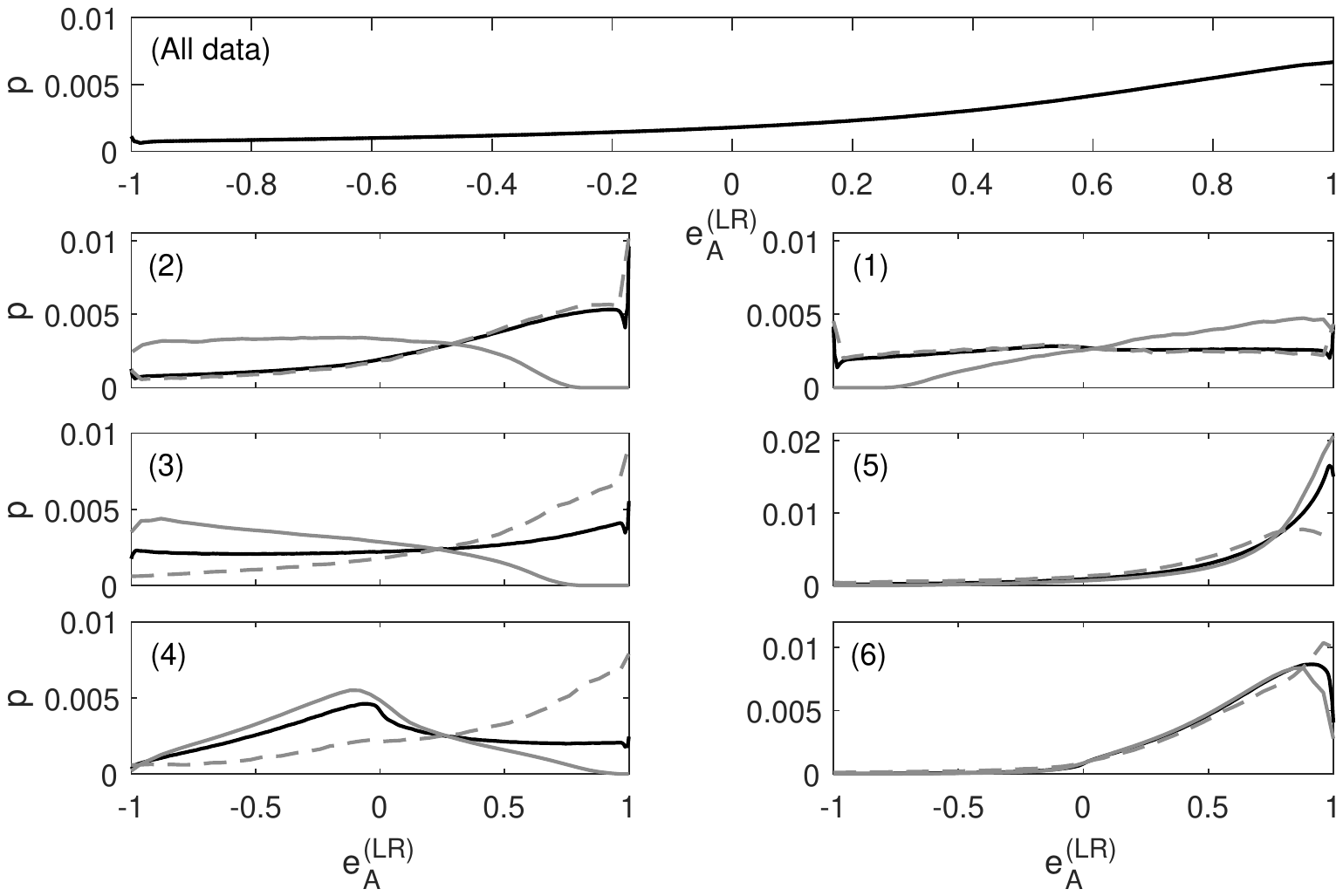}
\caption{Probability curves for $e^{(LR)}_{A}$ for all the data (top panel) and then by each region of the $\mbox{Q}_{A}-\mbox{R}_{A}$ diagram (black lines). We also show results conditioned on the sign of $\kappa_{Q}^{(S)}$, with $\kappa_{Q}^{(S)} > 0$ as a grey solid line and $\kappa_{Q}^{(S)} < 0$ as a grey dashed line.
}
\label{fig.eLRA}       
\end{figure*}
The well-known properties of the Lund and Rogers normalization, $e^{(LR)}_{A}$, are shown in the upper panel of Fig. \ref{fig.eLRA}. The tendency for HIT to form disc-like structures is very evident. Sub-dividing the results by region of the $\mbox{Q}_{A}-\mbox{R}_{A}$ diagram shows that this tendency is driven by regions 2, 5 and 6. However, given that $\mbox{R}_{A} > 0$ in regions 1, 5 and 6, and positive strain production, $-\mbox{det}(\mathsf{S}_{A}) > 0$ means there are two positive strain eigenvalues, it is not intuitive that it is region 2, rather than region 1,  where $e^{(LR)}_{A} \to 1$. Noting in section \ref{sect.ResKappa} that only $\sim 8\%$ of data in regions 1 and 2 have $\kappa_{Q}^{(S)} > 0$, then this property of HIT can be understood when we look at $e^{(LR)}_{A}$ conditioned on the sign of $\kappa_{Q}^{(S)}$ (the grey lines in Fig. \ref{fig.eLRA}). Where $\kappa_{Q}^{(S)} > 0$ (solid grey lines) we see the anticipated behaviour of a tendency for $e^{(LR)}_{A} \to +1$ in region 1 and $e^{(LR)}_{A} \to -1$ in region 2. Consequently, the reason for the observed behaviour is that the non-normal contribution to the strain tensor exceeds the normal part and exhibits a very different distribution to that driven by the eigenvalues. Given that it is in regions 1 and 2 that enstrophy exceeds strain and, according to the Q-criterion \citep{hunt88, dubief00}, is where there is a coherence to flow motion, we can see that non-normality is crucial for the evolution of disc-like structures. Given our earlier result that contributions from $\mathsf{C}$ do not feature in the restricted Euler formulation of the VGT dynamics, it is clear that the anistropic contributions from the pressure Hessian are extremely important for this evolution of disc-like structures. From Table \ref{table.Q_wws_sss}, region 2 is frequented more than twice as often as region 1 (26.5\% to 11.1\%), which also helps explain the strong tendency for $e^{(LR}_{A} \to 1$ seen in the top-most panel. 

From Fig. \ref{fig.kappaQ} we can determine that $\kappa_{Q}^{(S)} > 0$ occurs for 39.4\%, 68.6\%, 60.6\% and 82.2\% of occurrences in regions 3 to 6, respectively. In Fig. \ref{fig.eLRA} it is region 5 where $\kappa_{Q}^{(S)} > 0$ has the strongest tendency to produce values at $e^{(LR)}_{A} = 1$, while everywhere but region 1, one finds that $\kappa_{Q}^{(S)} < 0$ preferentially leads to positive values for $e^{(LR)}_{A}$. The complex case is region 4, where the normal contribution to the strain is strongly dominant, but has a mode at $e^{(LR)}_{A} = -0.08$, indicating that here the preferred state is close to isotropy (the ``blob'' in the terminology of \citet{corrsin72}); any tendency to form disc-like structures is a consequence of the non-normal dominant cases in this region.

\subsection{The disaggregation of the {L}und and {R}ogers normalization of the strain rate tensor}
\label{sect.res_eLRBC}

\begin{figure*}
  \includegraphics[width=0.85\textwidth]{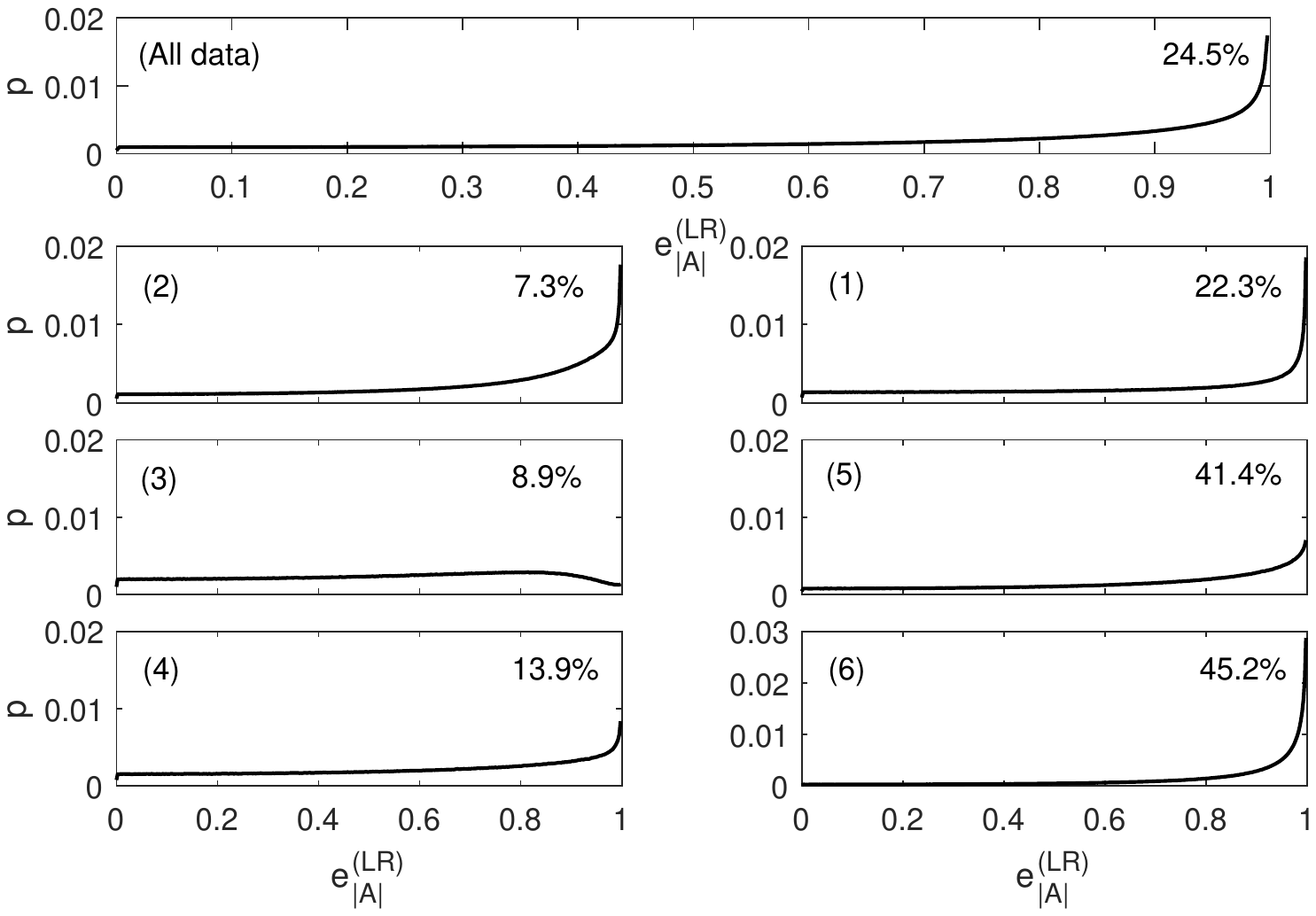}
\caption{Probability curves for $e^{(LR)}_{|A|}$ are shown for all the data and by region of the $\mbox{Q}_{A}-\mbox{R}_{A}$ diagram. Values are shown with a continuous line for $0 \le e^{(LR)}_{|A|} < 1$. The stated value in each panel is the percentage of values for which $e^{(LR)}_{|A|} = 1$, i.e. this is where none of the three terms contributing to the value for $e^{(LR)}_{A}$ have a sign different to that for $e^{(LR)}_{A}$.
}
\label{fig.eLR_AbsA}       
\end{figure*}

We established in Section \ref{sect.eLRABC} that $e^{(LR)}_{A} = e^{(LR)}_{B|A} + e^{(LR)}_{C|A} + e^{(LR)}_{C,B|A}$ and it is the component terms on the right-hand side that we investigate in this section. However, we first examine 
\begin{equation}
e^{(LR)}_{|A|} = \left \lvert \frac{e^{(LR)}_{A}}{|e^{(LR)}_{B|A}| + |e^{(LR)}_{C|A}| + |e^{(LR)}_{C,B|A}|} \right \rvert,
\end{equation}
to gain an insight into how common it is for some of these component terms to have the opposite sign to $e^{(LR)}_{|A|}$ and the relative magnitude of such effects as a function of $\mbox{Q}_{A}$ and $\mbox{R}_{A}$. Given the tendency in HIT for $e^{(LR)}_{A} \to 1$ and that, as explained in section \ref{sect.eLRABC}, $\mbox{sgn}(e^{(LR)}_{B|A}) = \mbox{sgn}(\mbox{R}_{A})$, we expect that $e^{(LR)}_{|A|} \ne 1$ more often in regions 2, 3, and 4, where $\mbox{R}_{A} < 0$. This is the case, particularly where $\Delta_{L} > 0$ in regions 2 and 3 (and to a certain extent in region 1 where $\mbox{R}_{A} > 0$ but $\mbox{Q}_{A} > 0$). Indeed, with the exception of region 1, the overall result that about 25\% of tensors in HIT have $e^{(LR)}_{|A|} = 1$ is simply not representative of the behaviour of the individual regions. 

\begin{figure*}
  \includegraphics[width=0.85\textwidth]{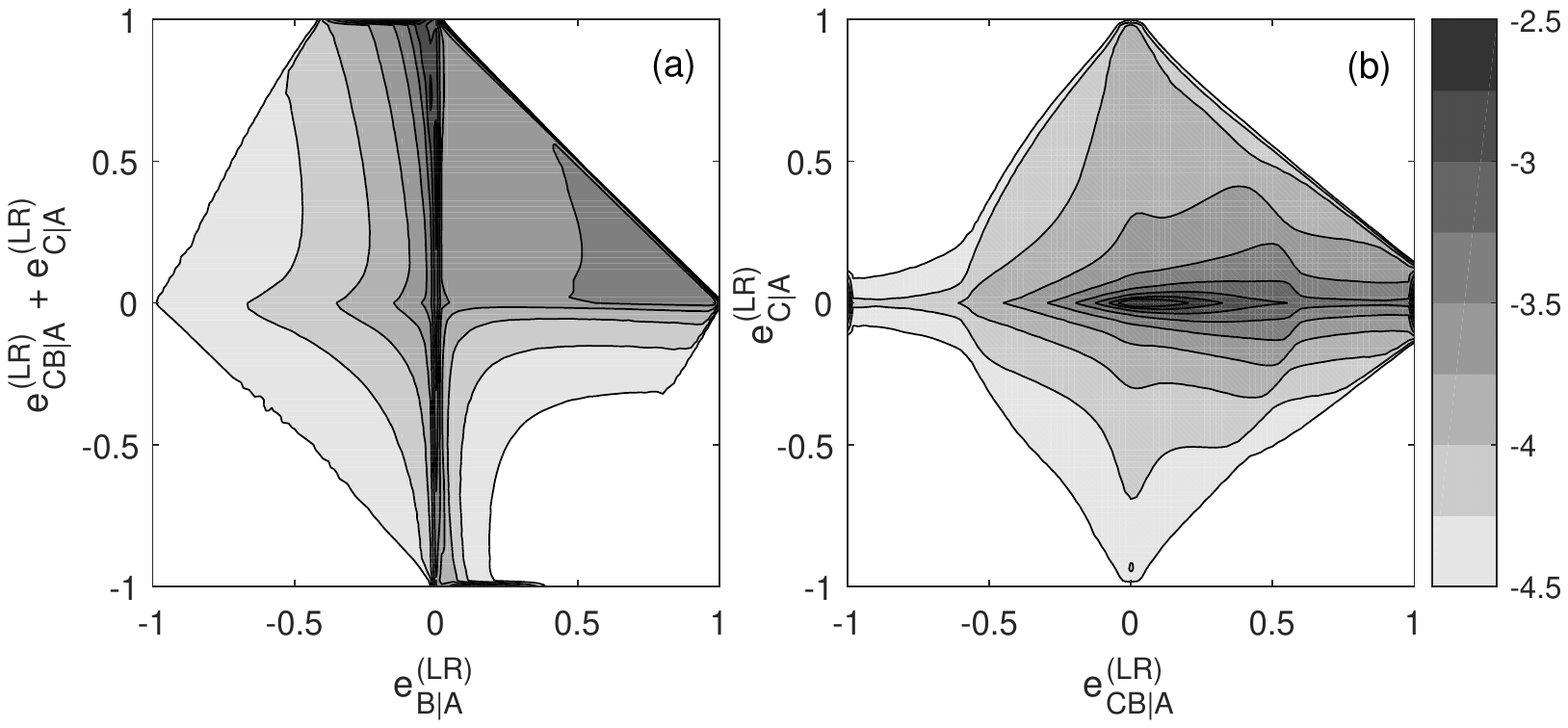}
\caption{Joint distribution functions for $e^{(LR)}_{B|A}$ and $e^{(LR)}_{C|A} + e^{(LR)}_{C,B|A}$ (a) as well as $e^{(LR)}_{C|A}$ and $e^{(LR)}_{C,B|A}$ (b). Contours are on a $\mbox{log}_{10}$ scale.
}
\label{fig.eLRB_C_CB}       
\end{figure*}

It follows from these results that the joint distribution function for the normal  term, $e^{(LR)}_{B|A}$, and the combined non-normal and interaction terms, $e^{(LR)}_{C|A} + e^{(LR)}_{C,B|A}$, should exhibit the strongest probability gradient across the frontier $e^{(LR)}_{B|A} + e^{(LR)}_{C|A} + e^{(LR)}_{C,B|A} = \pm 1$ where both $e^{(LR)}_{B|A}$ and $e^{(LR)}_{C|A} + e^{(LR)}_{C,B|A}$ are positive, with positive values for $e^{(LR)}_{B|A}$ indicating we are in region 1, 5 or 6. This tendency is clear from this joint distribution function in Fig. \ref{fig.eLRB_C_CB}a.

In this panel we see that where $e^{(LR)}_{B|A} < 0$, the combined behaviour of the non-normal and interaction terms behaves with some degree of symmetry about $e^{(LR)}_{C|A} + e^{(LR)}_{C,B|A} = 0$, although with a bias towards positive values as might be expected from the global result of a tendency for $e^{(LR)}_{A} \to 1$. Therefore, positive values for $e^{(LR)}_{A}$ arise even when $e^{(LR)}_{B|A} < 0$ as a consequence of the action of the non-normal and interaction terms as we have already shown in a less direct fashion from the conditioning on $\kappa_{Q}^{(S)}$ in Fig. \ref{fig.eLRA}. 

For positive $e^{(LR)}_{B|A}$ we see little symmetry about $e^{(LR)}_{C|A} + e^{(LR)}_{C,B|A} = 0$. Negative values for the combined term highly improbable with the exception of the limit of $e^{(LR)}_{C|A} + e^{(LR)}_{C,B|A} = -1$. Hence, while there is a weak mechanism for positive $e^{(LR)}_{B|A}$ to result in negative $e^{(LR)}_{A}$, it is rather different in nature and strength to the inverse case. From an inspection of Fig. \ref{fig.eLRB_C_CB}b, this state is most probably realised by $e^{(LR)}_{C|A} = 0$, $e^{(LR)}_{C,B|A} = -1$, again indicating the importance of the interaction term. That there is a peak to the distribution function for $e^{(LR)}_{B|A} > 0.5$ in Fig. \ref{fig.eLRB_C_CB}a that is clearly separated from the dominant ridge along $e^{(LR)}_{B|A} = 0$ highlights that for positive $e^{(LR)}_{B|A}$ both the normal and the other terms are interacting to produce $e^{(LR)}_{A}$ values that tend to 1. This may be contrasted with the negative $e^{(LR)}_{B|A}$ region where there is a much weaker tendency to converge on a particular value for $e^{(LR)}_{A}$. That there is a good degree of symmetry to Fig. \ref{fig.eLRB_C_CB}b about $e^{(LR)}_{C|A} = 0$ indicates that irrespective of the values for $e^{(LR)}_{C,B|A}$, there is no preferred contribution from $e^{(LR)}_{C|A}$. Hence, the term driving the positive bias in values for $e^{(LR)}_{A}$ is the interaction term, $e^{(LR)}_{C,B|A}$. 

\begin{figure*}
  \includegraphics[width=0.85\textwidth]{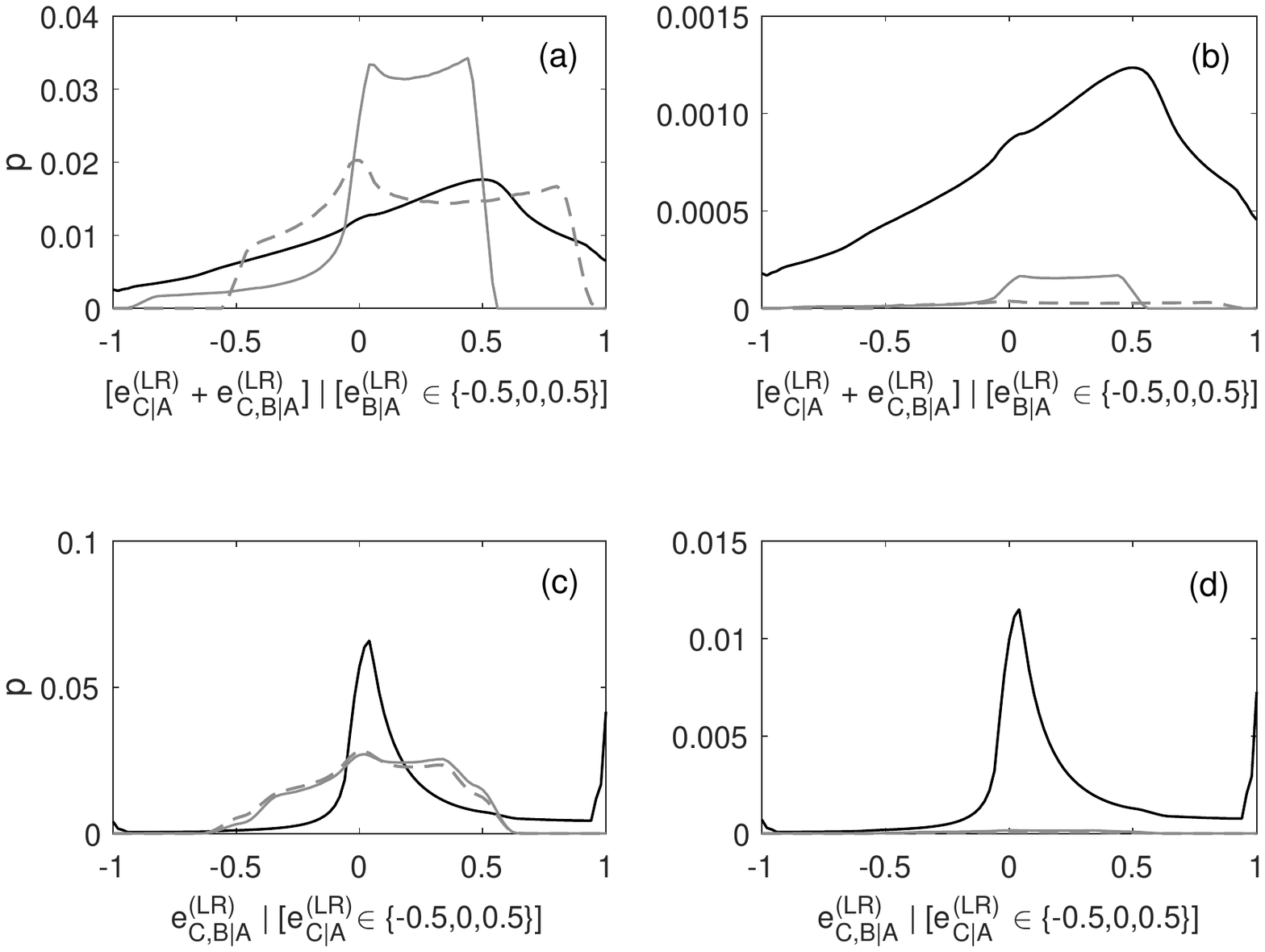}
\caption{Conditional distributions extracted from the joint distribution functions in Fig.\ref{fig.eLRB_C_CB}. The upper panels are related to Fig.\ref{fig.eLRB_C_CB}a and show probabilities for $e^{(LR)}_{C|A} + e^{(LR)}_{C,B|A}$ given $e^{(LR)}_{B|A} = 0$ (black line), $e^{(LR)}_{B|A} = 0.5$ (solid grey line), and $e^{(LR)}_{B|A} = -0.5$ (dashed grey line). The lower panels are extracted from Fig.\ref{fig.eLRB_C_CB}b and show probabilities for $e^{(LR)}_{C,B|A}$ given $e^{(LR)}_{C|A} = 0$ (black line), $e^{(LR)}_{C|A} = 0.5$ (solid grey line), and $e^{(LR)}_{C|A} = -0.5$ (dashed grey line). The left-hand panels are normalized such that the extracted values along this transect integrates to 1. The right-hand panels are normalized such that the full joint distribution integrates to 1. 
}
\label{fig.eLR_conditional}       
\end{figure*}

Some of the patterns described above are more readily discerned by extracting conditional distributions from the joint PDF, which are shown in Fig. \ref{fig.eLR_conditional}. The different normalization between panels (a) and (b) permits the relative and absolute nature of the these conditional distributions to be evaluated. Note, for example, that when $e^{(LR)}_{B|A} = 0$ (black line), the mode of the distribution for the non-normal and interaction terms is 0.5, leading to $e^{(LR)}_{A} = 0.5$. On the other hand, when $e^{(LR)}_{B|A} = 0.5$ (solid grey line), all values for $0 \le e^{(LR)}_{C|A} + e^{(LR)}_{C,B|A} \le 0.5$ are of similar probability, with the mode at $e^{(LR)}_{C|A} + e^{(LR)}_{C,B|A} = 0.5$ slightly larger than that at $e^{(LR)}_{C|A} + e^{(LR)}_{C,B|A} = 0$ and, thus $e^{(LR)}_{A} = 1$ somewhat more likely for these conditions than $e^{(LR)}_{A} = 0.5$. Where $e^{(LR)}_{B|A} = -0.5$ (dashed grey line), the tendency for positive values for $e^{(LR)}_{C|A} + e^{(LR)}_{C,B|A}$ alluded to above is also clear, providing a mechanism to bias the distribution for $e^{(LR)}_{A}$ towards positive values.

The symmetry about $e^{(LR)}_{C|A} = 0$ in Fig. \ref{fig.eLR_conditional}c can be seen in the similar nature of the values for $p(e^{(LR)}_{C,B|A} | [e^{(LR)}_{C|A} = \pm 0.5])$. Figure \ref{fig.eLR_conditional}d highlights the extent to which the mass of the joint distribution is concentrated along $e^{(LR)}_{C|A} = 0$ and both the lower panels show the strongly bimodal nature of $p(e^{(LR)}_{C,B|A} | [e^{(LR)}_{C|A} = 0])$, with maxima at 0 and 1. Hence, if we reduce the effect of $e^{(LR)}_{C|A}$ to a negligible, stochastic perturbation, these two modes provide the end-member states for the $e^{(LR)}_{B|A} + e^{(LR)}_{C,B|A} = 1$ frontier in the positive-positive quadrant of Fig. \ref{fig.eLRB_C_CB}a.

\begin{figure*}
  \includegraphics[width=0.85\textwidth]{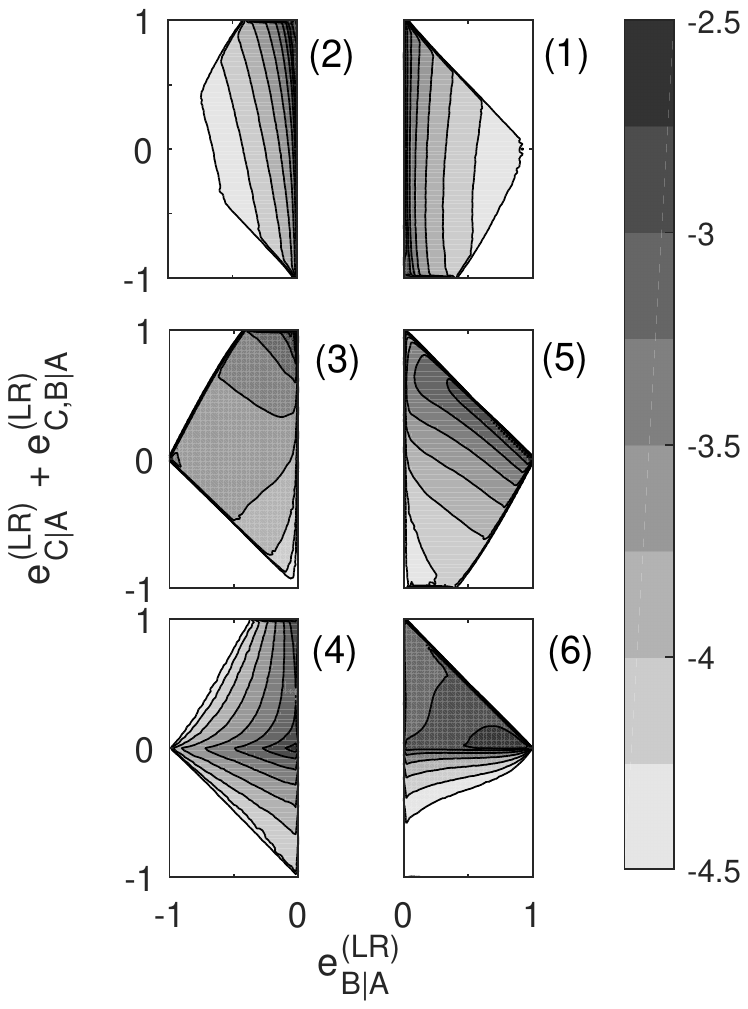}
\caption{Joint distribution functions for $e^{(LR)}_{B|A}$ and $e^{(LR)}_{C|A} + e^{(LR)}_{C,B|A}$ using the same logarithmic scaling for the contours as adopted in Fig. \ref{fig.eLRB_C_CB}, with results sub-divided by the six regions of the $Q_{A}-R_{A}$ diagram defined in Table \ref{table.Q_wws_sss}.
}
\label{fig.eLRB_C_CB_regions}        
\end{figure*}

\subsection{Results conditioned on $\mbox{Q}_{A}-\mbox{R}_{A}$ states}
In order to explore these results further, we look at joint PDFs similar to Fig. \ref{fig.eLRB_C_CB} conditioned on the different regions of $\mbox{Q}_{A}-\mbox{R}_{A}$ space, which are shown in Fig. \ref{fig.eLRB_C_CB_regions}. While the behaviour in regions 1 and 2 ($Q_{A} > 0$) is dominated by gradients in $e^{(LR)}_{B|A} = 0$, this cannot be said of the other regions that all behave very differently. Regarding $e^{(LR)}_{B|A}$, it is regions 5 and 6 where this term generates $e^{(LR)}_{A} = 1$ values as might be anticipated from the earlier analysis showing the importance of the normal tensor in this region. It is region 3 where the normal term generates $e^{(LR)}_{A} = -1$ on its own, and where the strong frontier shows that all terms combine to give this result. Hence,  when $e^{(LR)}_{B|A} = 0$, there is a strong positive bias for the non-normal and interaction term that drives a $e^{(LR)}_{A} = 1$ response. Region 5 has a concentration of values along and near to the frontier in the positive quadrant where the terms interact to give $e^{(LR)}_{A} = 1$, while region 4 can lead to $e^{(LR)}_{A} = -1$, but the dominant behaviour is for $e^{(LR)}_{B|A} = 0$ and for positive values to emerge on average because of the strong positive bias for $e^{(LR)}_{C|A} + e^{(LR)}_{C,B|A}$. Another final means to generate $e^{(LR)}_{A} = -1$ is from the combined effect of the non-normal and interaction terms in regions 1 and 2 when the normal term is expressed most weakly.

\begin{figure*}
  \includegraphics[width=0.85\textwidth]{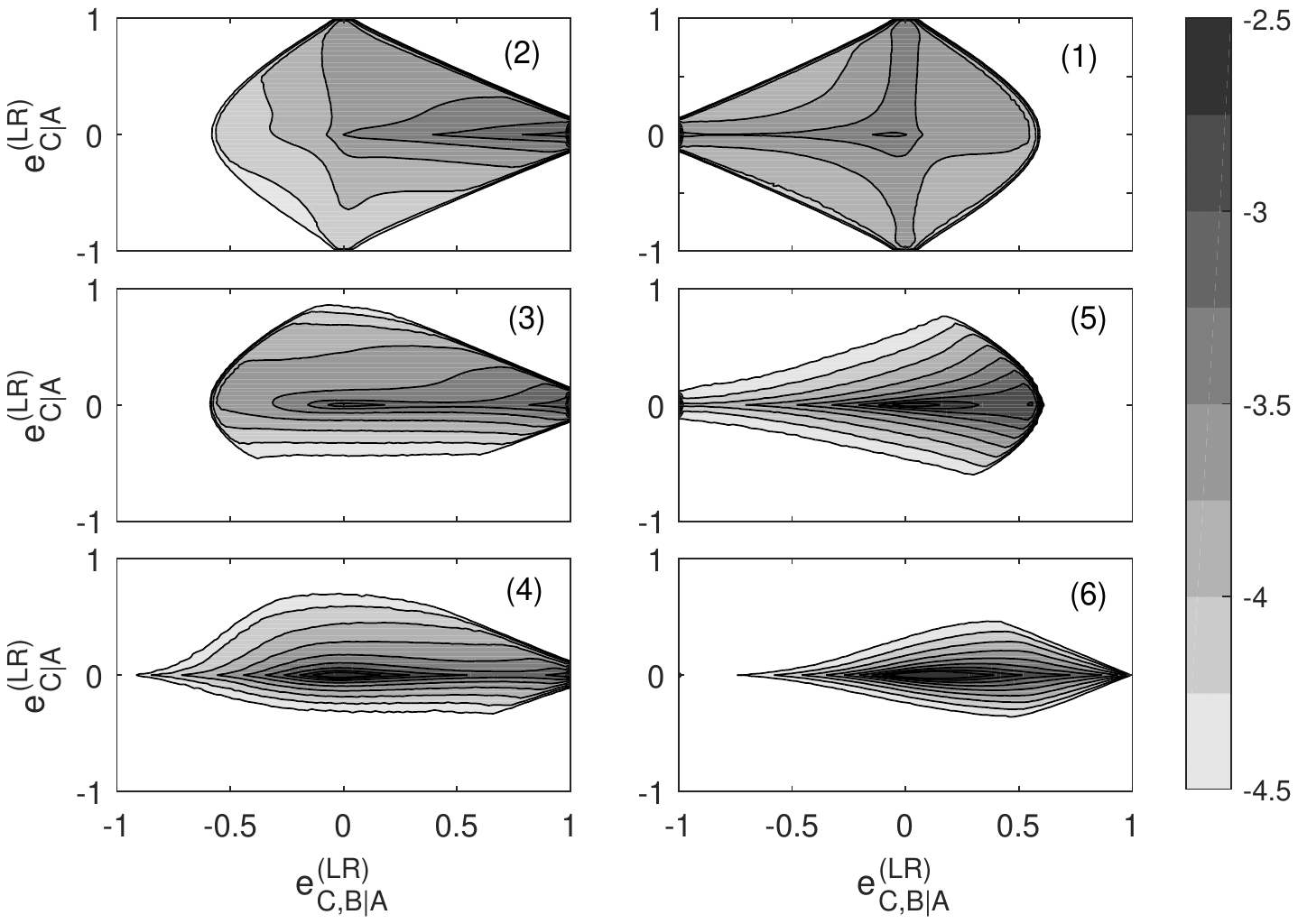}
\caption{Joint distribution functions for $e^{(LR)}_{C|A}$ and $e^{(LR)}_{C,B|A}$ using the same logarithmic scaling for the contours as adopted in Fig. \ref{fig.eLRB_C_CB}, with results sub-divided by the six regions of the $Q_{A}-R_{A}$ diagram defined in Table \ref{table.Q_wws_sss}.
}
\label{fig.eLRC1_C2_regions}        
\end{figure*}

\begin{figure*}
  \includegraphics[width=0.85\textwidth]{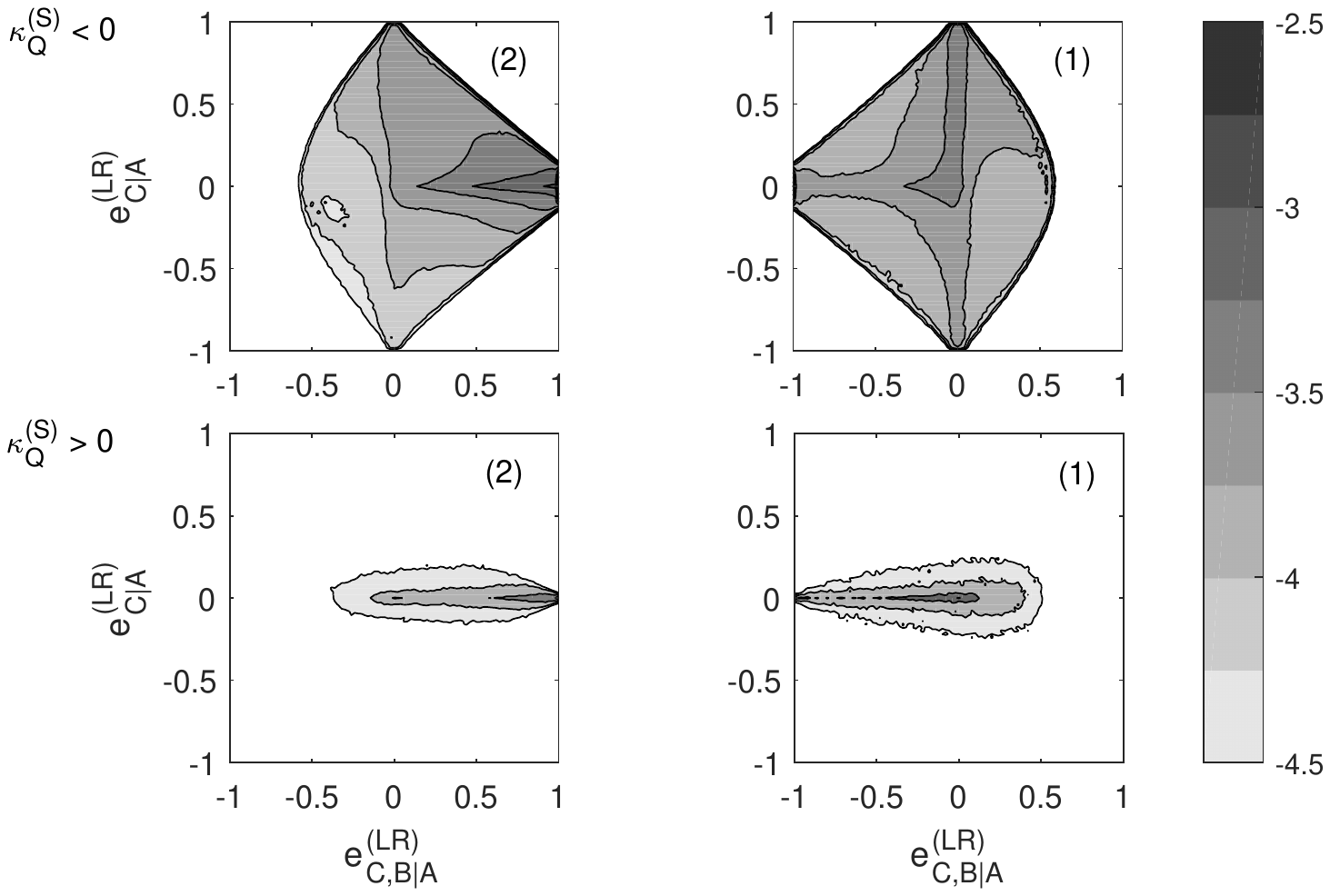}
\caption{Joint distribution functions for $e^{(LR)}_{C|A}$ and $e^{(LR)}_{C,B|A}$ using the same logarithmic scaling for the contours as adopted in Fig. \ref{fig.eLRB_C_CB} and \ref{fig.eLRC1_C2_regions}. Results are shown for regions 1 and 2 defined in Table \ref{table.Q_wws_sss} and subdivided by the sign for $\kappa_{Q}^{(S)}$.
}
\label{fig.eLRC1_C2_condKappa}        
\end{figure*}

The structure of the ordinate in Fig. \ref{fig.eLRB_C_CB_regions} is unpacked in Fig. \ref{fig.eLRC1_C2_regions} where we can again see a great variation in structure of the joint PDFs. Region 1 is particularly interesting as the distributions exhibits a ridge with a 90\degree bend so that in addition to a global maximum at the origin, we see that $e^{(LR)}_{C|A} = +1$ and $e^{(LR)}_{C,B|A} = -1$ have high probability. Such a complex behaviour is not anticipated from the behaviour of $e^{(LR)}_{B|A}$, meaning that in this region, it is the non-normal and the interaction terms rather than the normal term that are driving the topological states as could be discerned in Fig. \ref{fig.eLRA}. The complex behaviour is explained in Fig. \ref{fig.eLRC1_C2_condKappa}, where the joint PDFs for $e^{(LR)}_{C|A}$ and $e^{(LR)}_{C,B|A}$ in regions 1 and 2 are further conditioned on the sign for $\kappa_{Q}^{(S)}$. The relatively rare states where $\kappa_{Q}^{(S)} > 0$ (the bottom row of panels) have a weaker contribution from the straining of $\mathsf{C}$. This is reflected by the PDFs in regions 1 and 2 exhibiting little variance on the $e^{(LR)}_{C|A}$ axis and with a typical value for $e^{(LR)}_{C,B|A}$ with a sign opposite to that for $\mbox{R}_{A}$. However, when $\kappa_{Q}^{(S)} < 0$ (top row of panels), there is a dominance of the interaction term in region 2 and the non-normal term in region 1. Hence, in region 2, the strong tendency for the $e^{(LR)}_{C,B|A}$ distribution to dominate and to peak close to +1 for $\kappa_{Q}^{(S)} > 0$ is reinforced by the situation for the $\kappa_{Q}^{(S)} < 0$ case. In contrast, the two states act orthogonally in region 1, giving the 90\degree bend to the joint PDF, with the maximum along $e^{(LR)}_{C,B|A} = 0, e^{(LR)}_{C|A} > 0$ due to the mere $\sim 8\%$ of cases where $\kappa_{Q}^{(S)} > 0$.

The $\mbox{Q}_{A}-\mbox{R}_{A}$ diagram exhibits a concentration of values in regions 2 and 6 as shown in Table \ref{table.Q_wws_sss}. We have already seen in Fig. \ref{fig.eLRB_C_CB_regions} that the tendency to disc-like structures in region 6 is driven firstly by $e^{(LR)}_{B|A}$ and then by the combined interaction of all terms. Figure \ref{fig.eLRB_C_CB_regions} shows that region 2 also contributes effectively to this tendency to form disc-like structures, but is about $e^{(LR)}_{C,B|A}$ solely, or its interaction with $e^{(LR)}_{C|A}$ rather than the normal term.

In summary, there is a bias towards $\mbox{R}_{A} > 0$ states in the $\mbox{Q}_{A}-\mbox{R}_{A}$ diagram as shown in Table \ref{table.Q_wws_sss} and our approach highlights that the organization of the straining part of the normal tensor is constrained to be greater or equal to zero on this half of the diagram. However, this bias (54.6:45.4) is insufficient to explain the strong tendency to form disc-like structures. Our decomposition shows that this state emerges in a variety of different ways in different parts of $\mbox{Q}_{A}-\mbox{R}_{A}$ space as summarized in Table \ref{table.LRsummary}, and that the overall behaviour is dominated by the normal and interaction terms, with the non-normal term making, on average a weak contribution. However, in region 1, the non-normal term is of particular importance. It is also clear from both \ref{fig.eLRB_C_CB_regions} and \ref{fig.eLRB_C1_C2_regions} that our decomposition of this space into six regions is justified; while the topological analysis of this space prioritizes positive and negative values for $\Delta_{L}$ over the sign of $\mbox{Q}_{A}$, our results for the $\mbox{Q}_{A} > 0$ regions in particular are very different in nature.

\begin{table}
\caption{Summary of the behaviour of $e^{(LR)}_{A}$ in different regions of $\mbox{Q}_{A}-\mbox{R}_{A}$ space, based on our decomposition into its constitutive terms.}
\centering
\begin{tabular}{lccc}
\noalign{\smallskip}
Region & Normal term & Interaction term & Non-normal term\\
no. & ($e^{(LR)}_{B|A}$) & ($e^{(LR)}_{C,B|A}$) & ($e^{(LR)}_{C|A}$) \\
\noalign{\smallskip}
1 & 0 highly probable & -'ve values & +'ve values\\
2 & 0 highly probable & max. near +1 & with interaction \\
 & & & term gives +1\\
3 & maxima at 0 and -1; & maxima at 0 and +1 & 0 highly probable \\
 & coupling to +'ve values for & & \\
 & the sum of the other terms & &\\
4 & -'ve with weaker max. at 0 & +'ve with max. at 0 & 0 highly probable\\
5 & strong coupling to sum & 0 with negative tail & 0 highly probable\\
 & of other terms to give +1 & & \\
6 & tending to +1 & positive & essentially 0\\
\noalign{\smallskip}
\label{table.LRsummary}
\end{tabular}
\end{table}

\section{Results: Vorticity vector and strain eigenvector alignments}
\subsection{The existing relations for $\mathsf{A}$}
\begin{figure*}
  \includegraphics[width=0.9\textwidth]{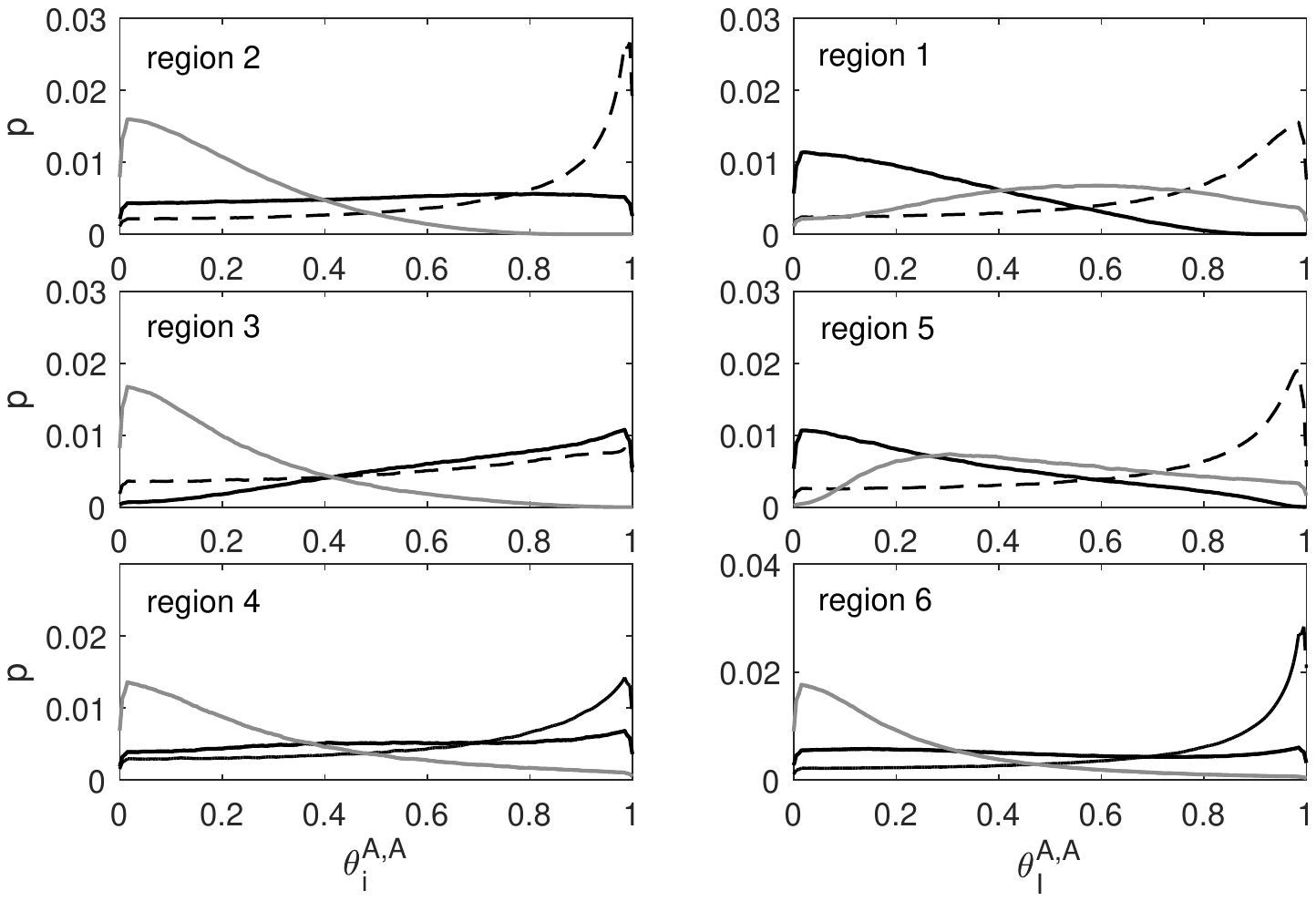}
\caption{The probability curves for the alignments, $\theta^{A,A}_{i}$ between the vorticity vector and the strain eigenvectors for $\mathsf{A}$. The black solid line is for $i =1$, the dashed line is for $i = 2$ and the grey line is for $i = 3$.
}
\label{fig.vsA}   
\end{figure*}

\begin{figure*}
  \includegraphics[width=0.8\textwidth]{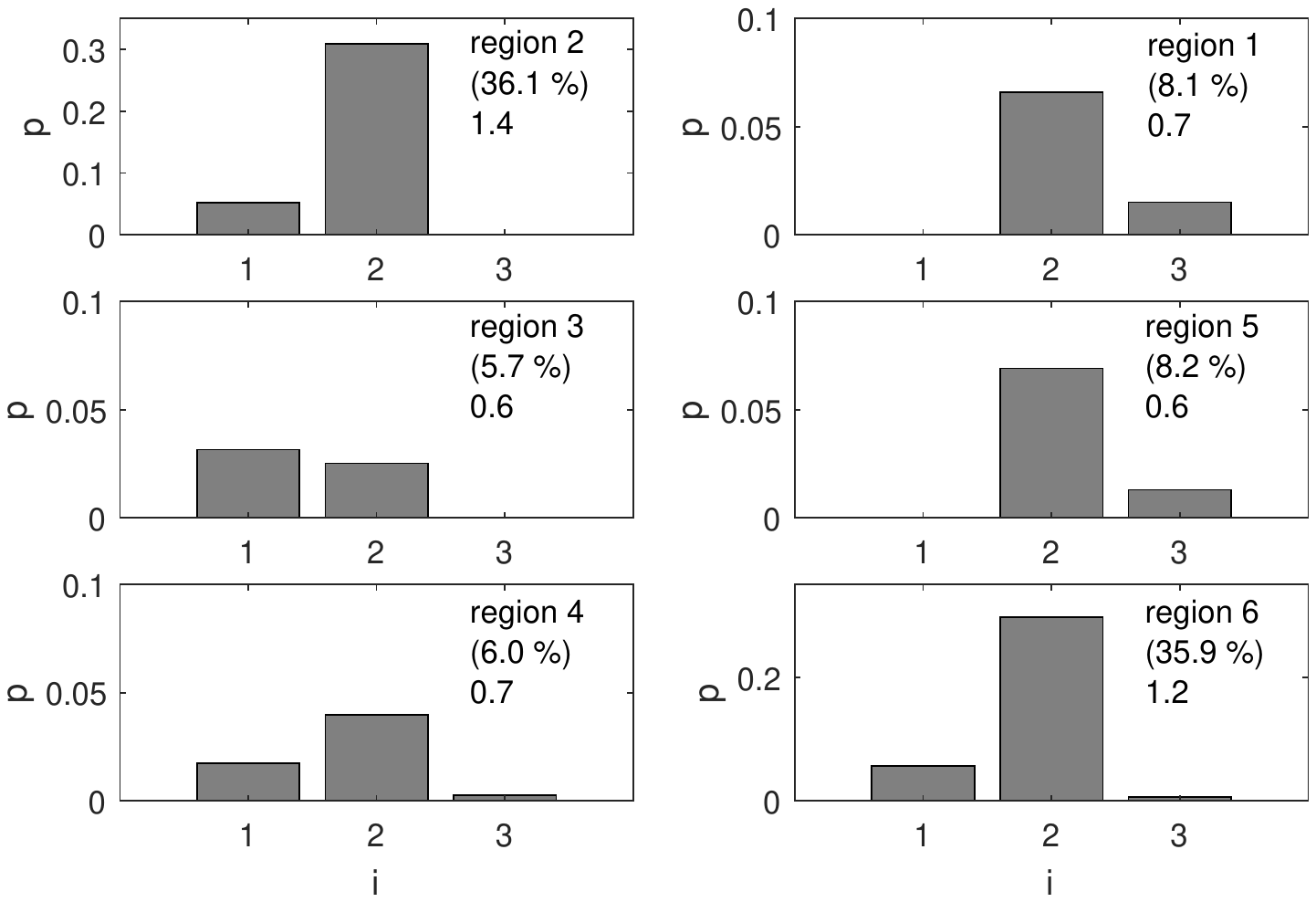}
\caption{Given $\theta^{A,A}_{i} > 0.985$, the bars show the probability of the alignments for particular \emph{i}.  The percentage value in brackets in each panel is the sum of the values in each panel. The number beneath this is the ratio between this percentage and that given for this region of the $\mbox{Q}_{A}-\mbox{R}_{A}$ diagram in Table \ref{table.Q_wws_sss}.
}
\label{fig.vvASA}   
\end{figure*}

As is well known, and is shown in Fig. \ref{fig.vsA}, $\boldsymbol{\omega}_{A}$ is typically most strongly aligned with $\mathbf{e}^{A}_{2}$. If we extract the cases where $\theta^{A,A}_{i} > 0.985$ (i.e. $\pm 10^{\circ}$), we obtain the results shown in Fig. \ref{fig.vvASA}. What is particularly notable in this figure is the high proportion of $\theta^{A,A}_{i} > 0.985$ occurrences in regions 2 and 6 ($\sim 36\%$ in both cases), the two regions that HIT occupies preferentially relative to random, synthetic tensors with appropriate bounds on their non-normality \citep{k17}. This relative occupancy is 1.4 times, and 1.2 times higher than is the case for all tensors (without conditioning on region of occurrence) as shown in Fig. \ref{fig.vvASA}. Overall, 81\% of cases where $\theta^{A,A}_{i} > 0.985$ were for $\theta^{A,A}_{2}$, with 16\% for $\theta^{A,A}_{1}$, and the propensity for alignment with $\theta^{A,A}_{1}$ and $\theta^{A,A}_{3}$ dictated by the sign of $\mbox{R}_{A}$ where $\Delta_{L} > 0$ and dominated by $\theta^{A,A}_{1}$ below the discriminant function. Region 3 is the only part of the $\mbox{Q}_{A}-\mbox{R}_{A}$ diagram where $\theta^{A,A}_{2}$ alignments are secondary.

\begin{figure*}
  \includegraphics[width=0.9\textwidth]{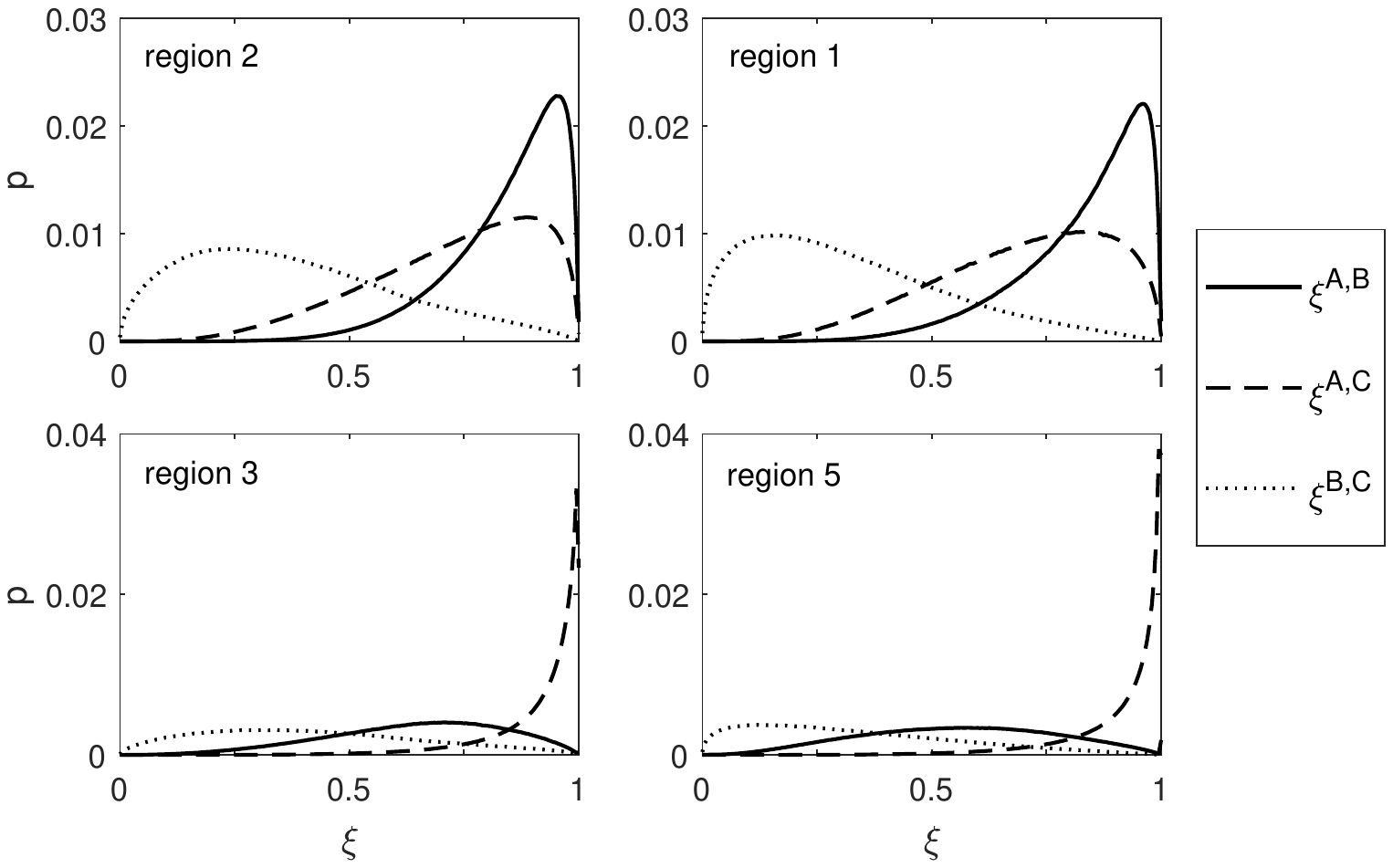}
\caption{The probability of attaining different values for $\xi$ as a function of the four regions of the $\mbox{Q}_{A}-\mbox{R}_{A}$ diagram where $\boldsymbol{\omega}^{B}$ is non-zero
}
\label{fig.vv}   
\end{figure*}

\begin{figure*}
  \includegraphics[width=0.7\textwidth]{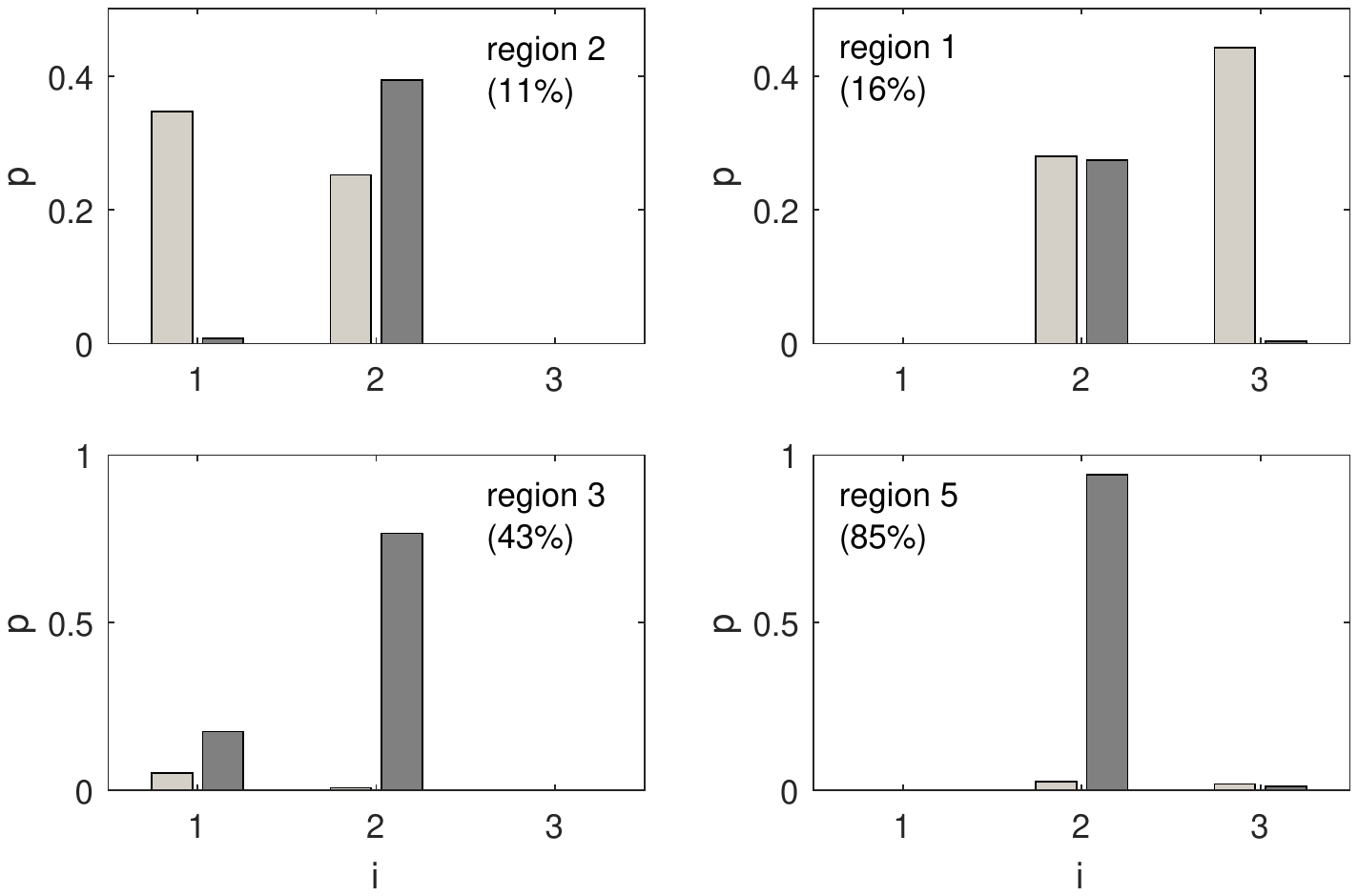}
\caption{Given $\theta^{A,A}_{i} > 0.985$ and that $\boldsymbol{\omega}_{B}$ exists, results are shown for cases where $\xi^{A,B} > 0.985$ (light grey) and $\xi^{A,C} > 0.985$ (grey) for the various $i$ in the $\theta^{A,A}_{i}$ alignment. The values in each panel sum to 1 and the percentage value in each panel gives the relative frequency that $\theta^{A,A}_{i} > 0.985$ leads to $\xi^{A,B} > 0.985$ or $\xi^{A,C} > 0.985$.
}
\label{fig.vAvB_vAvC_vvASA}   
\end{figure*}

\subsection{Vorticity-vorticity alignments}
The decomposition into components derived from $\mathsf{B}$ and $\mathsf{C}$ provides insights into the $\theta^{A,A}_{i}$ alignments discussed above. First, we examine the mutual vorticity vector alignments, above the discriminant function (because below it, $\xi^{A,C} = 1$ everywhere). Figure \ref{fig.vv} shows that immediately above the discriminant function, a strong tendency for the vorticity vector for $\xi^{A,C} \to 1$ is retained. However, when $\mbox{Q}_{A} > 0$ this is a weaker effect, with a stronger alignment for $\xi^{A,B}$. The accompanying curves for $\xi^{B,C}$ show that there is very limited strong alignment between $\boldsymbol{\omega}^{B}$ and $\boldsymbol{\omega}^{C}$, meaning there are two distinct sets of vorticity vectors in the positive $\mbox{Q}_{A}$ regions depending on these alignments.

When we condition these distributions on $\theta^{A,A}_{i} > 0.985$ (i.e. $\pm 10^{\circ}$) and then seek cases where $\xi^{A,B} > 0.985$ and $\xi^{A,C} > 0.985$, we obtain the results seen in Fig. \ref{fig.vAvB_vAvC_vvASA}. The percentages in each panel indicate the efficiency by which strong $\theta^{A,A}_{i}$ alignments lead to strong $\xi^{A,B}$ or $\xi^{A,C}$ alignments. Given that such values are at 100\% for $\xi^{A,C}$ where $\Delta_{L} < 0$, there is a clear decrease in this propensity as $\mbox{Q}_{A}$ increases. However, there is a major difference between region 5, which adjoins the Vieillefosse tail, and region 3, with the former approximately twice as effective at retaining strong alignments between the vorticity vectors for $\mathsf{A}$ and $\mathsf{C}$. 

It is clear from Fig. \ref{fig.vv} that in regions 3 and 5, $\xi^{A,C}$ dominates the strong alignments and this is clearly the case in Fig. \ref{fig.vAvB_vAvC_vvASA}, not only for where $\theta^{A,A}_{2} > 0.985$, but also for where $\theta^{A,A}_{1} > 0.985$ in region 3. However, an important contrast between Fig. \ref{fig.vAvB_vAvC_vvASA} and Fig. \ref{fig.vvASA} can be detected in region 3 in that the number of significant alignments for  $\xi^{A,B}$ and $\xi^{A,C}$ is far fewer for cases where $\theta^{A,A}_{1} > 0.985$, despite these cases being more numerous in Fig. \ref{fig.vvASA}. Hence, the observed $\theta^{A,A}_{1}$ alignments in region 3, and the low percentages in regions 1 and 2 show that mutual vorticity alignment is a less important explanation for the $\theta^{A,A}_{i}$ alignments than is the case in regions 4, 5, and 6. 

In regions 1 and 2 of Fig. \ref{fig.vAvB_vAvC_vvASA} we see that any degree of excess for $\theta^{A,A}_{2}$ compared to $\theta^{A,A}_{1}$ (region 2) or $\theta^{A,A}_{3}$ (region 1) is much less than is seen in Fig. \ref{fig.vvASA} - the opposite scenario to region 3. It is also the case that $\xi^{A,B}$ alignments completely dominate for $\theta^{A,A}_{1}$ and $\theta^{A,A}_{3}$ alignments. Thus, while mutual vorticity alignments are relatively rare where $\mbox{Q}_{A} > 0$, there is a clear difference regarding the types of alignments structures that occur: alignment between $\boldsymbol{\omega}_{A}$ and the largest or smallest strain eigenvector retains structure between $\boldsymbol{\omega}_{A}$ and $\boldsymbol{\omega}_{B}$ - it is a consequence of local interactions. In contrast, while local interactions are still relevant for alignments with the intermediate strain eigenvector, it is here than non-local interactions, as signified by $\xi^{A,C}$ alignments, are most important.

\begin{figure*}
  \includegraphics[width=0.85\textwidth]{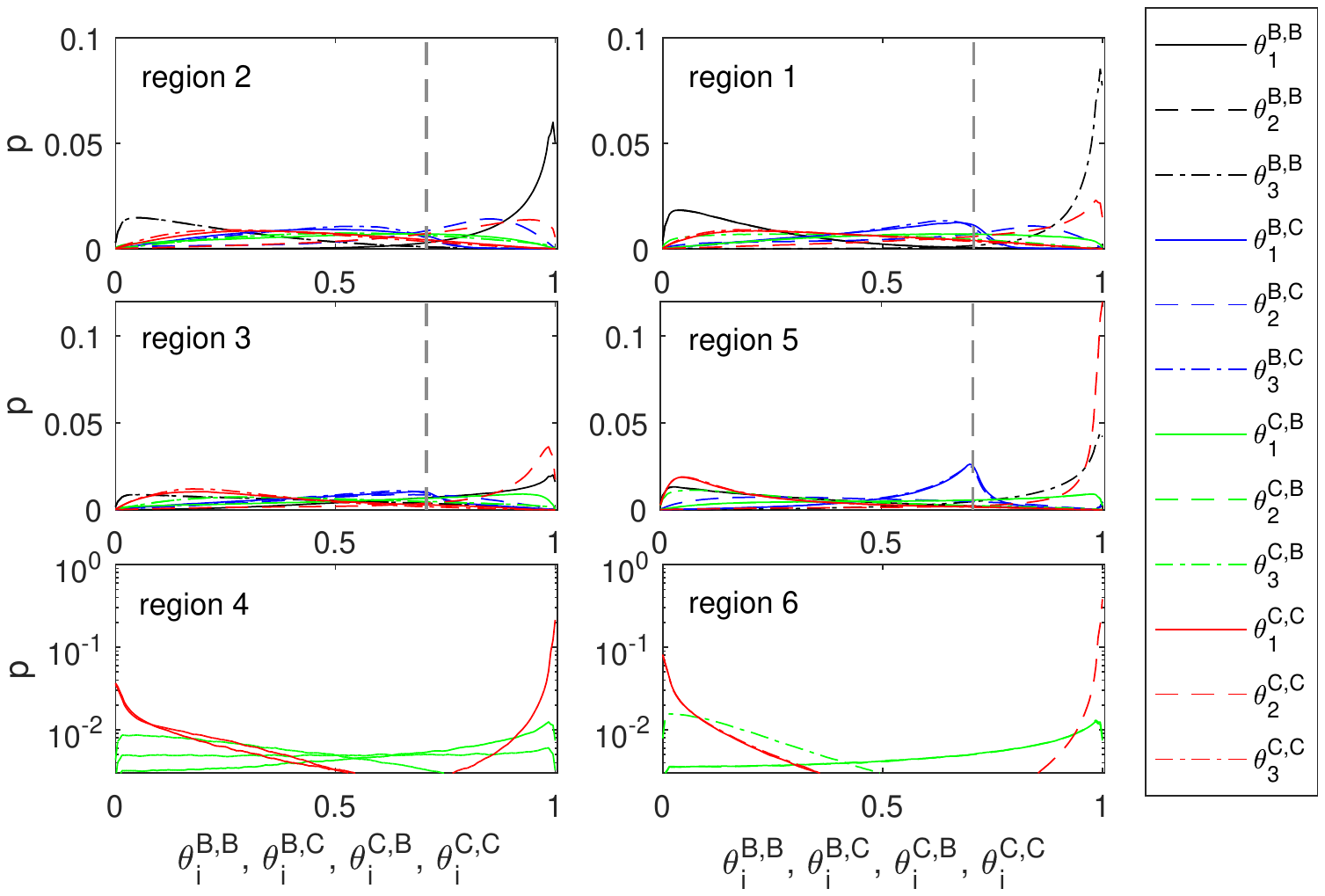}
\caption{The probability curves for the alignments, $\theta^{B,B}_{i}$ (black), $\theta^{B,C}_{i}$ (blue), $\theta^{C,B}_{i}$ (green) and $\theta^{C,C}_{i}$ (red) between the vorticity vector and the strain eigenvectors. The solid lines are for $i =1$, the dashed lines are for $i = 2$ and the dot-dashed lines are for $i = 3$. The vertical, dashed grey lines indicate an angle of $\pm 45^{\circ}$.
}
\label{fig.vsBC}   
\end{figure*}

\begin{figure*}
  \includegraphics[width=0.85\textwidth]{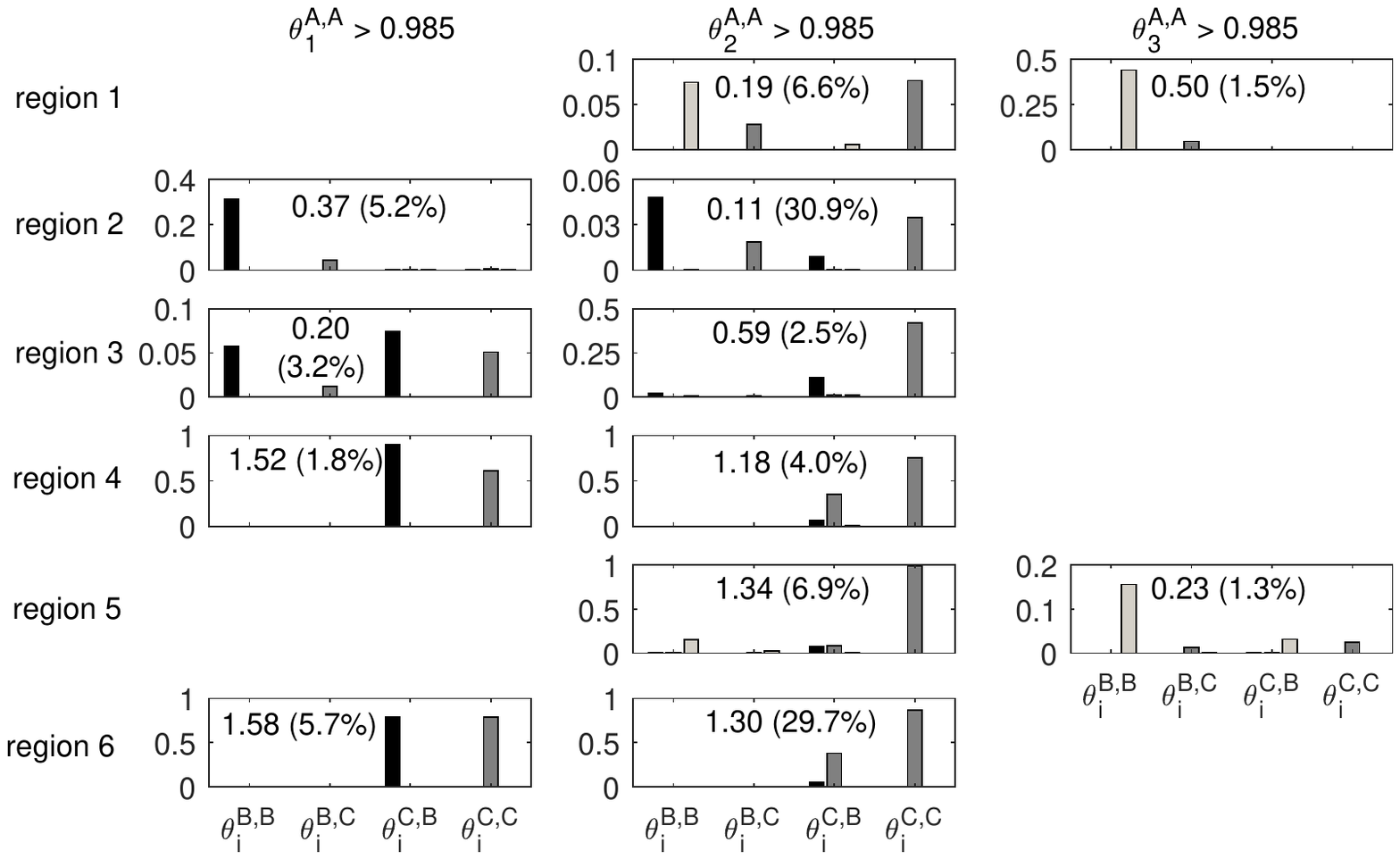}
\caption{Given $\theta^{A,A}_{i} > 0.985$, results are shown for cases where any of $\theta^{B,B}_{i}$, $\theta^{B,C}_{i}$, $\theta^{C,B}_{i}$ and $\theta^{C,C}_{i}$ exceed 0.985 as a function of the strain eigenvector that dictates the $\theta^{A,A}_{i} > 0.985$ state. Values for $i \in \{1, 2, 3\}$ are shown in black, grey and light grey, respectively. The number in each panel is the total of the values in that panel. Missing panels are those where the occurrences are less than 1\% of the total $\theta^{A,A}_{i} > 0.985$ alignments, as determined from Fig. \ref{fig.vvASA}. The percentages in each panel are the proportion of $\theta^{A,A}_{i} > 0.985$ alignments represented by each panel. Hence, the sum of the percentages in each row equates to the percentage in Fig. \ref{fig.vvASA}, minus the small values not included in this figure, which are in regions 4 and 6.
}
\label{fig.vBCSBC_vvASA}   
\end{figure*}

\subsection{Vorticity and strain alignments for $\mathsf{B}$ and $\mathsf{C}$}
Figure \ref{fig.vsBC} shows the distributions for $\theta_{i}^{B,B}$, $\theta_{i}^{B,C}$, $\theta_{i}^{C,B}$, and $\theta_{i}^{C,C}$, and there are several points of interest:
\begin{itemize}
\item Strong alignments for particular terms are more probable where $\mbox{R}_{A} > 0$;
\item Because of the structure of $\mathsf{B}$, for $\Delta_{L} > 0$, $\theta^{B,B}_{1}$ drives the alignments for $\mbox{R}_{A} < 0$ and $\theta^{B,B}_{3}$ where $\mbox{R}_{A} > 0$;
\item The $\theta^{C,C}_{2}$ alignment is so dominant in regions 4 and 6 that probabilities are shown on a logarithmic axis;
\item Strong alignments for $\theta^{C,C}_{2}$ and $\theta^{C,B}_{1}$ for $\Delta_{L} < 0$ are replaced by strong alignments for either $\theta^{B,B}_{1}$ or $\theta^{B,B}_{3}$, and also $\theta^{C,C}_{2}$ where $\Delta_{L} > 0$ and $\mbox{Q}_{A} < 0$ and then a clear dominance for $\theta^{B,B}_{1}$ or $\theta^{B,B}_{3}$ where $\mbox{Q}_{A} > 0$;
\item Where they are defined, $\theta^{B,C}_{1}$ and $\theta^{B,C}_{3}$ have clear maxima at $\pm 45^{\circ}$; if this is rather strictly the case as happens in region 5, then the mode for $\theta^{B,C}_{2} \to 0$, while in the other regions the mode for $\theta^{B,C}_{2} \to 1$;
\item In all regions, the modes for $\theta^{C,B}_{3}$ and $\theta^{C,C}_{1}$ are closer to orthogonal rather than aligned.
\end{itemize}

Results regarding strong alignments for these cases conditioned on strong alignments for $\theta^{A,A}_{i}$ are given in Fig. \ref{fig.vBCSBC_vvASA}, where the numbers in each panel are the sum of the values in that panel and the percentages relate directly to those in Fig. \ref{fig.vvASA}: the sum of the percentages in a row of Fig. \ref{fig.vBCSBC_vvASA} equals the percentages in Fig. \ref{fig.vvASA}, with the exception of regions 4 and 6, where the very small contributions from strong $\theta_{3}^{A,A}$ alignments have been excluded. It is notable that the sum of the values in the panels for regions 4 to 6 tend to exceed 1.0. That is, for a given tensor more than one vorticity vector - strain eigenvector pair is strongly aligned. It is also the case that in these regions, nearly all tensors exhibit a strong alignment for $\theta^{C,C}_{2}$ when $\theta^{A,A}_{2} > 0.985$. Given that $\xi^{A,C} = 1$ for regions 4 and 6 and the majority of alignments arise for $\theta^{A,A}_{2} > 0.985$, we have a physical explanation for the second eigenvector alignment here: it is driven by non-normal vorticity alignment, coupled to a strong alignment between the second eigenvector of the non-normal strain tensor and the second eigenvector for $\mathsf{S}_{A}$ with co-alignment with the second eigenvector for $\mathsf{S}_{B}$ also arising in a number of cases. Thus, while the analysis in the previous sections has highlighted the weak effects of non-normality in region 6, in particular, regarding strain tensor effects, because there is no normal rotation tensor in this region, non-normality plays a critical part in enstrophy production because $\xi^{A,C} = 1$.

In regions 1 and 2, there are not only fewer cases where $\theta^{A,A}_{2} > 0.985$ compared to $\theta^{A,A}_{3} > 0.985$ (region 1) and $\theta^{A,A}_{1} > 0.985$ (region 2), but the conversion rate of an $\theta^{A,A}_{i} > 0.985$ alignment to one for another vorticity vector - strain eigenvector is more efficient for $\theta^{A,A}_{1}$ and $\theta^{A,A}_{3}$ at 37\% and 50\%, respectively, compared to 19\% for $\theta^{A,A}_{2}$. For example, in region 2, $0.11 \times 30.9 = 3.4$\% for $\theta^{A,A}_{2} > 0.985$ is less than double $0.37 \times 5.2 = 1.9$\% for $\theta^{A,A}_{2} > 0.985$, despite the six times higher proportion of $\theta^{A,A}_{2} > 0.985$ instances. For $\theta^{A,A}_{1}$ and $\theta^{A,A}_{3}$, the dominant alignment is for $\theta^{B,B}_{i}$, with the value for $i$ equating to that in the $\theta^{A,A}_{i}$ alignment. Similar alignments are also important where $\theta^{A,A}_{2} > 0.985$ in regions 1 and 2. Thus, normal vorticity-strain interactions generated by local effects are the most important where $\mbox{Q}_{A} > 0$ and the overall result of $\theta^{A,A}_{2}$ dominating the alignment structure is really (in terms of direct effects) a consequence of the high values for $\theta^{C,C}_{2}$ in region 6 where normal vorticity does not exist.

The only real role played by the interaction terms are the $\theta^{C,B}_{1}$ alignments seen in regions 3, 4, and 6 where $\theta^{A,A}_{1} > 0.985$ . From the bottom row of Fig. \ref{fig.vsBC} it can be seen that while the mode for this curve lies close to 1, the peak is an order of magnitude smaller than that for $\theta^{C,C}_{2}$ and it is clear that when $\theta^{A,A}_{2} > 0.985$ this term is of minor importance. Hence, the non-normal vorticity vector aligning with the leading strain eigenvector for $\mathsf{S}_{B}$ is important for $\theta^{A,A}_{1} > 0.985$, although that the sums of the alignments in regions 4 and 6 are much greater than 1 shows that there is a co-alignment between the first eigenvector of $\mathsf{S}_{B}$ and the second for $\mathsf{S}_{C}$ in these regions. Clearly, therefore, the co-alignment between $\mathbf{e}^{B}_{1}$ and $\mathbf{e}^{A}_{1}$ must be stronger than that for $\mathbf{e}^{C}_{2}$ and $\mathbf{e}^{A}_{2}$ to produce this result. 

\subsection{Strain-strain eigenvector alignments}

\begin{figure*}
  \includegraphics[width=0.9\textwidth]{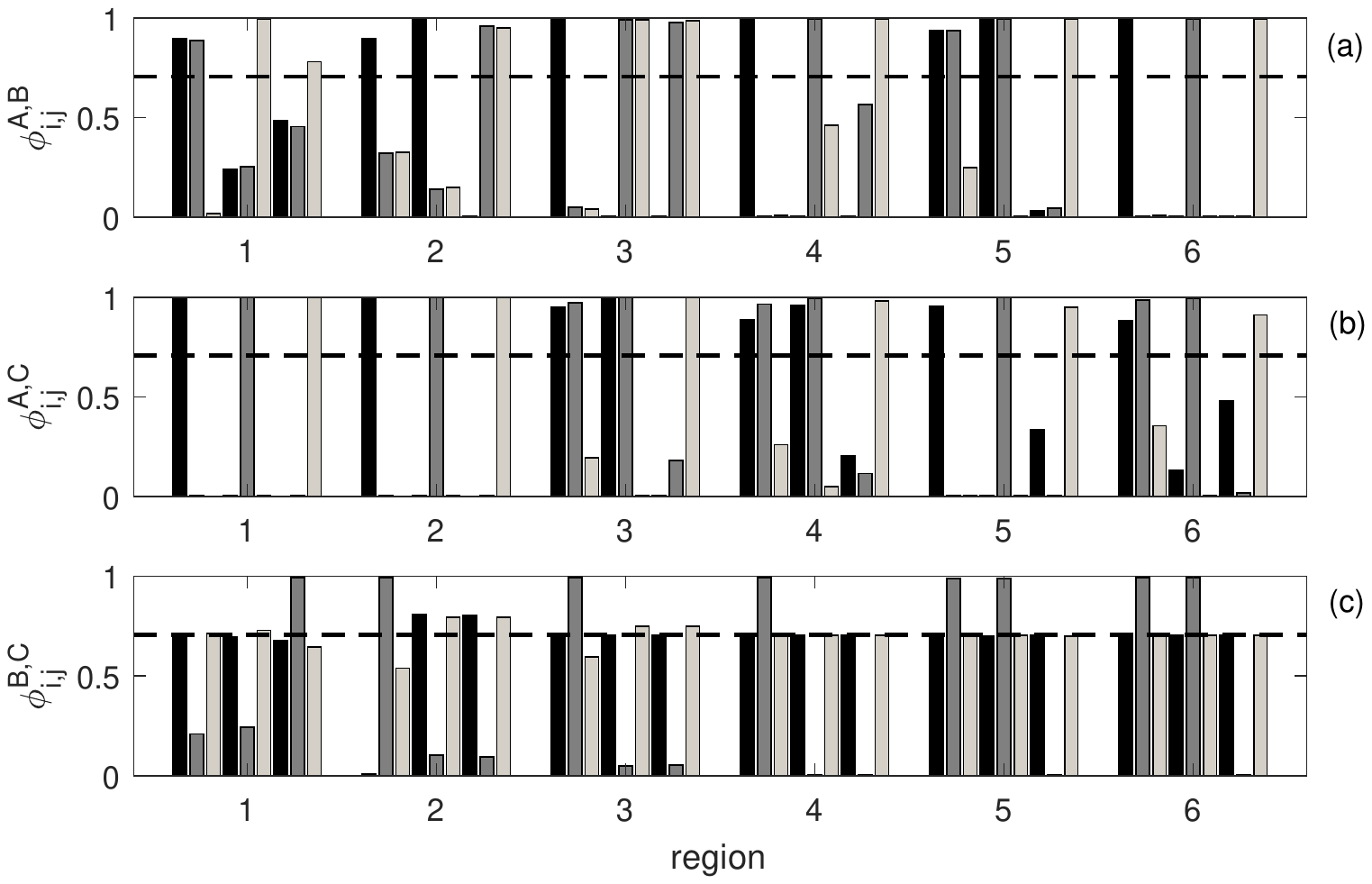}
\caption{The mode of the distributions for all nine combinations of $\{i,j\}$ for each of $\phi_{i,j}^{A,B}$, $\phi_{i,j}^{A,C}$, and $\phi_{i,j}^{B,C}$ and for each region of the   $\mbox{Q}_{A}-\mbox{R}_{A}$ diagram. The nine combinations in each region are coloured in three blocks (black, then grey, then near-white) for $i = \mbox{constant}, j$, with $i$ given by the position of each block ($i = 1$ is the leftmost block for each region and $i = 3$ is the right-most block) The dashed line in each panel is at $\sqrt{2} / 2$, i.e. $\pm 45^{\circ}$. 
}
\label{fig.ssmode}   
\end{figure*}

\begin{figure*}
  \includegraphics[width=0.85\textwidth]{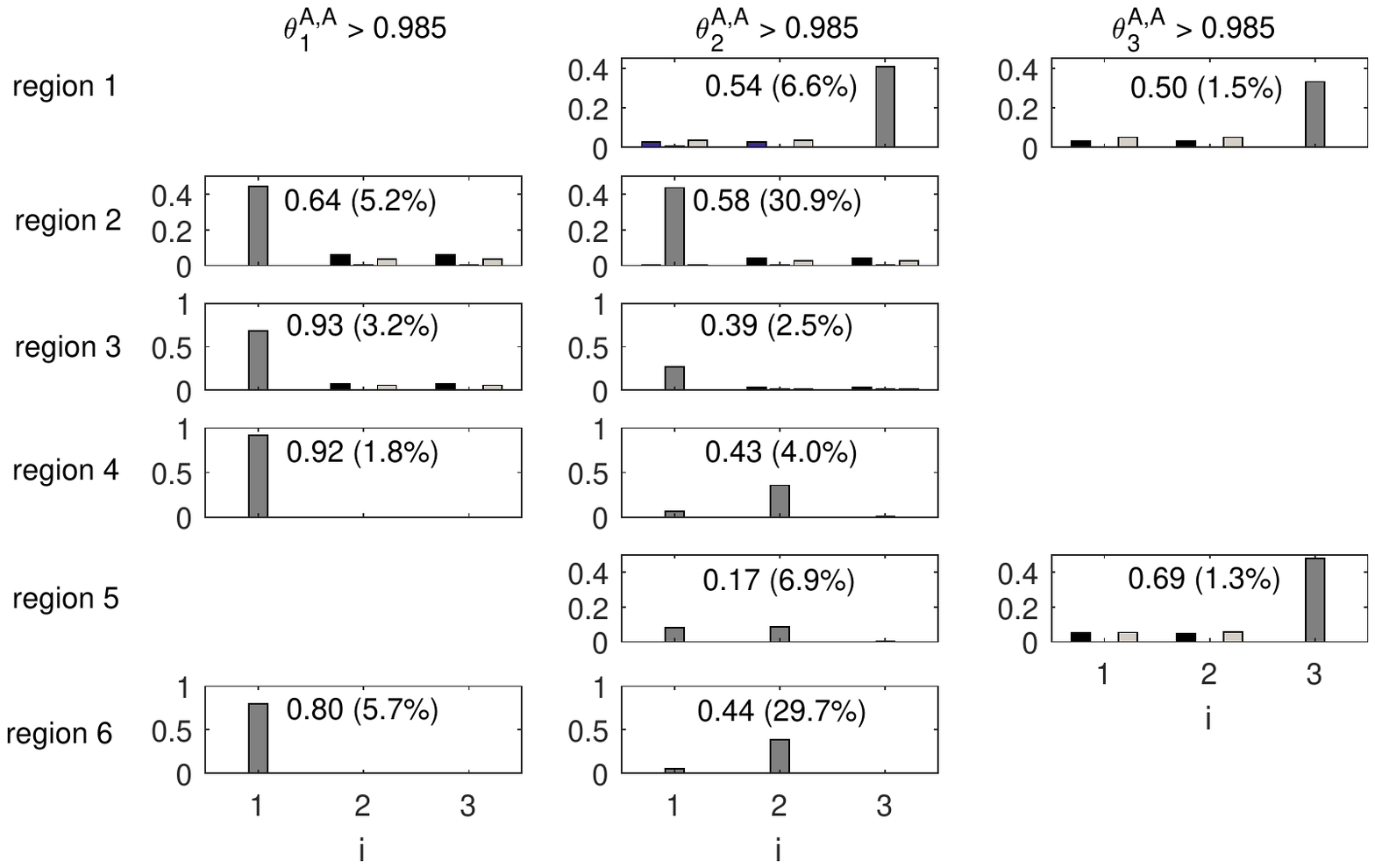}
\caption{Given $\theta^{A,A}_{i} > 0.985$, results are shown for where any $\phi^{B,C}_{i,j} > 0.985$ as a function of the strain eigenvector that dictates the $\theta^{A,A}_{i} > 0.985$ state. Values for $i$ are given by each group of bars, while $j \in \{1, 2, 3\}$ are shown in black, grey and light grey, respectively. The number in each panel is the total for the values in that panel.
}
\label{fig.SBC_vvASA}   
\end{figure*}

We can observe clear structure in the mutual strain alignments, $\phi_{i,j}^{A,B}$, $\phi_{i,j}^{A,C}$, and $\phi_{i,j}^{B,C}$, which are summarized by the modes for each distribution function in Fig. \ref{fig.ssmode}. In essence, regions 1 and 2 exhibit a perfect alignment between $\mathbf{e}^{A}_{i}$ and $\mathbf{e}^{C}_{i}$, while for regions 4 and 6 there is a close to perfect alignment between $\mathbf{e}^{A}_{i}$ and $\mathbf{e}^{B}_{i}$, that is particularly the case for the latter. In region 3 we see a mutual alignment for $\phi_{3,3}^{A,C}$, but with the other two eigenvectors aligning with each other such that different tensors have strong $\phi_{1,1}^{A,C}$, $\phi_{1,2}^{A,C}$, $\phi_{2,1}^{A,C}$ and $\phi_{2,2}^{A,C}$ alignments. This behaviour is inverted in the $\phi_{i,j}^{A,B}$ results, with a strong alignment for $\phi_{1,1}^{A,B}$, and different tensors having strong $\phi_{2,2}^{A,B}$, $\phi_{2,3}^{A,B}$, $\phi_{3,2}^{A,B}$ and $\phi_{3,3}^{A,B}$ alignments. Region 5 might be expected to be the mirror image of region 3 and this is approximately the case for the $\phi_{i,j}^{A,B}$ results. However, the $\phi_{i,j}^{A,C}$ results are much more similar to those for regions 1 and 2 with alignments at $\phi_{i = j}^{A,C}$.

The consequences of these alignments between the $\mathbf{e}^{A}_{i}$ and $\mathbf{e}^{B}_{i}$ and $\mathbf{e}^{C}_{j}$ strain tensors for the mutual relation between the $\mathbf{e}^{B}_{i}$ and $\mathbf{e}^{C}_{j}$ is shown in the bottom panel of Fig. \ref{fig.ssmode}. In region 4, $\phi_{1,2}^{B,C}$ is aligned well with the other $\mathbf{e}^{C}_{j}$ ($i \in \{1,3\}$) at $45^{\circ}$ to the $\mathbf{e}^{B}_{i}$. A similar result exists for the other regions where $\mbox{R}_{A} < 0$ (regions 2 and 3), although it is less coherent, with a weaker attraction to $45^{\circ}$ and (small) residual orientations for $\phi_{2,2}^{B,C}$ and $\phi_{3,2}^{B,C}$. Region 1 is, in turn, similar to these regions with the exception that, as a consequence of $\mbox{R}_{A} > 0$, the dominant alignment is $\phi_{3,2}^{B,C}$ not $\phi_{1,2}^{B,C}$. Regions 5 and 6 are very similar to one another: As with region 4, a mode at $45^{\circ}$ is strongly expressed, but in contrast, rather than just a strong alignment for $\phi_{1,2}^{B,C}$, two discrete sets of points exist with strong alignments for both $\phi_{1,2}^{B,C}$ and $\phi_{2,2}^{B,C}$.

Similar results to these but conditioned on $\theta^{A,A}_{i}$ alignments are shown in Fig. \ref{fig.SBC_vvASA}. In terms of the $\theta^{A,A}_{2} > 0.985$ cases in the central column our first result is that the mutual strain alignments are much more important in regions 1 and 2 (0.54 and 0.58 of 6.6\% and 30.9\%) than the vorticity-strain alignments in Fig. \ref{fig.vBCSBC_vvASA} (0.19 and 0.11 of 6.6\% and 30.9\%), with the opposite the case in regions 3 to 6 where $\mbox{Q}_{A} < 0$. Hence, we have the somewhat counterintuitive result that in enstrophy dominant regions, mutual strain alignments are of particular prominence for explaining vorticity-strain alignments. However, the reason is clearly a combination of $\xi^{A,C} = 1$ for $\Delta_{L} < 0$, while vorticity-vorticity alignments are more complex in regions 1 and 2 (Fig. \ref{fig.vv}), as well as the mutual alignments for $\phi^{A,C}_{i = j}$ for regions 1 and 2 in Fig. \ref{fig.ssmode}b. 

For the $\theta^{A,A}_{2} > 0.985$ cases we see that $\mathbf{e}^{B}_{3}$ and $\mathbf{e}^{C}_{2}$ are aligned in region 1 and $\mathbf{e}^{B}_{1}$ and $\mathbf{e}^{C}_{2}$ in regions 2 and 3. In regions 4 to 6, two sets of points arise with either alignments between $\mathbf{e}^{B}_{1}$ and $\mathbf{e}^{C}_{2}$ or $\mathbf{e}^{B}_{2}$ and $\mathbf{e}^{C}_{2}$. Such results for $\theta^{A,A}_{2} > 0.985$ are consistent with the patterns seen in Fig. \ref{fig.ssmode}c, with the exception of region 4. However, in regions 3 to 6 in particular, mutual strain alignments are more clearly expressed for the $\theta^{A,A}_{1} > 0.985$ and $\theta^{A,A}_{3} > 0.985$ cases, and here the results are readily explained by the sign of $\mbox{R}_{A}$, as they are for the $\theta^{A,A}_{2} > 0.985$ situation for regions 1 and 2: where $\mbox{R}_{A} > 0$, $\mathbf{e}^{B}_{3}$ and $\mathbf{e}^{C}_{2}$ are aligned; while, where $\mbox{R}_{A} < 0$, $\mathbf{e}^{B}_{1}$ and $\mathbf{e}^{C}_{2}$ are aligned.

\subsection{Summary}
\label{sect.alignsumm}
\begin{figure*}
  \includegraphics[width=0.85\textwidth]{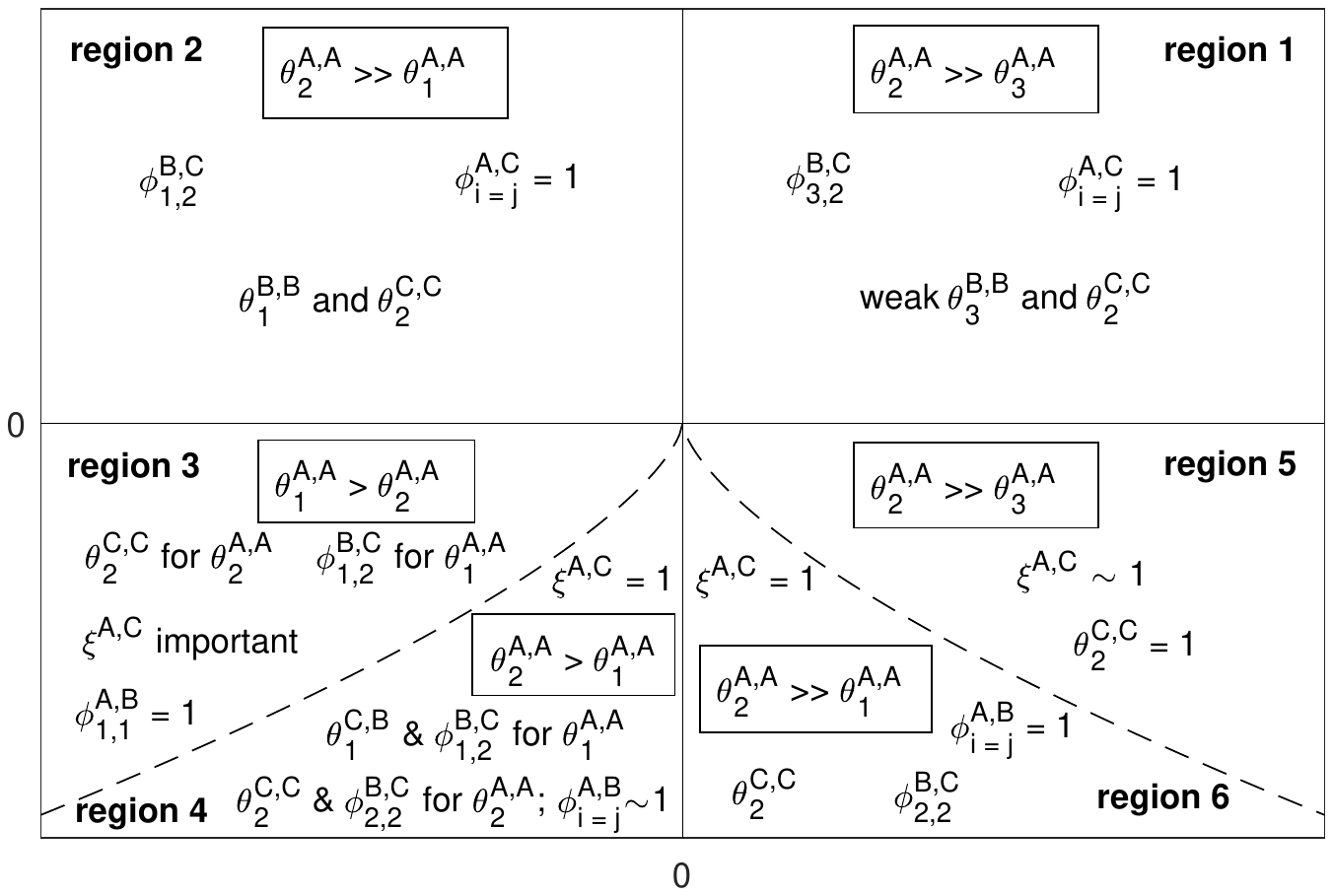}
\caption{A summary of the manner in which vorticity and strain alignments for $\mathsf{A}$, i.e. $\theta^{A,A}_{i}$ are generated in different regions of the $\mbox{Q}_{A} - \mbox{R}_{A}$ diagram. The result in the box with a border indicates how many alignments are considered in each region; if $\theta^{A,A}_{2} \gg \theta^{A,A}_{1}$, for example, then only the former is considered, while if $\theta^{A,A}_{2} > \theta^{A,A}_{1}$ both are considered.
}
\label{fig.sketch1}   
\end{figure*}
A summary of the results in this section regarding the nature of the $\theta^{A,A}_{2}$ alignment, in particular, is provided in Fig. \ref{fig.sketch1}. As noted in the caption for this figure, the description for each region concerns only one alignment if it much more frequent than the next most common, i.e. $\theta^{A,A}_{2} \gg \theta^{A,A}_{1}$, while both are described if the difference in relative frequencies is more minor, as arises in regions 3 and 4. All told 81\% of the $\theta^{A,A}_{i} > 0.985$ alignments were for $\theta^{A,A}_{2}$, with, in turn, 38\% and 37\% of these instances occuring in regions 2 and 6, respectively. Thus, focusing on these two regions in the first instance:
\begin{itemize}
\item Region 2 - All three strain eigenvectors for $\mathsf{A}$ are aligned very closely with those with the same order for $\mathsf{C}$. There is then two ways that the $\theta^{A,A}_{2} > 0.985$ state is generated: a direct route resulting from this strain eigenvector state combined with the $\theta^{C,C}_{2}$ alignment; and, an indirect route where the $\theta^{B,B}_{1}$ alignment and the $\phi^{A,C}_{i = j} = 1$ state, combines with the alignment between the leading eigenvector for $\mathsf{S}_{B}$ and the intermediate eigenvector for $\mathsf{S}_{C}$. It should be noted that at the end of section \ref{sect.inv2} it was suggested based on the results shown in Fig. \ref{fig.kappaQ} that, in effect, $\theta^{B,B}_{1}$ would be an important alignment in region 2 (and $\theta^{B,B}_{3}$ in region 1), which has been shown in the subsequent analysis;
\item Region 6 - Beneath the discriminant function $\xi^{A,C} = 1$ meaning that the alignment here is driven by the mutual vorticity vector alignment and, again, the $\theta^{C,C}_{2}$ alignment. an indirect path is also possible here as a consequence of $\phi^{A,B}_{i = j} = 1$ and the mutual alignment of the intermediate strain eigenvectors for $\mathsf{S}_{B}$ and $\mathsf{S}_{C}$, $\phi^{B,C}_{2,2}$.
\end{itemize}
These results help explain those in the remaining two regions on the positive $\mbox{R}_{A}$ side of the $\mbox{Q}_{A} - \mbox{R}_{A}$:
\begin{itemize}
\item Region 1 - As with region 2, region 1 has $\phi^{A,C}_{i = j} = 1$, and here the indirect route to alignment is of the opposite sense (via  $\theta^{B,B}_{3}$ and $\phi_{3,2}^{B,C}$). However, fewer events are generated because of weaker alignments for $\theta^{B,B}_{3}$ and $\theta^{C,C}_{2}$, reflecting the more unstable nature of the topology in this region \citep{chong90};
\item Region 5 - The alignment structure in region 5 is driven by similar mechanisms as region 6 with the primary being difference that $\xi^{A,C}$ is no longer set to 1. However, this slightly weaker mutual vorticity alignment is compensated by a more dominant $\theta^{C,C}_{2}$ alignment.
\end{itemize}

In regions 3 and 4 both $\theta^{A,A}_{1}$ and $\theta^{A,A}_{2}$ need to be considered, with the former actually dominant in region 3 (the only region where $\theta^{A,A}_{2}$ is of secondary importance):
\begin{itemize}
\item Region 3 - As with region 5, $\xi^{A,C}$ is still close to 1, although Fig. \ref{fig.vAvB_vAvC_vvASA} shows that this is of somewhat less importance for region 3. There is then the keenly expressed $\theta^{C,C}_{2}$ alignment as is found everywhere that  $\mbox{Q}_{A} < 0$ to generate the $\theta^{A,A}_{2}$ alignment. The $\theta^{A,A}_{1}$ alignments result more from mutual strain eigenvector alignments than vorticity-strain alignments such as $\theta^{C,C}_{2}$. In particular, there is a direct alignment, $\phi^{A,B}_{1,1} = 1$ and an indirect mechanism based on this alignment and those for $\phi^{B,C}_{1,2}$;
\item Region 4 - As with region 6, $\xi^{A,C} = 1$ and, combined with alignments for $\theta_{2}^{C,C}$ generates the $\theta^{A,A}_{2}$ structure in the typical way. A similar indirect mechanism is also possible because of $\phi^{A,B}_{i = j} \sim 1$ and the mutual alignment of the intermediate strain eigenvectors. The different aspect regarding the dynamics in this region is that the $\theta^{A,A}_{1}$ alignments are more strongly driven by the term, $\theta^{C,B}_{1}$ - the alignment between the vorticity vector for $\mathsf{C}$ and the leading strain eigenvector for $\mathsf{B}$ (a similar indirect mechanism to region 3 is also possible).
\end{itemize}

\section{Conclusion}
A radical interpretation of our approach to the additive decomposition of the velocity gradient tensor, $\mathsf{A}$ into normal and non-normal tensors, $\mathsf{B}$ and $\mathsf{C}$, is to state that while strain production takes place everywhere, enstrophy production does not occur beneath the discriminant function. Instead, from Table \ref{table.terms}, we have either the self-amplification of non-normality, or the normal straining of non-normality acting as a source of enstrophy production beneath the discriminant function. However, that these terms also act as an equal source for strain production means that in regions 4 and 6, (beneath the discriminant function) the third invariant for $\mathsf{A}$ ($\mbox{R}_{A}$ in our notation and simply $\mbox{R}$ conventionally) is just the strain production for $\mathsf{B}$.

As there is no normal rotation tensor in these regions, there is a perfect alignment between the vorticity vectors for $\mathsf{A}$ and $\mathsf{C}$, i.e. $\xi^{A,C} = 1$. From Fig. \ref{fig.vBCSBC_vvASA} we see that when $\theta^{A,A}_{2} > 0.985$, i.e. there is a strong alignment between the vorticity vector and the eigenvector for the intermediate strain eigenvalue for $\mathsf{A}$, there is an accompanying high frequency of occurrences where $\theta^{C,C}_{2} > 0.985$. The consequence of this ``non-normal'' alignment is important throughout all the $\mbox{Q}_{A}-\mbox{R}_{A}$ regions, but is crucial beneath the discriminant function where non-normality is the only source for enstrophy. This, coupled to the high proportion of the time that the flow spends in region 6 explains the alignment between vorticity and the intermediate eigenvector.

The importance of the non-local, non-normal contributions to the dynamics explains the difficulty of applying eigenvalue-focused thinking to the dynamics of complex systems such a turbulence. In a set of landmark papers in the 1990s it was shown that the difficulties of treating hydrodynamic stability in terms of the instability of the eigenvalues of the linearized problem could be overcome using pseudospectra \citep{reddy93,trefethen93}, aspects of which characterize the tensor non-normality \citep{trefethen05}. It was something of a surprise to fluid mechanics when the first direct numerical simulations of HIT showed a preferential alignment between the vorticity vector and the eigenvector for the second eigenvalue of the strain tensor rather than the first eigenvalue \citep{kerr85,ashurst87}. Our Schur decomposition-based approach to VGT dynamics clarifies this matter because much of the interesting dynamics resides in either the non-normal terms (the alignment between the non-normal vorticity and the second eigenvector of the non-normal straining) or is in the interaction between normal and non-normal effects (production by normal straining of the non-normal term). Clearly, these terms are not captured by the eigenvalues of $\mathsf{A}$, highlighting that, as with hydrodynamic stability, looking beyond the eigenvalues is crucial. The advantage of the Schur transform in this respect is that the non-normality, $\mathsf{N}$ is projected into the Schur matrix where it can be treated in a similar way to the eigenvalues, $\Lambda$. Hence, we could define $\mathsf{B} = \mathsf{U}\mathsf{\Lambda}\mathsf{U}^{*}$ and $\mathsf{C} = \mathsf{U}\mathsf{N}\mathsf{U}^{*}$. Thus, the rotations matrix, $\mathsf{U}$ retains a unitary form irrespective of the degree of non-normality. In contrast, the eigenvalue decomposition projects the non-normality into the departure of the eigenvectors from a unitary form, making it harder to directly compare normal (local) and non-normal (non-local) effects. 

It is only really along the Vieillefosse tail that normal effects, in particular, normal straining, dominate the dynamics. This helps explain the success of the restricted Euler model \citep{cantwell92} in approximating the dynamics of the VGT, but the more complex behaviour in region 1, in particular, demonstrates the difficulty of understanding the non-local effects that predominate without considering both the eigenvalues and the non-normal contributions. This work defines a suite of quantities that can serve as a new basis for evaluting existing models for the dynamics of the VGT \citep{wilczek14,johnson16}, and perhaps for formulating new models of this type. 

\appendix
\label{app.schur}
\section{Complex and real {S}chur transforms} 
The Schur transform \citep{schur1909} may be implemented in a complex or a real form, which we distinguish with bracketed superscripts \emph{c} and \emph{r} in this appendix:
\begin{align}
\notag{}
\mathsf{A} &= \mathsf{U}^{(c)}\mathsf{T}^{(c)}\mathsf{U}^{(c) *}\\
 &= \mathsf{U}^{(r)}\mathsf{T}^{(r)}\mathsf{U}^{(r) *}.
\end{align}
For the complex Schur decomposition or, equivalently, a real Schur decomposition of a tensor with real eigenvalues, the Schur matrix, $\mathsf{T}^{(c)} = \mathsf{\Lambda}^{(c)} + \mathsf{N}^{(c)}$, may be decomposed into a diagonal matrix of eigenvalues, $\mbox{diag}(\mathsf{\Lambda}^{(c)}) = \lambda_{1}, \ldots, \lambda_{3}$ and an upper triangular matrix, $\mathsf{N}^{(c)}$ that characterizes the non-normality of $\mathsf{A}$ \citep{GolubVanLoan13} as described in the main text of this paper. Hence, as we adopted the complex decomposition, we wrote (without the superscripts)
\begin{align}
\notag{}
\mathsf{A} &= \mathsf{B}^{(c)} + \mathsf{C}^{(c)}\\
\notag{}
\mathsf{B}^{(c)} &= \mathsf{U}^{(c)}\mathsf{\Lambda}^{(c)}\mathsf{U}^{(c) *}\\
\mathsf{C}^{(c)} &= \mathsf{U}^{(c)}\mathsf{N}^{(c)}\mathsf{U}^{(c) *}.
\label{eq.appenschur}
\end{align} 

However, if one wishes to impose that $\mathsf{B}, \mathsf{C} \in \Re$, but $\mathsf{A}$ has complex eigenvalues, then the real Schur transform should be adopted. In which case, $\mathsf{T}^{(r)}$ has a quasi-upper triangular, rather than triangular form, with the conjugate pair forming a $2 \times 2$ Jordan block, which complicates the decomposition in (\ref{eq.appenschur}).  In the case of a normal tensor, where $||\mathsf{N}^{(r)}|| = 0$, the eigenvalue-like tensor for the real decomposition, $\mathsf{\Lambda}^{(r)}$ is defined to be equal to the real Schur tensor, $\mathsf{T}^{(r)}$:
\begin{equation}
\mathsf{T}^{(r)} = \mathsf{\Lambda}^{(r)} = \biggl(\begin{smallmatrix} \Re(\lambda_{m,n}) & \mbox{sgn}(\chi)|\Im(\lambda_{m,n})| & 0\\ -\mbox{sgn}(\chi)|\Im(\lambda_{m,n})| & \Re(\lambda_{m,n}) & 0 \\ 0 & 0 & \lambda_{\ell}\end{smallmatrix} \biggr),
\label{eq.Schur_r}
\end{equation}  
where the subscripts, $m, n$ and $\ell$ indicate the positions of the conjugate pair eigenvalues and the real eigenvalue, respectively. The parameter $\chi$ ensures that the signs of the imaginary part are aligned correctly
\begin{equation}
\chi =  \biggl\{\begin{smallmatrix} 1 & \mbox{if} & T^{(r)}_{m,n} > T^{(r)}_{n,m}\\
-1 & \mbox{if} & T^{(r)}_{m,n} \le T^{(r)}_{n,m}\end{smallmatrix}
\end{equation}

We then may proceed as follows:
\begin{enumerate}
\item Determine the eigenvalues of $\mathsf{A}$ (a conjugate pair and a real-valued eigenvalue);
\item Perform a real-valued Schur decomposition, locate the Jordan block and denote these positions as $m$ and $n$ where $m < n$ and $m,n \in \{1,2,3\}$;
\item Complete the Jordan block, $\mathsf{J}$ and insert it, and $\lambda_{\ell}$, into the correct locations in a $3 \times 3$ zero matrix to form $\mathsf{\Lambda}^{(r)}$, as shown in (\ref{eq.Schur_r}); 
\item Define $\mathsf{N}^{(r)} = \mathsf{T}^{(r)} - \mathsf{\Lambda}^{(r)}$. We may then reconstruct using
\begin{align}
\notag{}
\mathsf{B}^{(r)} &= \mathsf{U}^{(r)}\mathsf{\Lambda}^{(r)}\mathsf{U}^{(r) *} \\
\mathsf{C}^{(r)} &= \mathsf{U}^{(r)}\mathsf{N}^{(r)}\mathsf{U}^{(r) *};
\end{align}
\end{enumerate}

With this form, while it is still true that $\mathsf{A} = \mathsf{S}^{(r)}_{B} + \mathsf{S}^{(r)}_{C} + \mathsf{\Omega}^{(r)}_{{B}} + \mathsf{\Omega}^{(r)}_{{C}}$ and $||\mathsf{S}_{{A}}||^{2} = ||\mathsf{S}^{(r)}_{{B}}||^{2} + ||\mathsf{S}^{(r)}_{{C}}||^{2}$, a difference emerges when one considers the additive property for $||\mathsf{\Omega}_{{A}}||^{2}$ and its constituent rotation tensors. Given the importance of enstrophy for the analyses in this paper, this explains why we adopt the complex variant of the transform throughout.

\begin{acknowledgements}
This research was supported by a Royal Academy of Engineering/Leverhulme Trust Senior Research Fellowship LTSRF1516-12-89 awarded to the author. 
\end{acknowledgements}


\end{document}